\documentclass[epj]{svjour}
\usepackage{graphics}
\usepackage{epsfig}
\title{Unitary Isobar Model - MAID2007}
\author{D. Drechsel\inst{1}, S. S.  Kamalov\inst{2}, L. Tiator\inst{1}}
\institute{Institut f\"ur Kernphysik, Universit\"at Mainz, D-55099 Mainz \and
JINR Dubna, 141980 Moscow Region, Russia }
\date{Received: date / Revised version: date}
\date{October 1, 2007}
\abstract{The unitary isobar model MAID2007 has been developed to
analyze the world data of pion photo- and electroproduction. The
model contains both a common background and several resonance
terms. The background is unitarized according to the K-matrix
prescription, and the 13 four-star resonances with masses below
2~GeV are described by appropriately unitarized Breit-Wigner
forms. The data have been analyzed by both single-energy and
global fits, and the transverse and longitudinal helicity
amplitudes have been extracted for the four-star resonances below
2~GeV. Because of its inherent simplicity, MAID2007 is well
adopted for predictions and analysis of the observables in pion
photo- and electroproduction.}
\begin{document}
\maketitle
\section{Introduction}
Our knowledge about the excitation spectrum of the nucleon was
originally provided by elastic pion-nucleon
scattering~\cite{Hohler79}. All the resonances listed in the
Particle Data Tables~\cite{PDG06} have been identified by
partial-wave analyses of this process with both Breit-Wigner and
pole extraction techniques. From such analyses we know the resonance
masses, widths, and branching ratios into the $\pi N$ and $\pi\pi N$
channels. These are reliable parameters for the resonances in the 3-
and 4-star tiers, with only few exceptions. In particular, there
remains some doubt about the structure of two prominent resonances,
the Roper $P_{11}(1440)$, which appears unusually broad, and the
$S_{11}(1535)$, where the pole can not be uniquely determined,
because it lies close to the $\eta N$ threshold.\\

On the basis of these relatively firm grounds, additional
information can be obtained for the electromagnetic (e.m.) $\gamma
N N^*$ couplings through pion photo- and electroproduction. These
couplings are described by electric, magnetic, and charge
transition form factors, $G_E^*(Q^2)$, $G_M^*(Q^2)$, and
$G_C^*(Q^2)$, or by linear combinations thereof as helicity
amplitudes $A_{1/2}(Q^2)$, $A_{3/2}(Q^2)$, and $S_{1/2}(Q^2)$. So
far we have some reasonable knowledge of the transverse amplitudes
$A_{1/2}$ and $A_{3/2}$ at the real photon point, which are
tabulated in the Particle Data Tables. For finite $Q^2$ the
information found in the literature is scarce and until recently
practically nonexistent for the longitudinal amplitudes $S_{1/2}$.
But even for the transverse amplitudes only few results have
remained  firm over the recent years, such as the $G_M^*$ form
factor of the $P_{33}(1232)$ or $\Delta(1232)$ resonance up to
$Q^2\approx 10$~GeV$^2$, the $A_{1/2}(Q^2)$ for the $S_{11}(1535)$
up to $Q^2\approx 5$~GeV$^2$, and the helicity asymmetry
${\mathcal {A}}(Q^2)$ for the resonances $D_{13}(1520)$ and
$F_{15}(1680)$ up to $Q^2\approx 3$~GeV$^2$~\cite{Boffi96}.
Frequently also data points for other resonances, e.g., the Roper
resonance, are shown together with quark model calculations.
However, the statistical errors are often quite large and the
model dependence of the analysis may be even larger. In this
context it is worth mentioning that also the notion of a `data
point' is somewhat misleading because the photon couplings and
amplitudes can only be derived indirectly by a partial-wave
analysis. It is in fact prerequisite to analyze a particular
experiment within a framework based on the ``world data''. The
only exception from this caveat is the $\Delta(1232)$ resonance.
For this lowest-lying and strongest resonance of the nucleon, the
analysis is facilitated by two important constraints: the validity
of (I) the Watson theorem at the 1~\% level and (II) the
truncation of the multipole series to $S$ and $P$ waves as a good
first-order approximation. With these assumptions the e.m.
couplings have been directly determined in the real photon limit
by a complete experiment with polarized photons and detecting both
neutral and charged pions in the final state, thus allowing also
for an isospin separation~\cite{Beck97}. Moreover, a nearly
complete separation of the possible polarization observables has
recently provided the basis to extend such a ``model-independent''
analysis also to electroproduction~\cite{Kelly}. However, we are
still far from such a situation for all the higher resonances.
Neither are the mentioned constraints valid nor are we close to a
complete experiment. Until recently the data base was rather
limited, the error bars were large, and no data were available
from target or recoil polarization experiments. Even now there
exist only very few data points from double-polarization
experiments at energies above the $\Delta(1232)$. However, the
situation for unpolarized $e+p\rightarrow e'+p+\pi^0$ reaction has
considerably improved, mainly by new JLab experiments in all three
halls A, B, and C. These data cover a large energy range from the
$\Delta(1232)$ up to the third resonance region with a wide
angular range in $\theta_\pi$. Furthermore, electron beam
polarization has been used in several experiments at JLab,
MIT/Bates, and MAMI/Mainz. Because of the large coverage in the
azimuthal angle by the modern large-acceptance detectors, a
separation of all 4 partial cross sections in the unpolarized
experiment becomes possible. But even without a Rosenbluth
separation of the transverse ($\sigma_T$) and longitudinal
($\sigma_L$) cross sections, there is an enhanced sensitivity to
the longitudinal amplitudes due to the interference terms
$\sigma_{LT}$ and $\sigma_{LT'}$. Such data are the basis of our
new partial-wave analysis with
an improved version of the Mainz unitary isobar model MAID.\\

We proceed by presenting a brief history of the unitary isobar
model in Sect.~2. The formalism of pion photo- and
electroproduction is summarized in Sect.~3. In the following
Sect.~4 we present our results for photoproduction as obtained
from the latest version MAID2007, and in Sect.~5 this analysis is
extended to electroproduction. We conclude with a short summary in
Sect.~6.
\section{History of MAID}
\begin{itemize}
\item
{\bf MAID98}\\
In 1998 the first version of the Unitary Isobar Model was
developed and implemented on the web to give an easy access for
the community. MAID98 was constructed with a limited set of
nucleon resonances described by Breit-Wigner forms and a
non-resonant background constructed from Born terms and t-channel
vector-meson contributions~\cite{Maid}. In order to have the right
threshold behavior and a reasonable description at the higher
energies, the Born terms were introduced with an energy-dependent
mixing of pseudovector and pseudoscalar $\pi NN$ coupling. Each
partial wave was unitarized up to the two-pion threshold by use of
Watson's theorem. Specifically, the unitarization was achieved by
introducing additional phases $\phi_R$ in the resonance amplitudes
in order to adjust the phase of the total amplitude. Only the
following 4-star resonances were included: $P_{33}(1232)$,
$P_{11}(1440)$, $D_{13}(1520)$, $S_{11}(1535)$, $S_{11}(1650)$,
$F_{15}(1680)$, and $D_{33}(1700)$. The e.m. vertices of these
resonances were extracted from a best fit to the VPI/GWU
partial-wave analysis~\cite{VPI97}. For the $P_{33}(1232)$
resonance, we determined the following ratios of transition
amplitudes: (I) electric quadrupole to magnetic dipole transition,
$R_{EM}=E2/M1=-2.2\%$, and (II) electric Coulomb to magnetic
dipole transition, $R_{CM}=C2/M1=-3.6\%$, independent of the
4-momentum transfer $Q^2$. The $Q^2$-dependence of the resonance
amplitudes in the second and third resonance regions was expressed
in terms of the quark electric and magnetic
multipoles~\cite{Burk}. The non-unitarized background
contributions were determined using standard Born terms and
vector-meson exchange. In order to preserve gauge invariance, the
Born terms were expressed by the usual dipole form for the Sachs
form factors, and both the pion and the axial form factor were set
equal to the isovector Dirac form factor, $ F_{\pi}(Q^2)=G_A(Q^2)=
F_1^p(Q^2)-F_1^n(Q^2)$.
\item
{\bf MAID2000}\\
In this version of MAID, the background contribution was
unitarized for the multipoles up to $F$ waves according to the
prescription of K-matrix theory. The $S$-wave multipoles $E_{0+}$
and $L_{0+}$ were modified in order to improve their energy
dependence in the threshold region. With the new unitarization
procedure, the pion photoproduction multipoles of SAID and some
selected data for pion photo- and electroproduction in the energy
range up to $W=1.6$ GeV were fitted~\cite{NSTAR2001}. The ratios
of the $\Delta(1232)$ multipoles were found to be $R_{EM}=-2.2\%$
and $R_{CM}=-6.5\%$, still independent of $Q^2$.
\item
{\bf MAID2003}\\
In accordance with results of Ref.~\cite{DMT}, the $Q^2$
dependence of the electric and Coulomb excitations of the
$\Delta(1232)$ resonance was modified. The ratio $R_{EM}$ was
found to change sign at $Q^2 \approx 3.3$~GeV$^2$ from negative to
positive values, and $R_{CM}$ decreased from $-6.5\%$ at $Q^2=0$
to $-13.5\%$ at $Q^2=4$~GeV$^2$. Moreover, the following 4-star
resonances were included in MAID2003: $S_{31}(1620)$,
$D_{15}(1675)$, $P_{13}(1720)$, $F_{35}(1905)$, $P_{31}(1910)$,
and $F_{37}(1950)$. In contrast to previous versions, the helicity
amplitudes of all 13 resonances were input parameters and their
$Q^2$ dependence was parameterized by polynomials. With this new
version of MAID we directly analyzed all the pion photo- and
electroproduction data available since 1960, and for the first
time we made local (single energy) and global (energy dependent)
fits, independent of the GWU/SAID group.
\item
{\bf MAID2005}\\
The $Q^2$ dependence of the Sachs form factors in the Born terms
was replaced by the more recent parameterization of
Ref.~\cite{Kelly04}, and at the e.m. vertices of the pion-pole and
seagull terms, realistic pion and axial form factors were
introduced. As a result the description of charged pion
electroproduction was much improved. On the basis of a large
amount of new data from MIT/Bates, ELSA/Bonn, Grenoble, Mainz, and
Jefferson Lab, we performed new local and global as well as
single-$Q^2$ fits and obtained a better description of the data in
the energy range 1.6~GeV$<W<$2~GeV~\cite{MAID05}.
\item
{\bf MAID2007}\\
The present version of MAID is presented in some detail in the
following sections. As far as pion photoproduction is concerned,
this version is identical to MAID2005. The main changes are
related to the $Q^2$ evolution of e.m. form factors. In
particular, the $Q^2$-dependence of the $\Delta(1232)$ transition
form factors are remodeled to be consistent with the Siegert
theorem. As a result the ratio $R_{CM}$ decreases sharply if $Q^2$
approaches zero. Furthermore, our analysis of the recent
high-$Q^2$ data~\cite{Ungaro06} led to the conclusion that
$R_{EM}$ remains negative in the range of the existing
experiments.
\end{itemize}
\section{Formalism for pion photo- and electroproduction}
Let us first define the kinematics of pion photo- and electroproduction on a
nucleon,
\begin{equation}
\gamma^{\ast}(k) + N(p) \to \pi(q) + N'(p')\,,\label{eq:3.1}
\end{equation}
where the variables in brackets denote the 4-momenta of the participating
particles. In the pion-nucleon center-of mass (c.m.) system, we define
\begin{equation}
k^{\mu}  =  (\omega_{\gamma}, {\vec k})\,,\; q^{\mu}  = (\omega_{\pi}, {\vec
q})\,,\;p^{\mu}  =  (E_N, -{\vec k}) \,,\label{eq:3.2}
\end{equation}
where
\begin{eqnarray}
k(W,Q^2)& = & |\vec k|=
\sqrt{\left( \frac{W^2-m^2-Q^2}{2W}\right)^2+Q^2}\,,\label{eq:3.3}\\
q(W)& = & |\vec q|=
\sqrt{\left(\frac{W^2-m^2+m_{\pi}^2}{2W}\right)^2-m_{\pi}^2}\,, \label{eq:3.4}
\end{eqnarray}
with $W=\omega_{\gamma}+E_N$ the total c.m. energy and
$Q^2=k^2-\omega_{\gamma}^2>0$ the square of 4-momentum deposited by the photon
at the nucleon vertex, also referred to as the virtuality of the photon. In
order to simplify the notation, we use the following abbreviations:
\begin{equation}
k_W  =k (W,0)=  \frac{W^2-m^2}{2W}\,,\label{eq:3.5}\,
\end{equation}
describing the momentum of a real photon, and
\begin{equation}
k_R  =k (M_R,0)\,,\; q_R  =q (M_R)\,,\label{eq:3.6}\,
\end{equation}
for the real photon and pion momenta at the resonance position, $W=M_R$.\\

The basic equations used for MAID2007 are taken from the dynamical
Dubna-Mainz-Taipei (DMT) model~\cite{DMT,Yang85,KY99}. In this approach the
t-matrix for pion photo- and electroproduction takes the form
\begin{equation}
t_{\gamma\pi}(W)=v_{\gamma\pi}(W)+v_{\gamma\pi}(W)\,g_0(W)\,t_{\pi N}(W)\,,
\label{eq:3.7}
\end{equation}
with $v_{\gamma\pi}$ the transition potential for the reaction $\gamma^{\ast}N
\rightarrow \pi N$, $t_{\pi N}$ the $\pi N$ scattering matrix, and $g_0$ the
free $\pi N$ propagator. In a resonant channel the transition potential $v_{\gamma\pi}$
consists of two terms,
\begin{equation}
v_{\gamma\pi}(W)=v_{\gamma\pi}^B(W) +v_{\gamma\pi}^R(W)\,,
\label{eq:3.8}
\end{equation}
with $v_{\gamma\pi}^B$ the background transition potential and
$v_{\gamma\pi}^R$ the contribution of the ``bare'' resonance excitation. The
resulting t-matrix can be decomposed into two terms~\cite{KY99}
\begin{equation}
t_{\gamma\pi}(W)=t_{\gamma\pi}^B(W) + t_{\gamma\pi}^{R}(W)\,,\label{eq:3.9}
\end{equation}
where
\begin{eqnarray}
t_{\gamma\pi}^B(W) &=&
v_{\gamma\pi}^B(W)+v_{\gamma\pi}^B(W)\,g_0(W)\,t_{\pi N}(W)\,,
\label{eq:3.10}\\
t_{\gamma\pi}^R(W) &=& v_{\gamma\pi}^R(W)+v_{\gamma\pi}^R(W)\,g_0(W)\,t_{\pi N}(W)\,,
\label{eq:3.11}
\end{eqnarray}
with $t_{\gamma\pi}^B$ including the contributions from both the non-resonant
background and the $\gamma^*NR$ vertex renormalization.
The decomposition in resonance and background contributions is not unique,
however, our definition has the advantage that all the processes
starting with the e.m. excitation of a bare resonance are summed up in
$t_{\gamma\pi}^R$.
\subsection{Unitarized background}
The multipole decomposition of Eq.~(\ref{eq:3.10}) yields the background
contribution to the physical amplitudes in the channels
$\alpha=(\xi,\ell,j,I)$~\cite{Yang85}, where $\ell$, $j$ and $I$ denote the
orbital momentum, the total angular momentum, and the isospin of the
pion-nucleon final state, and $\xi$ stands for the magnetic ($\xi=M$), electric
($\xi=E$), and Coulomb or ``scalar'' ($\xi=S$) transitions,
\begin{eqnarray}
t^{B,\alpha}_{\gamma\pi}(q,k,Q^2)& = &
v_{\gamma\pi}^{B,\alpha}(q,k,Q^2)\label{eq:3.12}\\
&+& \int_0^{\infty} dq' \frac{q'^2 t_{\pi
N}^{\alpha}(q,q';W)\,v_{\gamma\pi}^{B,\alpha}(q',k,Q^2)} {W-W_{\pi
N}(q')+i\epsilon}\,,\nonumber
\end{eqnarray}
where $W_{\pi N}(q')$ is the hadronic c.m. energy in the intermediate state.
The pion electroproduction potential $v^{B,\alpha}_{\gamma\pi}$ is constructed
as in Ref.~\cite{Maid} and contains contributions from the Born terms described
by an energy-dependent mixing of pseudovector~(PV) and pseudoscalar~(PS) $\pi
NN$ coupling as well as t-channel vector meson exchange. The quasi-potential
$v^{B,\alpha}_{\gamma\pi}$ depends on 5 parameters: the PV-PS mixing parameter
$\Lambda_m$ as defined in Eq.~(12) of Ref.~\cite{Maid} and 4 coupling constants
for the vector-meson exchange. The on-shell parts of $v^{B,\alpha}_{\gamma\pi}$
and $t^{B,\alpha}_{\gamma\pi}$ depend on two variables only, i.e.,
\begin{eqnarray}
v^{B,\alpha}_{\gamma\pi}(q,k,Q^2) &=& v^{B,\alpha}_{\gamma\pi}(W,Q^2)\label{eq:3.13}\\
t^{B,\alpha}_{\gamma\pi}(q,k,Q^2) &=&
t^{B,\alpha}_{\gamma\pi}(W,Q^2)\,.\label{eq:3.14}
\end{eqnarray}
The $Q^2$ evolution of the s- and u-channel nucleon pole terms of the
background is described by the form factors of Ref.~\cite{Kelly}. At the e.m.
vertices of the pion-pole and seagull terms we apply a monopole form for the
pion form factor and a dipole form for the axial form factor, while the
standard dipole form factor is used for the vector-meson exchange.\\

We note that the background contribution of MAID98 was defined by
$t_{\gamma\pi}^{B,\alpha}({\rm MAID}98)=v_{\gamma\pi}^{B,\alpha}(W,Q^2)$ and
assumed to be a real and smooth function. The unitarization of the total
amplitude was then provided by an additional phase $\phi_{\alpha}$ in the
resonance contribution such that the phase of the total amplitude had the phase
$\delta_{\alpha}$ of the respective $\pi N$ scattering state. In MAID2007,
however, the background contributions are complex functions defined according
to K-matrix theory,
\begin{equation}
 t^{B,\alpha}_{\gamma\pi}(W,Q^2)=v^{B,\alpha}_{\gamma\pi}(W,Q^2)\,[1+it_{\pi N}^{\alpha}(W) ]\, ,
\label{eq:3.15}
\end{equation}
where the pion-nucleon elastic scattering amplitudes, $t^{\alpha}_{\pi
N}=[\eta_{\alpha} \exp(2i\delta_{\alpha})-1]/2i$, are described by the phase
shifts $\delta_{\alpha}$ and the inelasticity parameters $\eta_{\alpha}$ taken
from the GWU/SAID analysis~\cite{VPI}. The assumed structure of the background
corresponds to neglecting the principal value integral in the pion-rescattering
term of Eq.~(\ref{eq:3.12}). Our previous studies of the $P$-wave multipoles in
the (3,3)~channel~\cite{DMT,KY99} showed that the ``pion cloud'' contributions
of the principal value integral are effectively included by the
dressing of the $\gamma N N^{\ast}$~vertex.\\

Furthermore, the threshold behavior of the $S$ waves was improved. The results
of the dynamical approaches~\cite{DMTthr} show that the pion cloud
contributions are very important to obtain a good description of the
$E_{0+}$~multipole in the $\pi^0 p$~channel. For this purpose we have
introduced the following phenomenological term:
\begin{equation}
E^{\rm {corr}}_{0+}(W,Q^2)= \frac{A}{(1 + B^2 q^2)^2}\,G_D(Q^2)\,,
\label{eq:3.16}
\end{equation}
with $A$ and $B$ free parameters fixed by fitting the low-energy $\pi^0$
photoproduction data, and $G_D$ the standard nucleon dipole form factor.
The threshold correction for the $L_{0+}$
multipole we will consider later in Sect. 5.3.
As a result the background contribution of MAID now depends on 8
parameters. We furthermore account for the cusp effect in the $\pi^0 p$~channel
appearing at the $\pi^+n$ threshold by the term~\cite{Laget,Bernard}
\begin{equation}
E^{\rm {cusp}}_{0+}= - a_{\pi N} \,\omega_c\,Re
E_{0+}^{\gamma\pi^+}\,\sqrt{1-\frac{\omega_{\pi}^2}{\omega_c^2}}\,,
\label{eq:3.17}
\end{equation}
where $\omega_c=140$~MeV is the $\pi^0$ c.m. energy at the cusp and $a_{\pi
N}=0.124/m_{\pi^+}$ the pion charge-exchange amplitude.
\subsection{Resonance contributions}
For the resonance contributions we follow Ref.~\cite{Maid} and assume
Breit-Wigner forms for the resonance shape,
\begin{equation}
t_{\gamma\pi}^{R,\alpha}(W,Q^2)\,=\,{\bar{\cal A}}_{\alpha}^R(W,Q^2)\,
\frac{f_{\gamma N}(W)\Gamma_{tot}\,M_R\,f_{\pi N}(W)}{M_R^2-W^2-iM_R\, \Gamma_{tot}}
\,e^{i\phi_R}\,, \label{eq:3.18}
\end{equation}
where $f_{\pi N}(W)$ is the usual Breit-Wigner factor describing the decay of a
resonance with total width $\Gamma_{tot}(W)$, partial $\pi N$ width $\Gamma_{\pi N}$,
and spin $j$,
\begin{equation}
f_{\pi N}(W)=C_{\pi N}\left[\frac{1}{(2j+1)\pi}\frac{k_W}{q}
\frac{m}{M_R}\frac{\Gamma_{\pi N}}{\Gamma_{tot}^2}\right]^{1/2}\,, \label{eq:3.19}
\end{equation}
with $C_{\pi N}=\sqrt{3/2}$ and $-1/\sqrt{3}$ for isospin $\frac {3}{2}$ and
$\frac {1}{2}$, respectively. The energy dependence of the partial width is
given by
\begin{equation}
\Gamma_{\pi N}(W)=\beta_{\pi}\,\Gamma_R\,\left(\frac{q}{q_R}
\right)^{2l+1}\,\left(\frac{X^2_R+q_R^2}{X^2_R+q^2}\right)^{\ell} \,
\frac{M_R}{W}\,, \label{eq:3.20}
\end{equation}
with $\Gamma_R=\Gamma_{tot}(M_R)$ and $X_R$ a damping parameter and $\beta_{\pi}$
the single-pion branching ratio. The expression for the total width $\Gamma_{tot}$
is given in Ref.~\cite{Maid}. The $\gamma NN^*$ vertex is assumed to have the following
dependence on $W$:
\begin{equation}
f_{\gamma N}(W)=\left(\frac{k_W}{k_R}\right)^n\,
\left(\frac{X^2_R+k_R^2}{X^2_R+k_W^2}\right)\,, \label{eq:3.21}
\end{equation}
where $n$ is obtained from a best fit to the real photon data, and with the
normalization condition $f_{\gamma N}(M_R)=1$. The phase $\phi_R(W)$ in
Eq.~(\ref{eq:3.18}) is introduced to adjust the total phase such that the
Fermi-Watson theorem is fulfilled below two-pion threshold. For the $S$- and
$P$-wave multipoles we extend this unitarization procedure up to $W=1400$ MeV.
Because of a lack of further information, we assume that the phases $\phi_R$
are constant at the higher energies. In particular we note that the phase
$\phi_R$ for the $P_{33}(1232)$ excitation vanishes at $W=M_R=1232$~MeV for all
values of $Q^2$. For this multipole we may even apply the Fermi-Watson theorem
up to $W \approx 1600$~MeV because the inelasticity parameter $\eta_{\alpha}$
remains close to 1. For the $D$- and $F$-wave resonances, the phases
$\phi_R$ are assumed to be constant and determined from the best fit.\\

Whereas MAID98~\cite{Maid} included only the 7 most important nucleon
resonances, essentially with only transverse e.m. couplings, our present
version contains all 13 resonances of the 4-star tier below 2~GeV with
transverse electric (${\bar{\cal A}}_{\alpha}^R=\bar{E}_{l\pm}$), transverse
magnetic (${\bar{\cal A}}_{\alpha}^R=\bar{M}_{l\pm}$), and Coulomb (${\bar{\cal
A}}_{\alpha}^R=\bar{S}_{l\pm}$) couplings: $P_{33}(1232)$, $P_{11}(1440)$,
$D_{13}(1520)$, $S_{11}(1535)$, $S_{31}(1620)$, $S_{11}(1650)$, $D_{15}(1675)$,
$F_{15}(1680)$, $D_{33}(1700)$, $P_{13}(1720)$, $F_{35}(1905)$, $P_{31}(1910)$,
and $F_{37}(1950)$. Because we determine the isovector amplitudes from the
proton channels, the number of the e.m. couplings is 34 for the proton and 18
for the neutron channels, that is 52 parameters altogether. These are taken to
be constant in a single-Q$^2$ analysis, e.g., in photoproduction but also at
any fixed $Q^2$ if sufficient data are available in the chosen energy and
angular range. Alternatively, the couplings have also been parameterized as
functions of $Q^2$, as is discussed in Sec.~5.\\
\begin{table*}[ht]
\caption{The reduced e.m. amplitudes ${\bar{\cal A}}_{\alpha}$ defined by
Eq.~(\ref{eq:3.18}) in terms of the helicity amplitudes.}\label{tab:amplitudes}
\begin{center}
\begin{tabular}{|c|ccc|ccc|c|}
\hline
$N^{\ast}$        &      & $\bar{E}$  & & & $\bar{M}$  &  & $\bar{S}$  \\
\hline
$S_{11}$/$S_{31}$ &  & $-A_{1/2}$  &    &   & ---  &  & $-\sqrt{2}S_{1/2}$\\
$P_{13}$/$P_{33}$ &  & $\frac{1}{2}(\frac{1}{\sqrt{3}}A_{3/2}-A_{1/2})$ & & &
$-\frac{1}{2}(\sqrt{3}A_{3/2}+A_{1/2})$ & &  $-\frac{1}{\sqrt{2}}S_{1/2}$\\
$P_{11}$/$P_{31}$ &  & ---  &    &   & $A_{1/2}$  & &$-\sqrt{2}S_{1/2}$ \\
$D_{13}$/$D_{33}$ &  & $-\frac{1}{2}(\sqrt{3}A_{3/2}+A_{1/2})$ & & &
$-\frac{1}{2}(\frac{1}{\sqrt{3}}A_{3/2}-A_{1/2})$ & & $-\frac{1}{\sqrt{2}}S_{1/2}$\\
$D_{15}$/$D_{35}$&  & $\frac{1}{3}(\frac{1}{\sqrt{2}}A_{3/2}-A_{1/2})$ & & &
$-\frac{1}{3}(\sqrt{2}A_{3/2}+A_{1/2})$ & & $-\frac{\sqrt{2}}{3}S_{1/2}$\\
$F_{15}$/$F_{35}$ &  & $-\frac{1}{3}(\sqrt{2}A_{3/2}+A_{1/2})$ & & &
$-\frac{1}{3}(\frac{1}{\sqrt{2}}A_{3/2}-A_{1/2})$ & & $-\frac{\sqrt{2}}{3}S_{1/2}$\\
$F_{17}$/$F_{37}$ &  & $\frac{1}{4}(\sqrt{\frac{3}{5}}A_{3/2}-A_{1/2})$ & & &
$-\frac{1}{4}(\sqrt{\frac{5}{3}}A_{3/2}+A_{1/2})$ & & $-\frac{1}{2\,\sqrt{2}}S_{1/2}$\\
\hline
\end{tabular}
\end{center}
\end{table*}

The more commonly used helicity amplitudes $A_{1/2}$, $A_{3/2}$, and $S_{1/2}$
are given by linear combinations of the e.m. couplings $\bar{\cal
A}_{\alpha}^R$. These relations take the form
\begin{eqnarray}
A^{\ell +}_{1/2} &=& -\frac{1}{2} [(\ell +2) \bar{E}_{\ell+} + \ell \bar{M}_{\ell+} ]
\,, \nonumber\\
A^{\ell +}_{3/2} &=& \frac{1}{2}\sqrt{\ell(\ell+2)} (\bar{E}_{\ell+} - \bar{M}_{\ell+})
\,, \label{3.22}\\
S^{\ell +}_{1/2} &=& -\frac{\ell+1}{\sqrt{2}} \bar{S}_{\ell+} \nonumber
\end{eqnarray}
for resonances with total spin $j=\ell +\frac {1}{2}$, and
\begin{eqnarray}
A^{\ell -}_{1/2} &=& \frac{1}{2} [(\ell +1) \bar{M}_{\ell -} - (\ell -1) \bar{E}_{\ell -}]\,, \nonumber\\
A^{\ell -}_{3/2} &=& -\frac{1}{2}\sqrt{(\ell -1)(\ell+1)}(\bar{E}_{\ell -} + \bar{M}_{\ell -})\,, \label{3.23}\\
S^{\ell -}_{1/2} &=& -\frac{\ell}{\sqrt{2}} \bar{S}_{\ell -} \nonumber
\end{eqnarray}
for total spin $j=\ell - \frac {1}{2}$. The inverse relations for the partial
waves are listed in Table~\ref{tab:amplitudes}.
The helicity amplitudes are related to matrix elements of the e.m. current
$J_{\mu}$ between the nucleon and the resonance states, e.g., as obtained in
the framework of quark models,
\begin{eqnarray}
A_{1/2} &=& -\sqrt{\frac{2\pi\alpha_{\rm {em}}}{k_W}}
<R,\frac{1}{2}\,|\,J_{+}\,|\,N,-\frac{1}{2}>\, \zeta\,, \nonumber\\
A_{3/2} &=& -\sqrt{\frac{2\pi\alpha_{\rm {em}}}{k_W}}
<R,\frac{3}{2}\,|\,J_{+}\,|\,N,\frac{1}{2}>\,\zeta\,, \label{3.24}\\
S_{1/2} &=& -\sqrt{\frac{2\pi\alpha_{\rm {em}}}{k_W}}
<R,\frac{1}{2}\,|\,\rho\,|\,N,\frac{1}{2}>\, \zeta \,, \nonumber
\end{eqnarray}
where $J_{+}=-\frac{1}{\sqrt{2}}(J_x+iJ_y)\,$ and $\alpha_{\rm {em}}=1/137$.
However, these equations define the couplings only up to a phase $\zeta$, which
in principle can be obtained from the pionic decay of the resonance calculated
within the same model. Because this phase is ignored in most of the literature,
the comparison of the sign is not always meaningful, especially in critical
cases such as the Roper resonance whose correct sign is not obvious from the
data. In contrast with MAID98 and MAID2000, our present version uses the
helicity amplitudes $A_{1/2}$, $A_{1/2}$, and $S_{1/2}$ for photoproduction as
input parameters, except for the $P_{33}(1232)$ resonance which is directly
described by the 3 e.m. amplitudes $\bar{\cal A}_{\alpha}$.
\section{Partial-wave analysis of pion photoproduction data}
The unitary isobar model MAID2007 has been developed to analyze the world data
of pion photo- and electroproduction. In this section we fix (I) the background
parameters and the helicity amplitudes for pion photoproduction ($Q^2=0$) and
(II) the dependence of the resonance contributions on the c.m. energy $W$.
These results are then generalized to pion electroproduction in the next
section.
\subsection{Data base for pion photoproduction and fit procedure}
The main part of the photoproduction data was taken from the GWU/SAID
compilation of SAID2000, which includes the data published between 1960 and
2000, a total of 14700 data points. A separation of these data in different
physical channels and observables is given in Table~\ref{tab:data_SAID}. In the
following years the data base was extended by including recent results from
MAMI~(Mainz)~\cite{Leukel,GDH01,Ilia,GDH06},
GRAAL~(Grenoble)~\cite{GRAAL02,GRAAL05}, LEGS~(Brookhaven)~\cite{LEGS04}, and
ELSA~(Bonn)~\cite{Bonn05} as listed in Table~\ref{tab:data_new}. Altogether
4976 more data points were added. As a result our full data base contains
19676 points within the energy range 140~MeV$<E_{\gamma}<$1610~MeV.
\begin{table}[htbp]
\caption{Number of data points from the SAID2000 data base for differential
cross sections ($d\sigma$), photon asymmetries ($\Sigma$), target asymmetries
($T$), and recoil asymmetries ($P$).}\label{tab:data_SAID}
\begin{center}
\begin{tabular}{|c|cccc|c|}
\hline
channel & $d\sigma$ &  $\Sigma$ & $T$ & $P$  & total  \\
\hline
 $n\pi^+$  &  4646  & 760  & 645 & 205 & 6256 \\
 $p\pi^0$  &  4936  & 673  & 353 & 540 & 6502 \\
 $p\pi^-$  &  1554  & 206  & 94 &  88 &  1942 \\
\hline
\end{tabular}
\end{center}
\end{table}
\begin{table}[htbp]
\caption{Number of data points collected after 2000 for differential cross
sections ($d\sigma$), photon asymmetries ($\Sigma$), and helicity asymmetries
$\Delta\sigma=d\sigma_{1/2}-d\sigma_{3/2}$.}\label{tab:data_new}
\begin{center}
\begin{tabular}{|c|c|c|c|}
\hline
channel & range $E_{\gamma}$(MeV) &data points~(observable) & Ref. \\
\hline
$p\pi^0$ & 202-790 & 1129 ($d\sigma$) + 357 ($\Sigma$) & \cite{Leukel} \\
$p\pi^0$ & 310-780 & 174 ($d\sigma$) + 138 ($\Delta\sigma$)    & \cite{GDH01}   \\
$p\pi^+$ & 180-450 & 205 ($d\sigma$) + 129 ($\Delta\sigma$)    & \cite{Ilia}    \\
$p\pi^+$ & 463-783 & 204 ($d\sigma$) + 102 ($\Delta\sigma$)    & \cite{GDH06}   \\
$p\pi^+$ & 800-1454& 237 ($\Sigma$)                & \cite{GRAAL02} \\
$p\pi^0$ & 555-1541& 861 ($d\sigma$) + 469 ($\Sigma$)      & \cite{GRAAL05} \\
$n\pi^-$ & 285-769 & 300 ($d\sigma$)               & \cite{LEGS04}  \\
$p\pi^0$ & 513-1575& 671 ($d\sigma$)               & \cite{Bonn05}  \\
\hline
total        &     & 4976  &    \\
\hline
\end{tabular}
\end{center}
\end{table}

Our strategy for the data analysis is as follows. First, we try to find a
global (energy dependent) solution by fitting all the data in the range
140~MeV$\le E_{\gamma} \le$1610~MeV. This allows us to determine the phase of
the multipoles, i.e., the ratio $ {\rm {Im}}\, t_{\gamma\pi}^{\alpha}/{\rm
{Re}}\, t_{\gamma\pi}^{\alpha}$ above the two-pion threshold. At the lower
energies this phase is constrained by the $\pi N$ scattering phase. In a second
step we perform local (single energy) fits to the data, in energy bins of 10
MeV in the range 140~MeV$\le E_{\gamma} \le$460~MeV and of 20 MeV for the
higher energies, by varying the absolute values of the multipoles but keeping
the phase as previously determined. Similar to the prescription of the SAID
group we minimize the modified $\chi^2$ function
\begin{equation}
\chi^2 = \sum_i^{N_{\rm {data}}} \left(\frac{\Theta_i-\Theta_i^{\rm
{exp}}}{\delta\Theta_i}\right)^2 + \sum_j^{N_{\rm {mult}}}
\left(\frac{X_j-1}{\Delta}\right)^2\,. \label{eq:4.1}
\end{equation}
The first term on the r.h.s. of this equation is the standard $\chi^2$
function with $\Theta_i$ the calculated and $\Theta_i^{\rm {exp}}$ the measured
observables, $\delta\Theta_i$ the statistical errors, and $N_{\rm {data}}$ the
number of data points. In the second term, $N_{\rm {mult}}$ is the number of
the varied multipoles and  $X_j$ describes the deviation from the global fit.
The fitting procedure starts with the initial value $X_j=1$ corresponding to
the global solution, and the quantity $\Delta$ enforces a smooth energy
dependence of the single-energy solution. In the limit of
$\Delta\rightarrow\infty$ we obtain the standard $\chi^2_{\rm {std}}$, and for
$\Delta\rightarrow 0$ the single-energy and the global solutions become
identical. The optimum value for $\Delta$ is chosen from the condition
$1<\chi^2/\chi^2_{\rm {std}}<1.05$. The described two-step fitting procedure
can be repeated several times by adjusting the energy dependence of the global
solution, for example by changing the parameters $X_R$ and $n$ in Eqs.
(\ref{eq:3.20}-\ref{eq:3.21}) in order to improve the agreement between the
global and local solutions.
\subsection{Results for pion photoproduction}
Our results for $\chi^2$ are summarized in Table~\ref{tab:chisq} by comparing
the local and global solutions for different energy ranges and channels. We
recall that the number of varied multipoles $N_{\rm {mult}}$ in the proton and
neutron channels is different. Since the number of data points in the proton
channels ($\gamma\pi^0$ and $\gamma\pi^+$) is about one order of magnitude
larger than for the neutron channel ($\gamma\pi^-$), we proceed as follows.
First, we analyze the proton channel and extract the multipoles
$_pE_{l\pm}^{1/2}$, $_pM_{l\pm}^{1/2}$, $E_{l\pm}^{3/2}$, and $M_{l\pm}^{3/2}$
as defined in Ref.~\cite{Maid}. Second, with the thus obtained values for the
isospin 3/2 multipoles, we extract the multipoles $_nE_{l\pm}^{1/2}$ and
$_nM_{l\pm}^{1/2}$ from the neutron channel. In this way we minimize the
pressure from the large number of proton data on the results in the neutron
channel. The number of varied multipoles also depends on the energy. For
$E_{\gamma}<$450~MeV we vary all the $S$- and $P$-wave multipoles plus
$_{p,n}E_{2-}^{1/2}$ and $E_{2-}^{3/2}$. At the higher energies we include
all the multipoles up to the $F$ waves.\\

\begin{table}[htbp]
\caption{Results for $\chi^2$ from single-energy (se) and global (gl)
 solutions.}\label{tab:chisq}
\begin{center}
\begin{tabular}{|c|cccc|}
\hline
&  &  proton &  & \\
\hline
$E_{\gamma}$[MeV]  & $N_{\rm {mult}}$ & $N_{\rm {data}}$
& $\chi^2_{\rm {se}}$& $\chi^2_{\rm {gl}}$ \\
\hline
140---200   & 10 & 990  & 787  & 2346   \\
200---450   & 10 & 5622 & 5454 & 14236  \\
450---850   & 24 & 6403 & 6996 & 22700  \\
850---1210  & 24 & 2965 & 4133 & 24990  \\
1210---1610 & 24 & 1454 & 4737 & 12594  \\
\hline
total   &   & 17434& 22107& 76866 \\
\hline
& & neutron &  & \\
\hline
$E_{\gamma}$[MeV]  & $N_{\rm {mult}}$ & $N_{\rm {data}}$
& $\chi^2_{\rm {se}}$& $\chi^2_{\rm {gl}}$ \\
\hline
140---200    & 5  & 51   & 113   & 151  \\
200---450    & 5  & 872  & 1613  & 2203 \\
450---850    & 12 & 902  & 1748  & 3414 \\
850---1210   & 12 & 334  & 651   & 2311 \\
1210---1610  & 12 & 83   & 107   & 583  \\
\hline
total  &   & 2242 & 4222  & 9262 \\
\hline
\end{tabular}
\end{center}
\end{table}

The best fit for the background and the resonance parameters yields the results
listed in Tables~\ref{tab:par_backgr} and \ref{tab:par_res}, respectively. The
PS-PV mixing parameter and the vector-meson coupling constants are defined as
in Ref.~\cite{Maid}. However we note that in the present version we do not use
form factors at the hadronic vertices involving vector-meson exchange.
\begin{table}[htbp]
\caption{Masses and coupling constants for vector mesons, PS-PV mixing
parameter $\Lambda_m$, and parameter $A$ for the low-energy correction of
Eq.~(\ref{eq:3.16}).}\label{tab:par_backgr}
\begin{center}
\begin{tabular}{|c|cccc|}
\hline \multicolumn{1}{|c|}{} & $m_{V}$[MeV] & $\lambda_V$ &
$\tilde g_{V1}$ & $\tilde g_{V2}/\tilde g_{V1}$ \\
\hline
$\omega$ & 783 & 0.314 &  16.3  & -0.94  \\
$\rho$   & 770 & 0.103 &  1.8  &  12.7 \\
\hline \multicolumn{4}{|l}{$\Lambda_m=423$ MeV
\vline
\hspace{1mm}$ A= 1.9 \times 10^{-3}/m_{\pi}^+$}
\vline
\hspace{1mm}$B= 0.71 fm$ \\
\hline
\end{tabular}
\end{center}
\end{table}
\begin{table*}[htp]
\caption{Resonance masses $M_R$, widths $\Gamma_R$, single-pion branching
ratios $\beta_{\pi}$, and angles $\phi_R$ as well as the parameters $X_R, n_E$,
and $n_M$ of the vertex function Eq.~(\ref{eq:3.21}).}\label{tab:par_res}
\begin{center}
\begin{tabular}{|c|ccccc|cc|cc|}
\hline
&  &  &  & & & \quad \quad proton &  &\quad \quad  neutron   \\
$N^{\ast}$  &$M_R$[MeV]& $\Gamma_R$[MeV]& $\beta_{\pi}$ & $\phi_R$[deg]
& $X_R$[MeV]& $n_E$ & $n_M$  & $n_E$ & $n_M$ \\
\hline $P_{33}(1232)$ & 1232 & 130 &
1.0  & 0.0  & 570& -1  & 2  & -1  & 2 \\
 $P_{11}(1440)$ & 1440 &
350 & 0.70 &  -15  & 470 &  ---   &  0  &  ---   &  -1 \\
$D_{13}(1520)$ &
1530 & 130 & 0.60 &  32  & 500 &  3  & 4 &  7  & 2  \\
$S_{11}(1535)$ & 1535 & 100& 0.40 & 8.2 & 500&  2  &
--- &  2  & ---  \\
$S_{31}(1620)$ & 1620 & 150 & 0.25 & 23  & 470  &
5 &  ---  & 5 &  ---  \\
 $S_{11}(1650)$ & 1690 & 100 & 0.85 &   7.0  & 500 &
 4 &  --- &  4 &  ---  \\
$D_{15}(1675)$ & 1675 & 150 & 0.45 &
20 & 500 &  3  &  5 &  3  &  4 \\
$F_{15}(1680)$ & 1680 & 135 &
0.70 &  10  & 500 &  3  &  3 &  2  &  2  \\
 $D_{33}(1700)$ & 1740 &
450 & 0.15 & 61  & 700 &  4  &  5 &  4  &  5 \\
 $P_{13}(1720)$ &
1740 & 250 & 0.20 &  0.0  & 500 &  3 &  3 &  3 & 3 \\
$F_{35}(1905)$ & 1905 & 350 & 0.10 &  40  & 500 & 4  &
 5 &  4  & 5 \\
$P_{31}(1910)$ & 1910 & 250 & 0.25 &  35   & 500 &   ---
& 1 &---& 1 \\
 $F_{37}(1950)$ & 1945 & 280 & 0.40 &  30  & 500 &
6  &  6 & 6  &  6 \\
\hline
\end{tabular}
\end{center}
\end{table*}
In Tables~\ref{tab:helicity_p} and \ref{tab:helicity_n} we compare the helicity
amplitudes obtained from MAID2003 and MAID2007 with the results of the
PDG~\cite{PDG06} and GWU/SAID~\cite{SAID02,SAID06} analysis. As is very typical for a
global analysis with about 20,000 data points fitted to a small set of 20-30
parameters, the fit errors appear unrealistically small. However, one should
realize that these errors only reflect the statistical uncertainty of the
experimental error, whereas the model uncertainty can be larger by an order of
magnitude. We therefore do not list our fit errors, which in fact are very
similar in the GW02 or GW06 fits of the SAID group~\cite{SAID02,SAID06}.
The only realistic error estimate is obtained by comparing different analysis, such as
SAID, MAID, and coupled-channels approaches.\\

\begin{figure}[ht]
\centerline{\epsfig{file=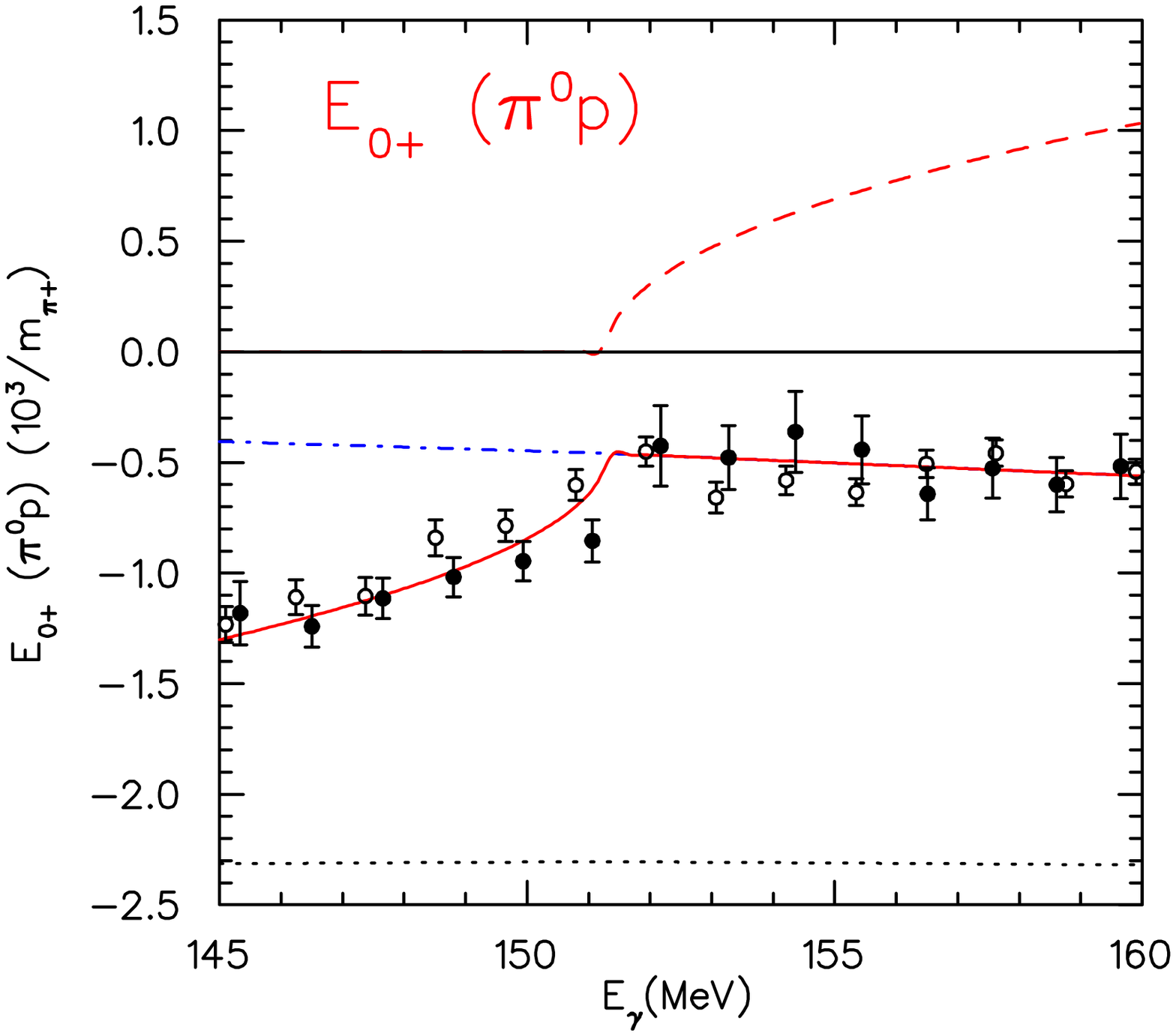, width=8 cm} }
\caption{The $E_{0+}$ multipole for $\gamma p\rightarrow \pi^0 p$. Real part:
the results of MAID98 (dotted line) and MAID2007 without the cusp
effect (dashed-dotted line) as well as the full MAID2007
calculation (red solid line). The red dashed line is the imaginary
part of the full MAID2007 solution. The data points are from
Refs.~\cite{Bergstrom}($\bullet$) and \cite{Schmidt}($\circ$).
\label{fig:E0thr}}
\end{figure}

\begin{table}[htbp]
\caption{Proton helicity amplitudes at $Q^2=0$ for the major nucleon
resonances, in units $10^{-3}$~GeV$^{-1/2}$. The results with MAID2003 and
MAID2007 are compared to the PDG~\cite{PDG06} and GWU/SAID~\cite{SAID06}
analysis.}\label{tab:helicity_p}
\begin{center}
\small
\begin{tabular}{|l l|c c c c|} \hline
&  &  PDG & GW06 & 2003 &  2007\\
\hline
$P_{33}(1232)$& $A_{1/2}$ & -135$\pm$6 & -139.1$\pm$3.6 & -140 & -140\\
              & $A_{3/2}$ & -250$\pm$8 & -257.6$\pm$4.6 & -265 & -265\\
$E2/M1$ (\%)  &           & -2.5$\pm$0.5&          & -2.2& -2.2\\
\hline
$P_{11}(1440)$& $A_{1/2}$ & -65 $\pm$4 & -50.6 $\pm$1.9 & -77 &  -61\\
\hline
$D_{13}(1520)$& $A_{1/2}$ & -24 $\pm$9 & -28.0 $\pm$1.9 & -30 & -27\\
              & $A_{3/2}$ & 166 $\pm$5 & 143.1 $\pm$2.0 & 166  & 161\\
\hline
$S_{11}(1535)$& $A_{1/2}$ &  90 $\pm$30 & 91.0 $\pm$2.2 & 73 &  66 \\
\hline
$S_{31}(1620)$& $A_{1/2}$ &  27 $\pm$11 & 49.6$\pm$2.2 & 71  & 66 \\
\hline
$S_{11}(1650)$& $A_{1/2}$ &  53$\pm $16 & 22.2 $\pm$7.2 & 32 & 33 \\
\hline
$D_{15}(1675)$& $A_{1/2}$ &  19 $\pm$8 &  18.0 $\pm$2.3 &  23  & 15\\
              & $A_{3/2}$ &  15 $\pm$9 &  21.2 $\pm$1.4 &  24  & 22\\
\hline
$F_{15}(1680)$& $A_{1/2}$ & -15 $\pm$6 & -17.3 $\pm$1.4 & -25  & -25\\
              & $A_{3/2}$ & 133 $\pm$12& 133.6 $\pm$1.6 & 134  & 134\\
\hline
$D_{33}(1700)$& $A_{1/2}$ & 104 $\pm$15 & 125.4 $\pm$3.0 & 135 & 226\\
              & $A_{3/2}$ & 85  $\pm$22& 105.0  $\pm$3.2 & 213   &  210\\
\hline
$P_{13}(1720)$& $A_{1/2}$ &  18 $\pm$30&   96.6 $\pm$3.4   &  55  &  73\\
              & $A_{3/2}$ & -19 $\pm$20&  -39.0 $\pm$3.2  & -32  &  -11\\
\hline
$F_{35}(1905)$& $A_{1/2}$ & 26  $\pm$11 & 21.3  $\pm$3.6 & 14  &  18\\
              & $A_{3/2}$ & -45 $\pm$20 & -45.6$\pm$4.7 & -22  & -28\\
\hline
$F_{37}(1950)$& $A_{1/2}$ & -76 $\pm$12 &   & -78  & -94\\
              & $A_{3/2}$ & -97 $\pm$10 &   & -101  &  -121\\
\hline
\end{tabular}
\end{center}
\end{table}
\begin{table}[htbp]
\caption{Neutron helicity amplitudes at $Q^2=0$ for the major nucleon
resonances. GW02 are the results GWU/SAID analysis~\cite{SAID02}.
Further notation as in Tab.~\ref{tab:helicity_p}.}\label{tab:helicity_n}
\begin{center}
\small
\begin{tabular}{|l l|c c c c|} \hline
&  &  PDG & GW02 & 2003 & 2007 \\
\hline
$P_{11}(1440)$& $A_{1/2}$ & 40$\pm$10 & 47$\pm$5 & 52  & 54 \\
\hline
$D_{13}(1520)$& $A_{1/2}$ & -59 $\pm$9 & -67 $\pm$4 & -85  & -77\\
              & $A_{3/2}$ &-139 $\pm$11&-112 $\pm$3 &-148  & -154\\
\hline
$S_{11}(1535)$& $A_{1/2}$ & -46$\pm$27 & -16$\pm$5 & -42   & -51 \\
\hline
$S_{11}(1650)$& $A_{1/2}$ & -15$\pm$21 &-28$\pm$4 & 27     & 9\\
\hline
$D_{15}(1675)$& $A_{1/2}$ & -43 $\pm$12& -50 $\pm$4 & -61  & -62\\
              & $A_{3/2}$ & -58 $\pm$13& -71 $\pm$5 & -74  & -84\\
\hline
$F_{15}(1680)$& $A_{1/2}$ &  29 $\pm$10&  29 $\pm$6 &  25  & 28\\
              & $A_{3/2}$ & -33 $\pm$9 & -58 $\pm$9 & -35  &-38\\
\hline
$P_{13}(1720)$& $A_{1/2}$ &   1 $\pm$15&            &  17  & -3\\
              & $A_{3/2}$ & -29 $\pm$61&            & -75  & -31\\
\hline
\end{tabular}
\end{center}
\end{table}
Next we present our results for the multipoles starting with the threshold
region. In Fig.~\ref{fig:E0thr} we demonstrate the effects of the low-energy
correction and the cusp effect for $\pi^0$ photoproduction, as described by
Eqs.~(\ref{eq:3.16}) and (\ref{eq:3.17}), respectively. The prediction of
MAID98 for $\pi^0$ photoproduction at threshold (dotted lines) lies
substantially below the data. In accordance with Ref.~\cite{DMTthr}, the
phenomenological term $E^{corr}_{0+}$ simulates the pion off-shell rescattering
or pion-loop contributions of ChPT. The cusp term of Eq.~(\ref{eq:3.17})
describes the strong energy dependence near $\pi^+$ threshold, which has its
origin in the pion mass difference and the strong coupling with the $\pi^+n$
channel. The figure shows that the off-shell pion rescattering substantially
improves the agreement with the data. However, one problem still remains in the
threshold region. The experimental photon asymmetry $\Sigma$ in $\pi^0$
photoproduction at $E_{\gamma}\approx 160$ MeV takes positive values, whereas
the MAID results are negative in this region. As has been demonstrated in
Refs.~\cite{DMTthr,HDT}, this observable is very sensitive to the $M_{1-}$
multipole which strongly depends on the details of the low-energy behavior of
Roper resonance, vector meson and off-shell pion rescattering contributions.
Therefore, a slight modification of one or all of these
mechanisms can drastically change the photon asymmetry.\\

\begin{figure*}[ht]
\centerline{\epsfig{file=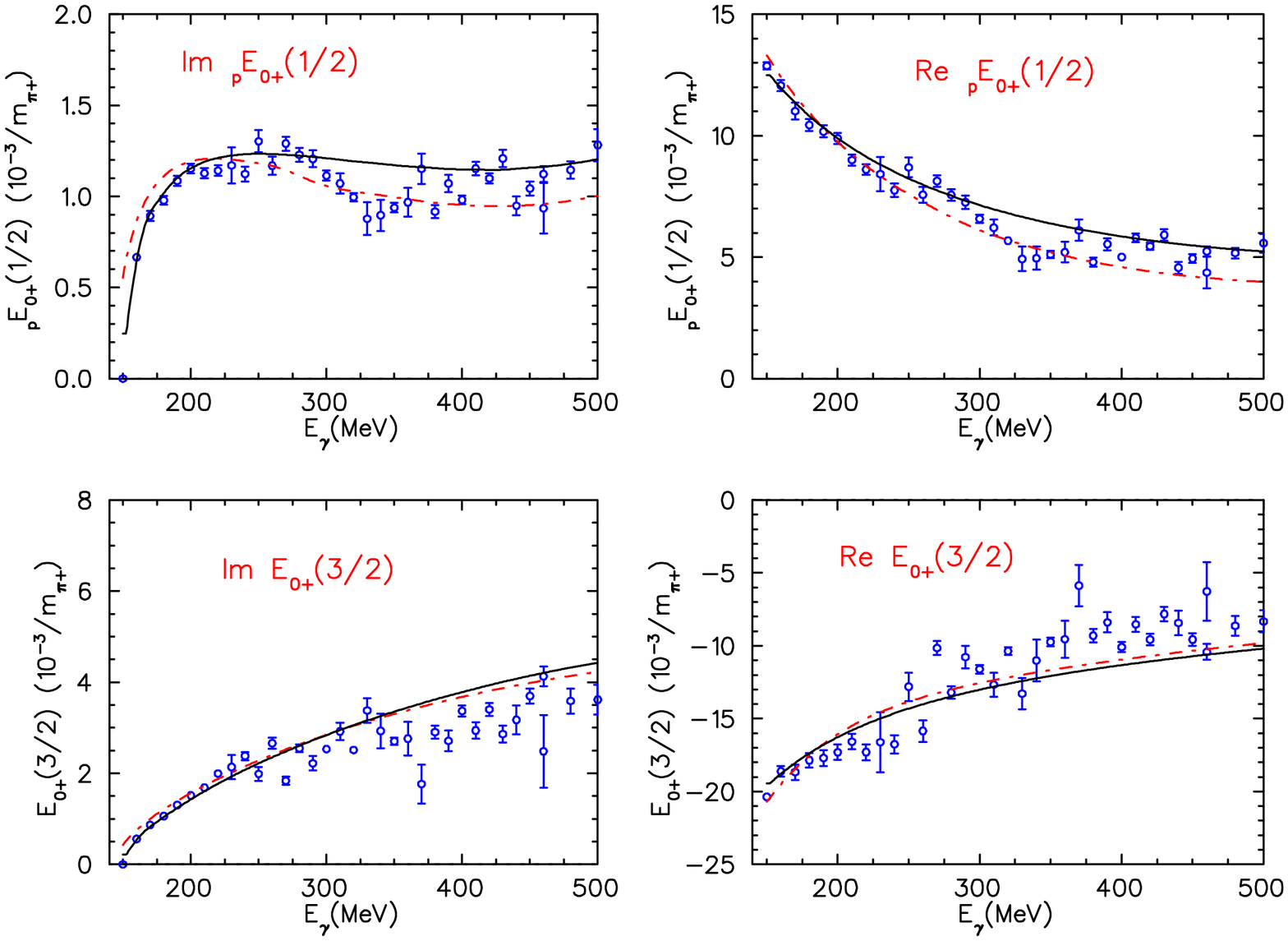, width=10 cm, angle=90} }
\vspace{0.5cm}
\centerline{\epsfig{file=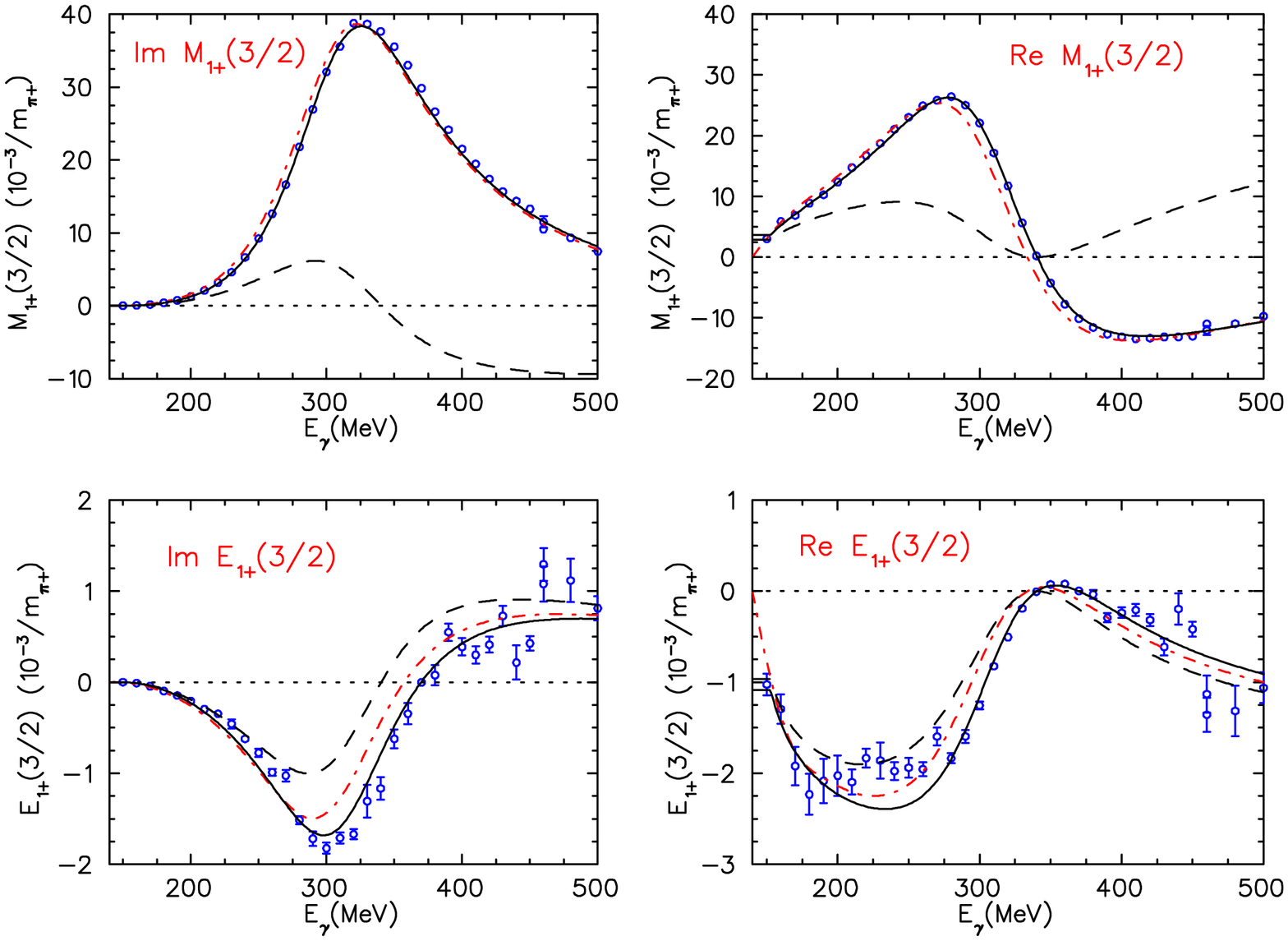, width=10cm, angle=90} }
\caption{The global solutions of MAID2007 (solid
lines) and GWU/SAID~\cite{SAID06} (red dashed-dotted lines,
solution FA06K) for the multipoles $_pE_{0+}^{1/2}$,
$E_{0+}^{3/2}$, $M_{1+}^{3/2}$, and $E_{1+}^{3/2}$ as function of
the photon lab energy $E_{\gamma}$ in the first resonance region.
The blue open circles show our single-energy solution. The dashed
lines represent our unitarized background contributions in the
$M_{1+}^{3/2}$ and $E_{1+}^{3/2}$ multipoles.}
\label{fig:2_photo}
\end{figure*}

\begin{figure*}[ht]
\centerline{\epsfig{file=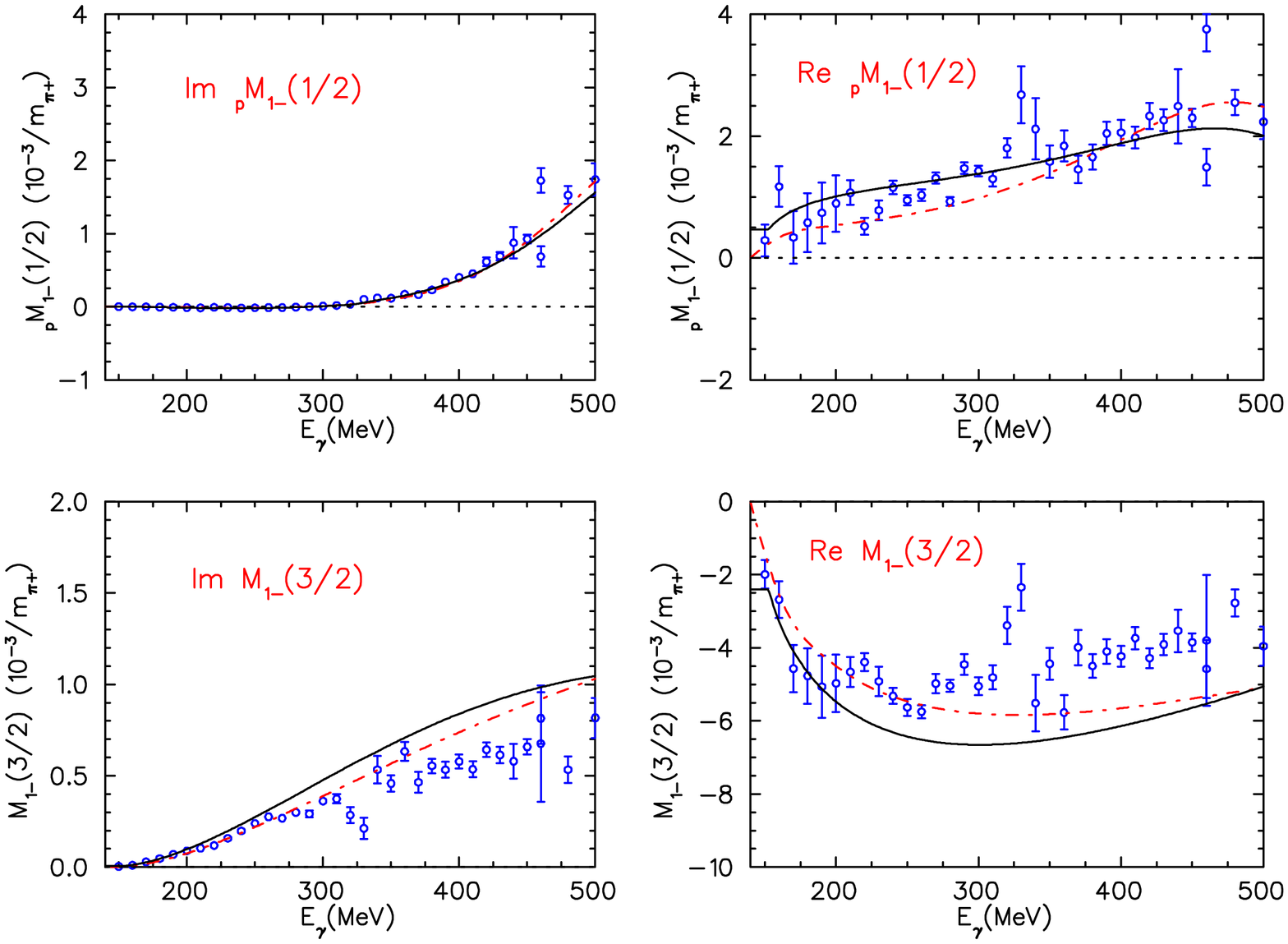, width=10 cm, angle=90}}
\vspace{0.5cm}
\centerline{\epsfig{file=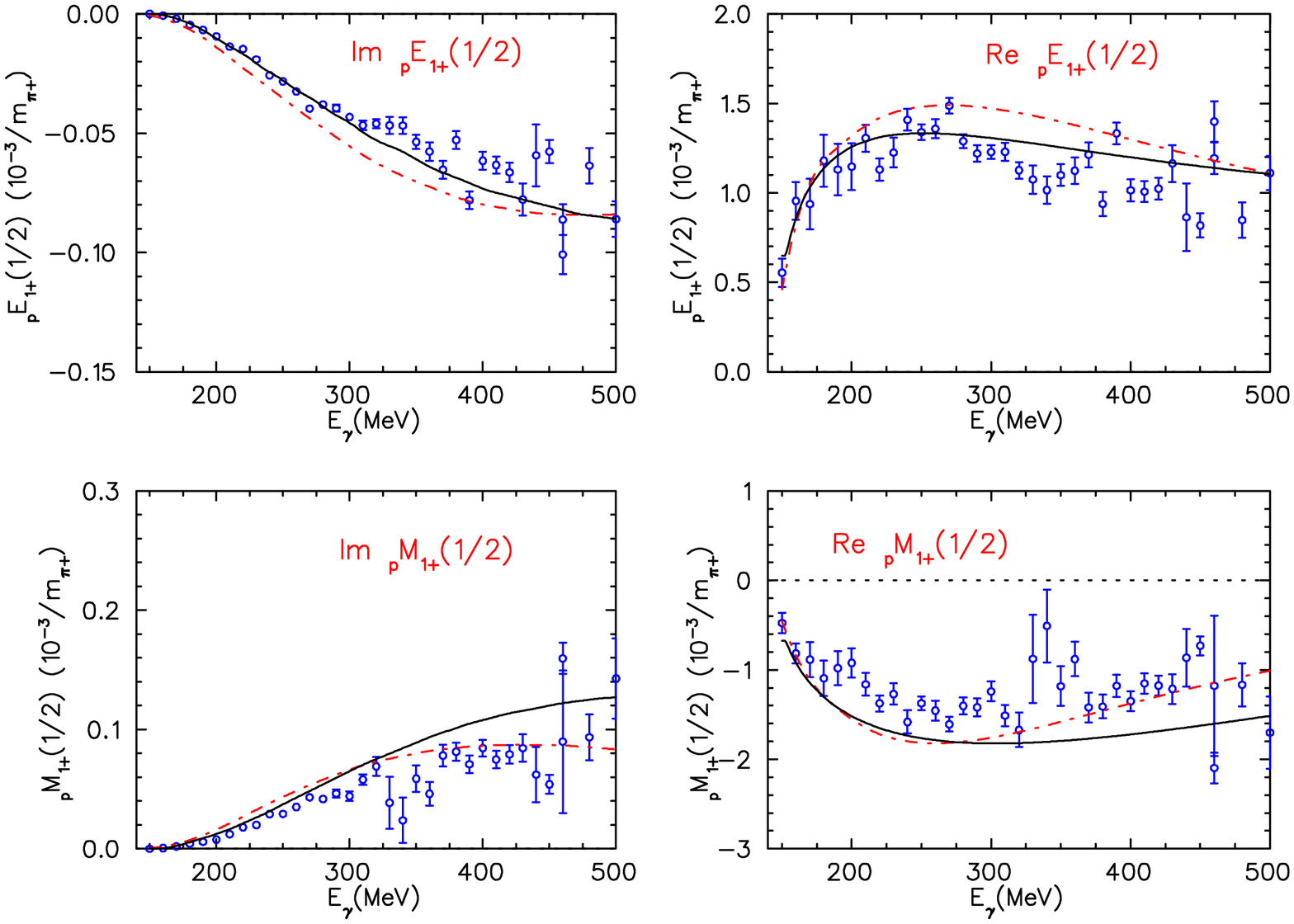, width=10 cm, angle=90}}
\caption{The multipoles $_pM_{1-}^{1/2}$, $M_{1-}^{3/2}$, $_pE_{1+}^{1/2}$, and
$_pM_{1+}^{1/2}$ as function of the photon lab energy $E_{\gamma}$. Further
notation as in Fig.~\ref{fig:2_photo}.}
\label{fig:3_photo}
\end{figure*}

\begin{figure*}[ht]
\centerline{\epsfig{file=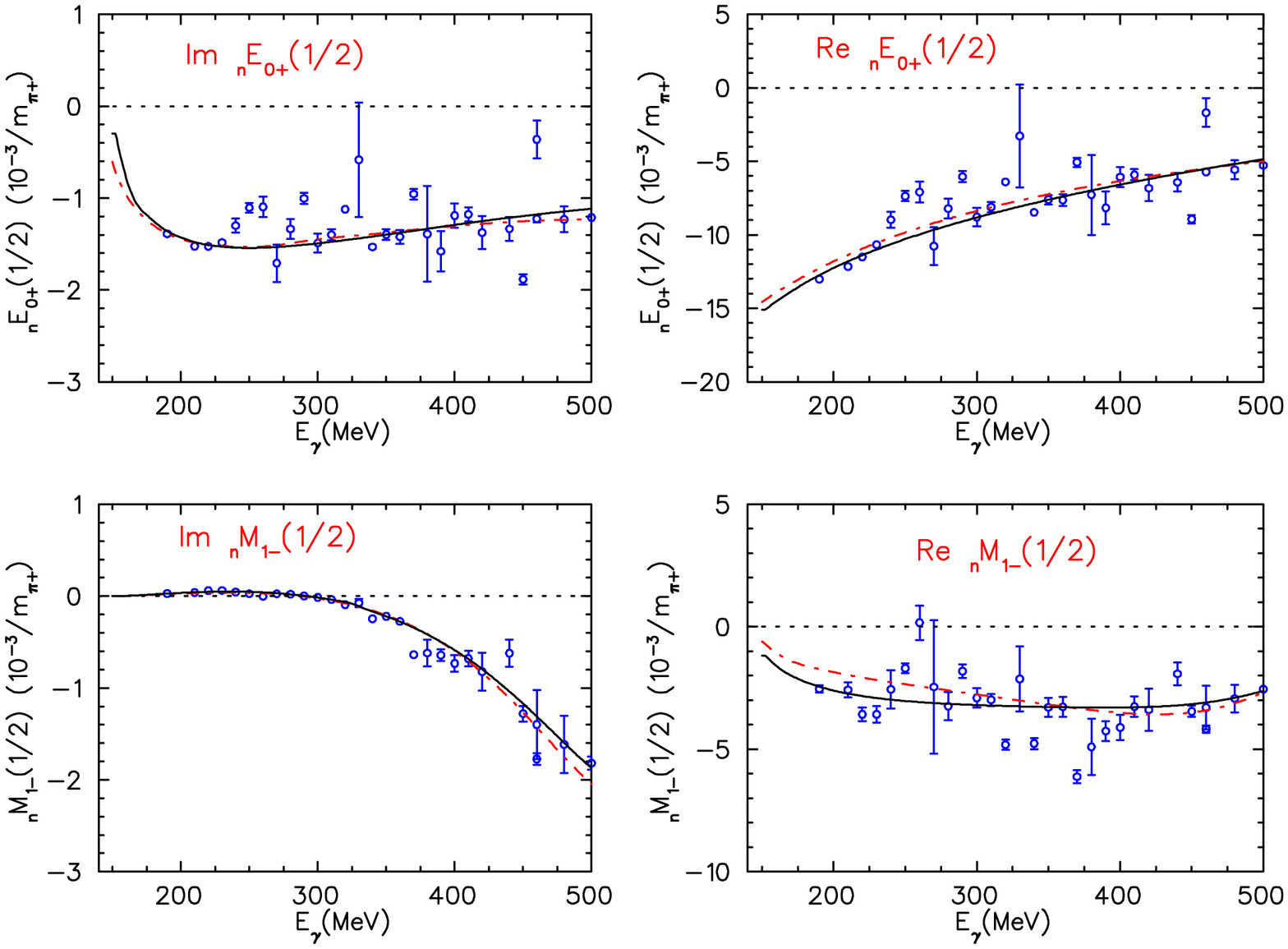, width=10 cm, angle=90}}
\vspace{0.5cm}
\centerline{\epsfig{file=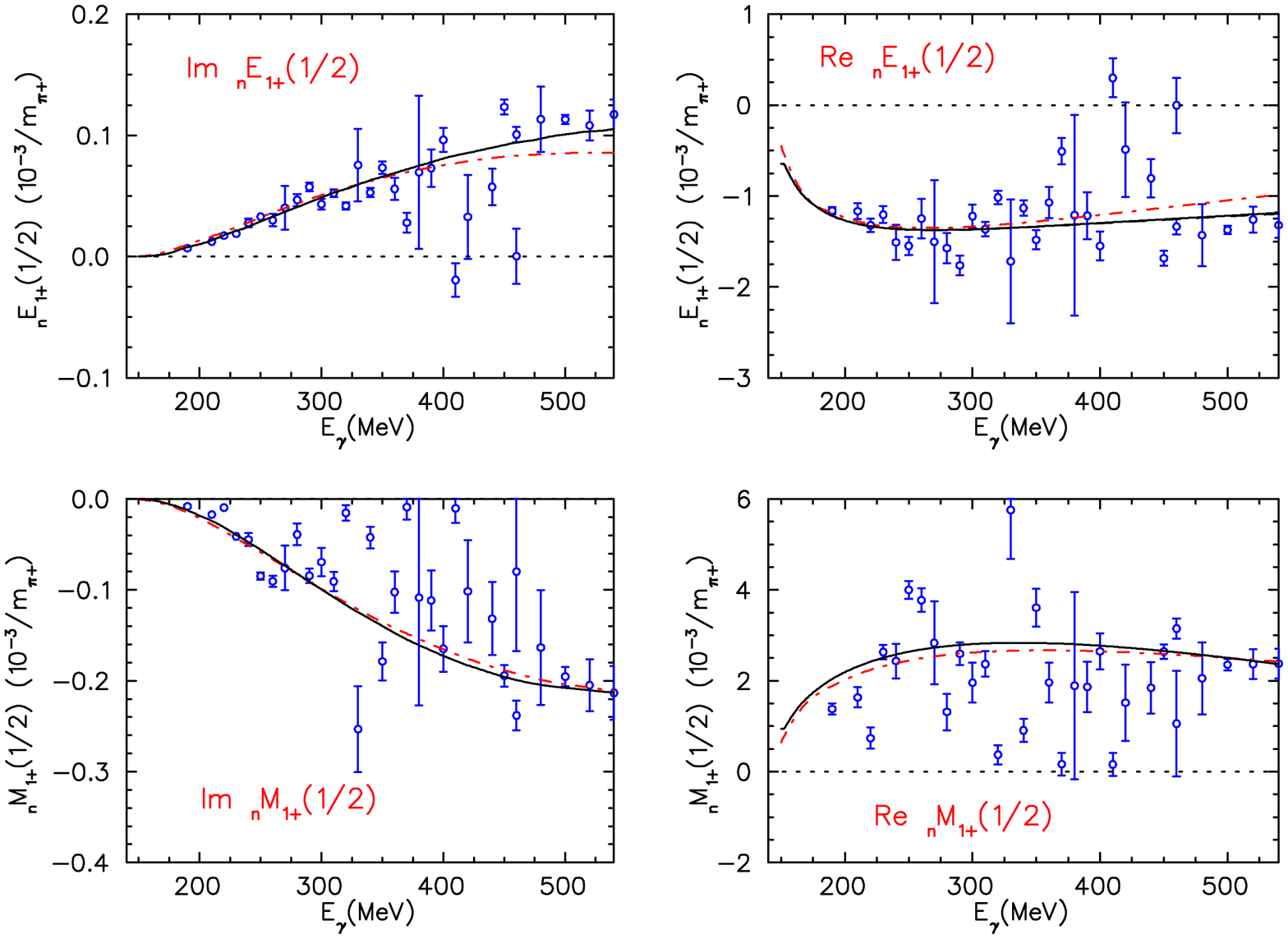, width=10 cm, angle=90}}
\caption{The multipoles $_nE_{0+}^{1/2}$, $_nM_{1-}^{1/2}$, $_nE_{1+}^{1/2}$,
and $_nM_{1+}^{1/2}$ as function of the photon lab energy $E_{\gamma}$. Further
notation as in Fig.~\ref{fig:2_photo}.}
\label{fig:4_photo}
\end{figure*}

\begin{figure*}[ht]
\centerline{\epsfig{file=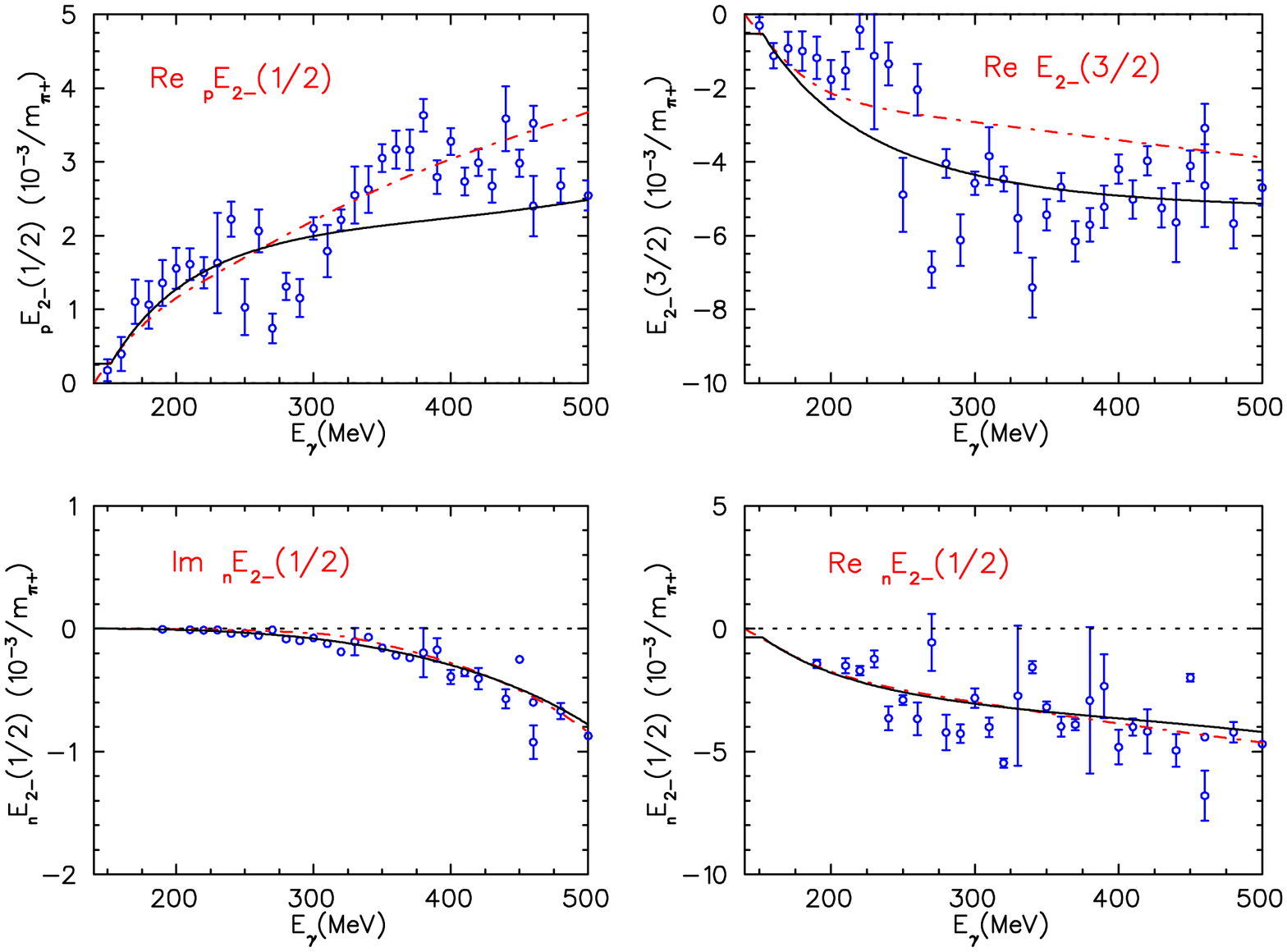, width=10 cm, angle=90}}
\caption{The multipoles $_pE_{2-}^{1/2}$, $E_{2-}^{3/2}$,
and $_nE_{2-}^{1/2}$ as function
of the photon lab energy $E_{\gamma}$. Further notation as in
Fig.~\ref{fig:2_photo}.}
\label{fig:5_photo}
\end{figure*}

Figures~\ref{fig:2_photo}-\ref{fig:4_photo} display the results for the most
important $S$ and $P$ waves in the $\Delta$(1232) region. However, a look at
Fig.~\ref{fig:5_photo} shows that also the $D$-wave amplitudes
$_pE_{2-}^{1/2}$, $E_{2-}^{3/2}$, and $_nE_{2-}^{1/2}$, give sizable
contributions in this region, in particular through their real parts. In these
figures, we present the MAID and SAID global (energy dependent) solutions,
together with our local (single energy) fit obtained for energy bins of 10 MeV.
In general the MAID and SAID results are close, which is not too surprising
because the phases are constrained by the Fermi-Watson theorem. However, there
are much larger differences in the $_pE_{0+}^{1/2}$ and $_pE_{2-}^{1/2}$
amplitudes, which indicates that the present data base is still too limited to
determine these background amplitudes in a reliable way.\\

\begin{figure*}[ht]
\centerline{\epsfig{file=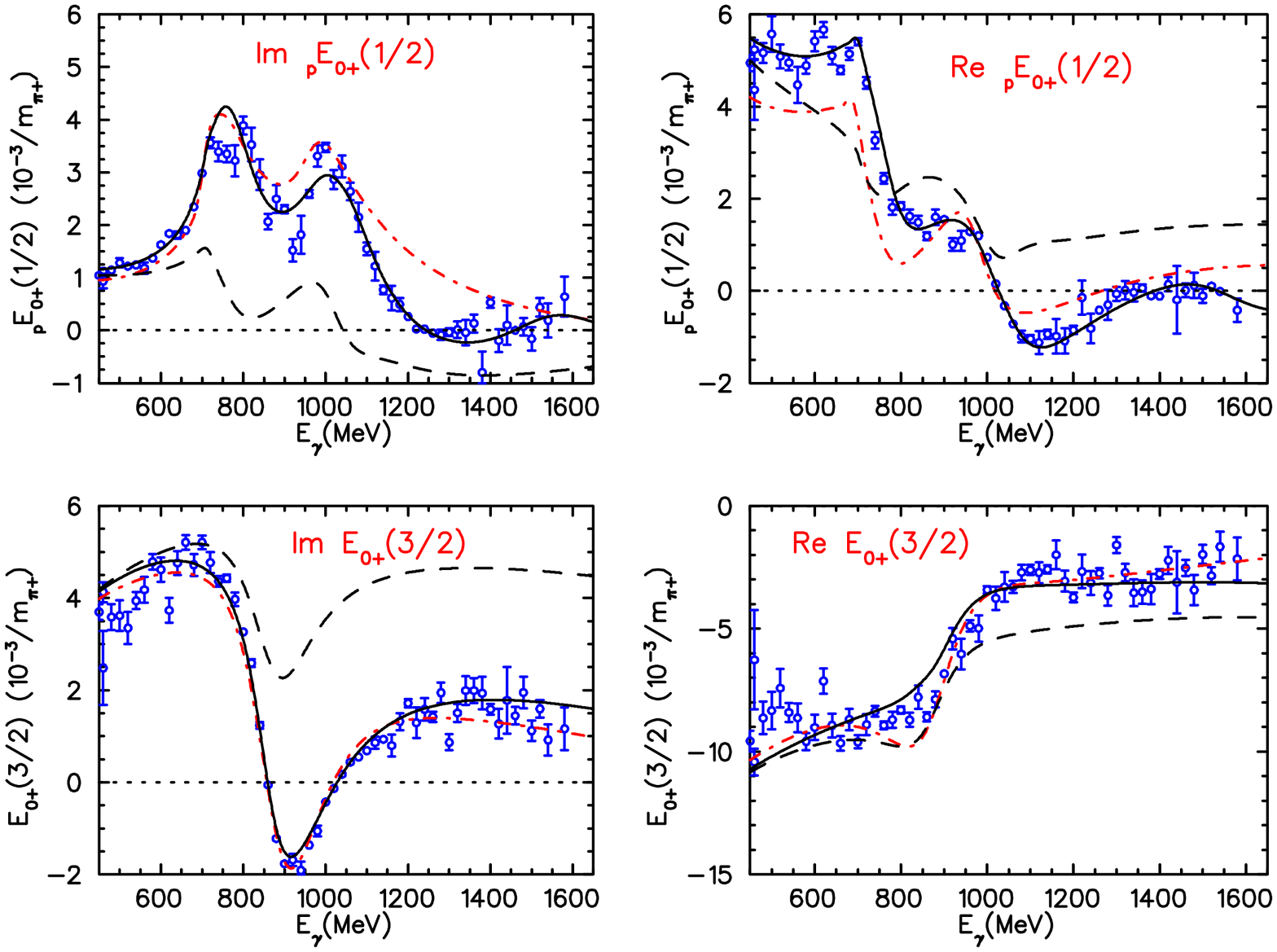, width=10 cm, angle=90}}
\vspace{0.5cm}
\centerline{\epsfig{file=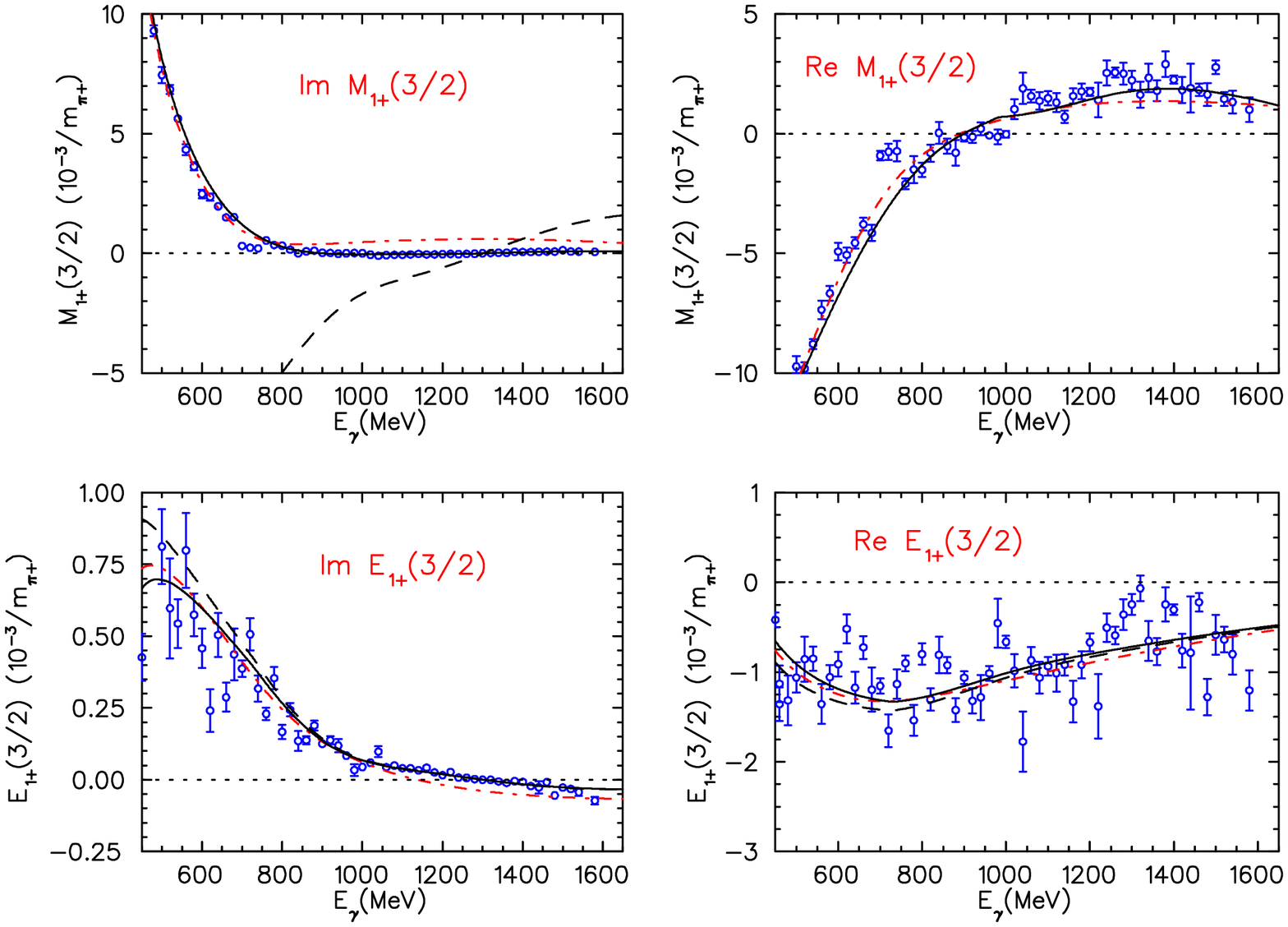, width=10 cm, angle=90}}
\caption{The global solutions of MAID2007 (solid lines) and
GWU/SAID~\cite{SAID06} (red dashed-dotted lines, solution FA06K) for the
multipoles $_pE_{0+}^{1/2}$, $E_{0+}^{3/2}$, $M_{1+}^{3/2}$, and $E_{1+}^{3/2}$
as function of the photon lab energy $E_{\gamma}$ in the second and third
resonance regions. The blue open circles show our single-energy solution. The
dashed lines represent our unitarized background given by Eq.~(\ref{eq:3.15}).
Note that the background for Re $M_{1+}^{3/2}$ is out of scale.}
\label{fig:6_photo}
\end{figure*}

\begin{figure*}[ht]
\centerline{\epsfig{file=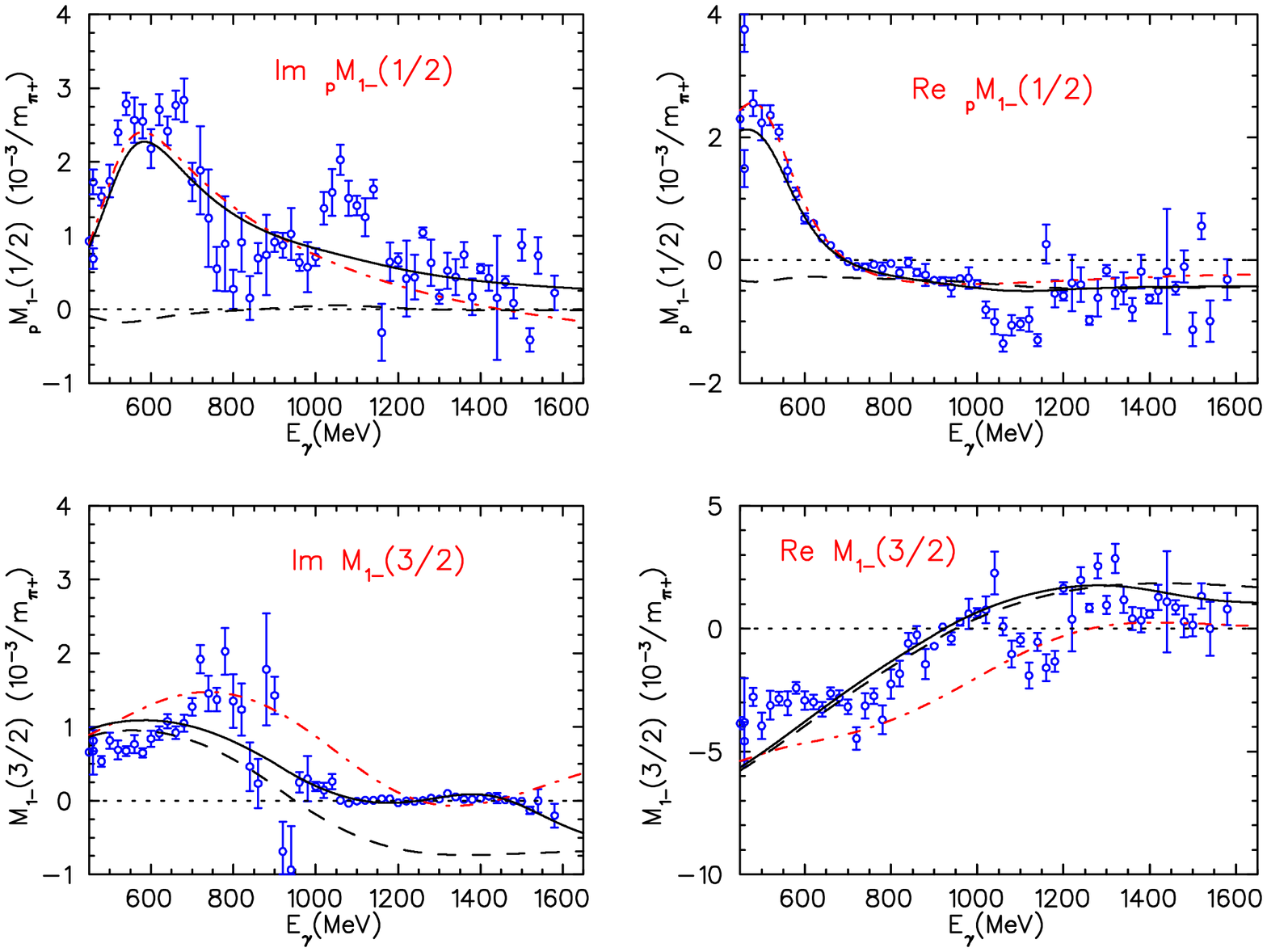, width=10 cm, angle=90}}
\vspace{0.5cm}
\centerline{\epsfig{file=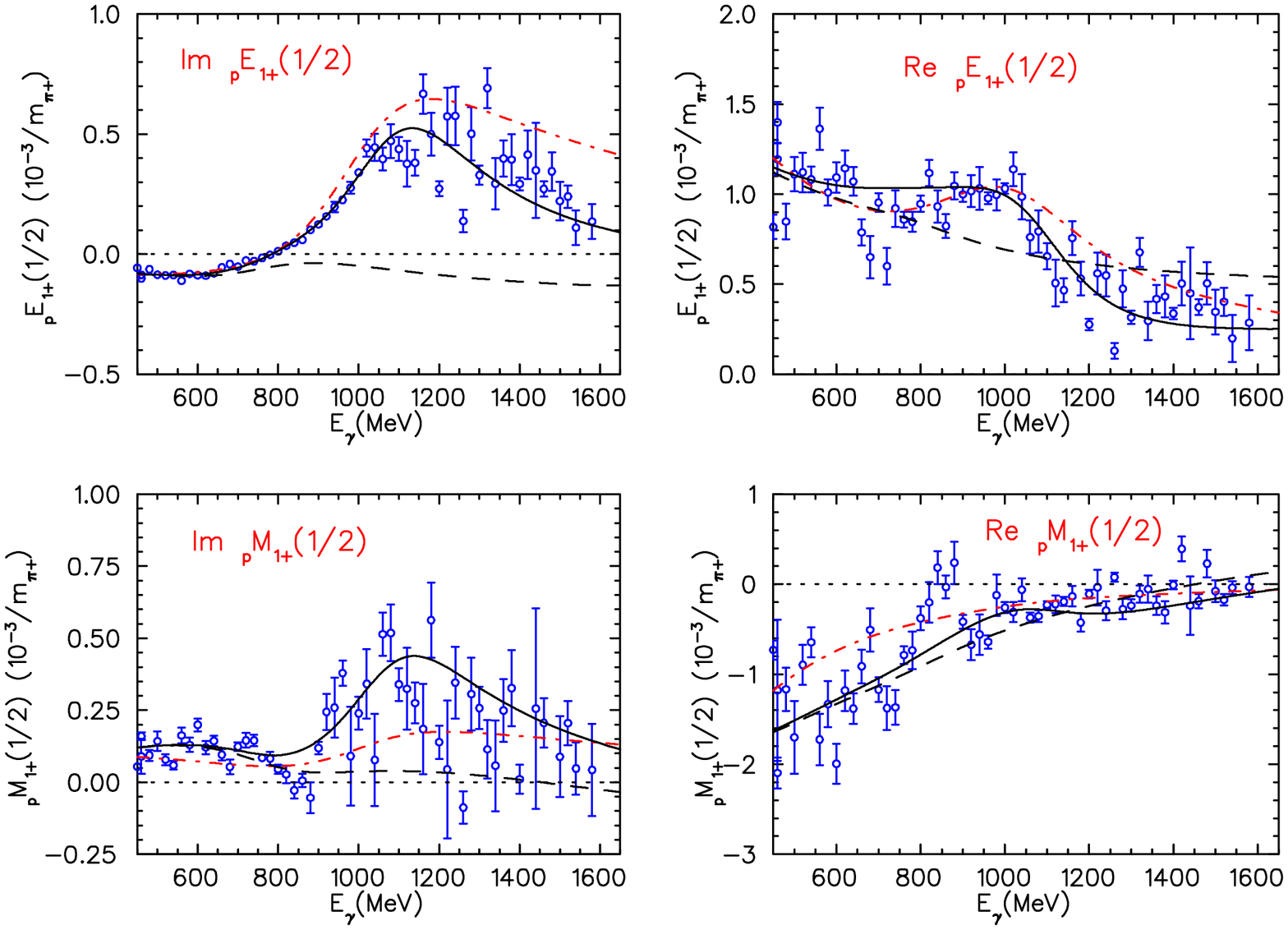, width=10 cm, angle=90}}
\caption{The global solutions of MAID2007 (black solid lines) and
GWU/SAID~\cite{SAID06} (red dashed-dotted lines) for the multipoles
$_pM_{1-}^{1/2}$, $M_{1-}^{3/2}$, $_pE_{1+}^{1/2}$, and $_pM_{1+}^{1/2}$ as
function of the photon lab energy $E_{\gamma}$ in the second and third
resonance regions. Further notation as in Fig.~\ref{fig:6_photo}.}
\label{fig:7_photo}
\end{figure*}

\begin{figure*}[ht]
\centerline{\epsfig{file=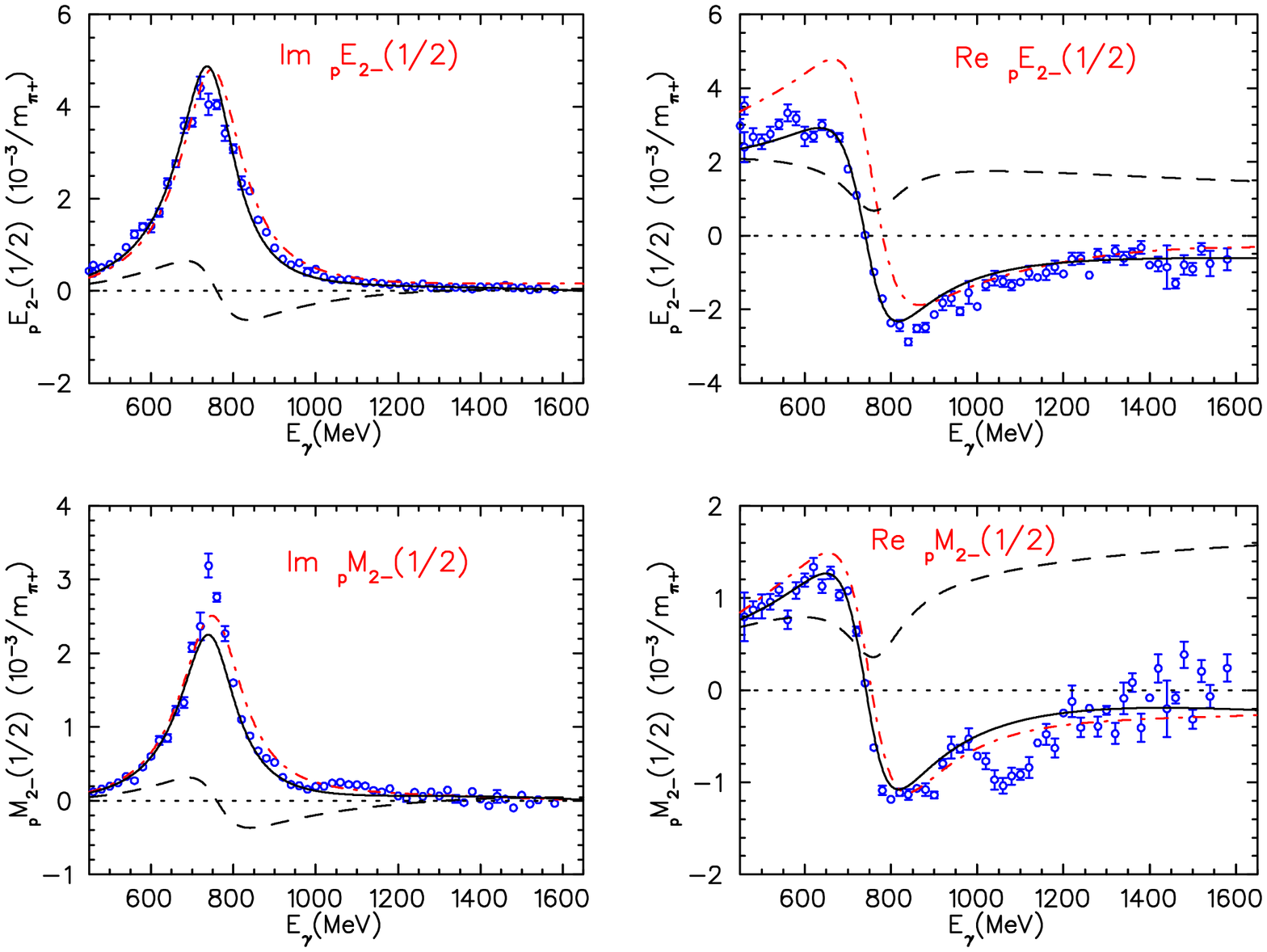, width=10 cm, angle=90}}
\vspace{0.5cm}
\centerline{\epsfig{file=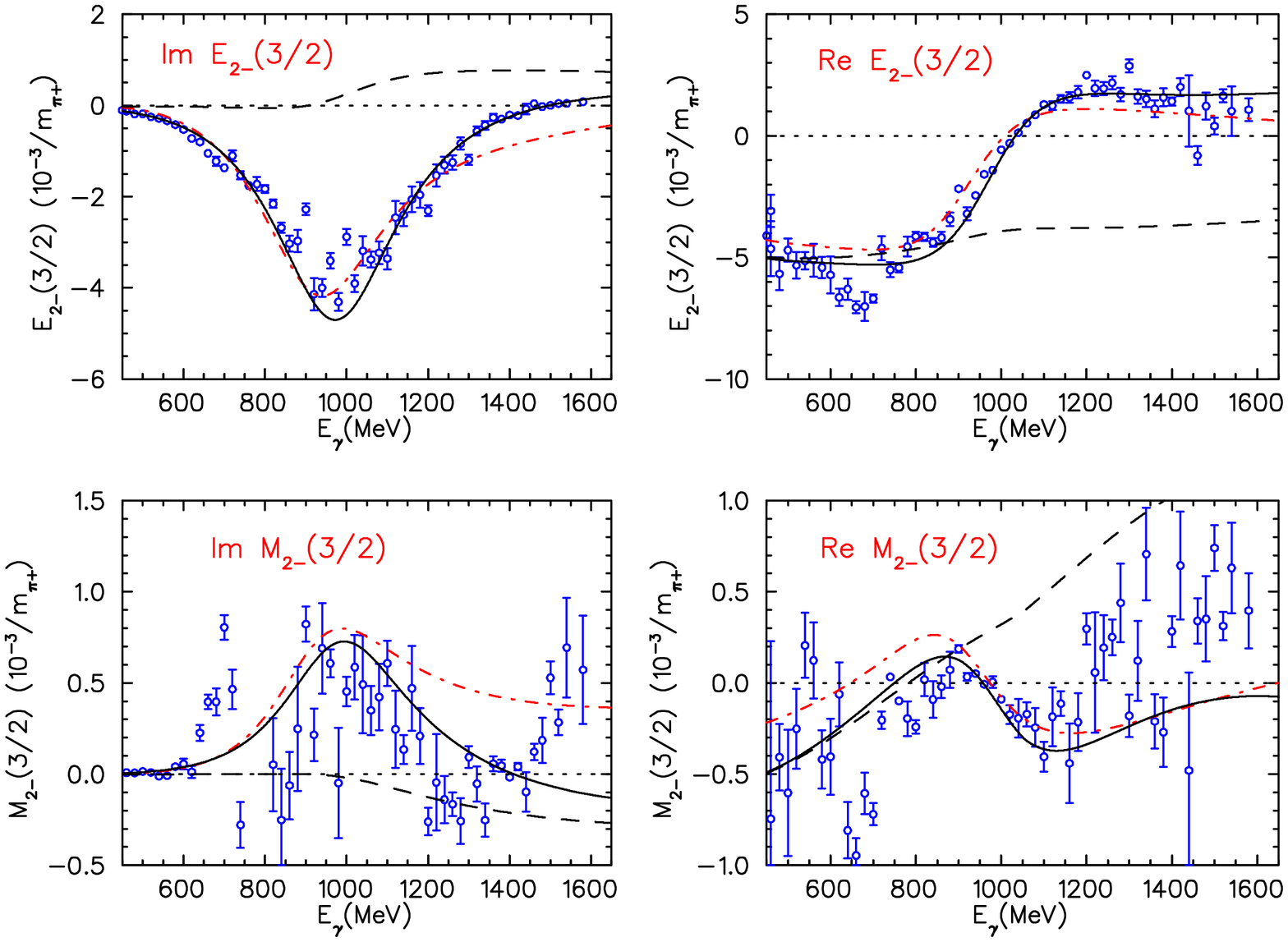, width=10 cm, angle=90}}
\caption{The global solutions of MAID2007 (black solid lines) and GWU/SAID~\cite{SAID06}
(red dashed-dotted lines) for the multipoles $_pE_{2-}^{1/2}$,
$_pM_{2-}^{1/2}$, $E_{2-}^{3/2}$, and $M_{2-}^{3/2}$ as function of the photon
lab energy $E_{\gamma}$ in the second and third resonance regions. Further
notation as in Fig.~\ref{fig:6_photo}.} \label{fig:8_photo}
\end{figure*}

\begin{figure*}[ht]
\centerline{\epsfig{file=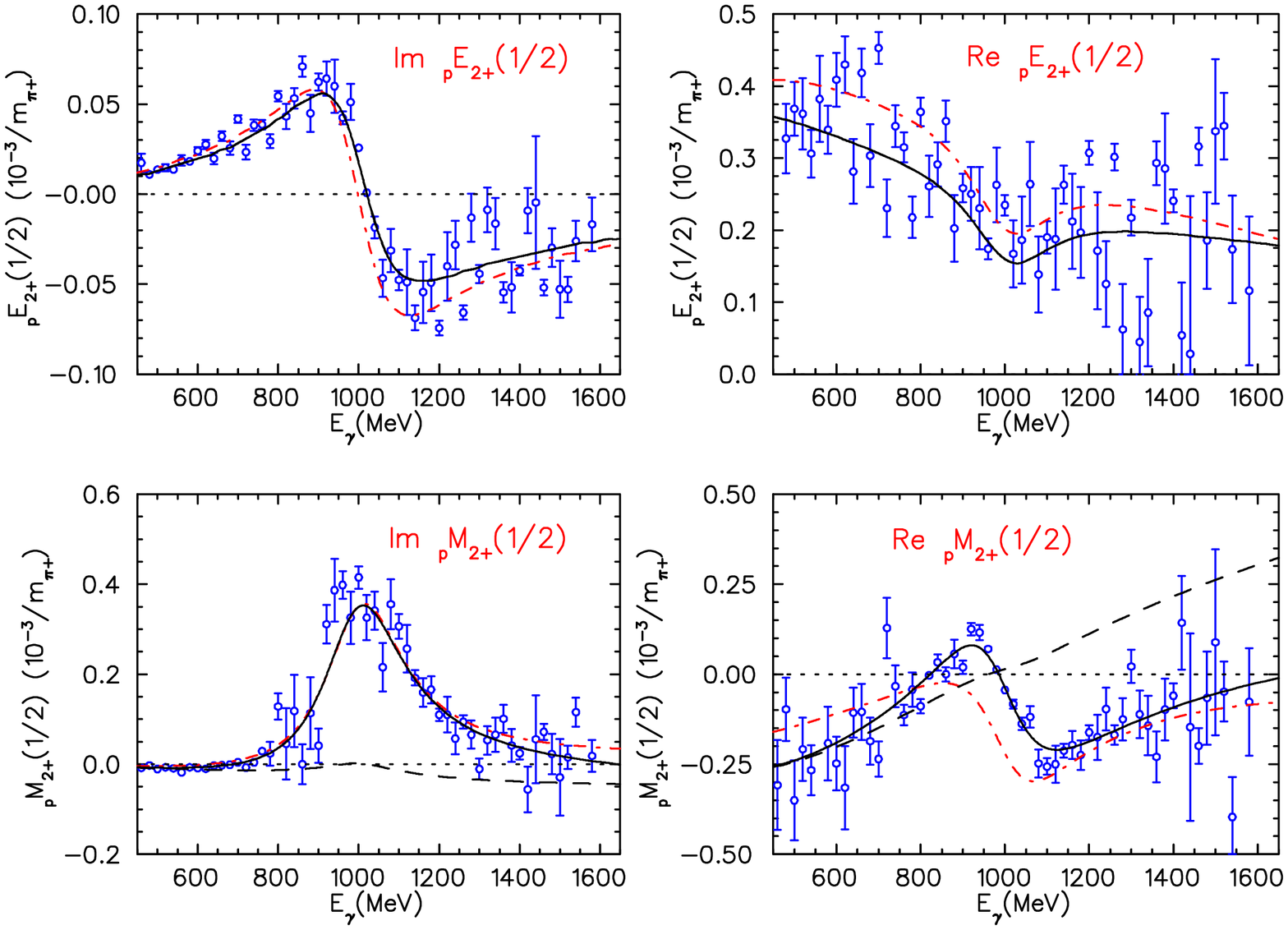, width=10 cm, angle=90}}
\vspace{0.5cm}
\centerline{\epsfig{file=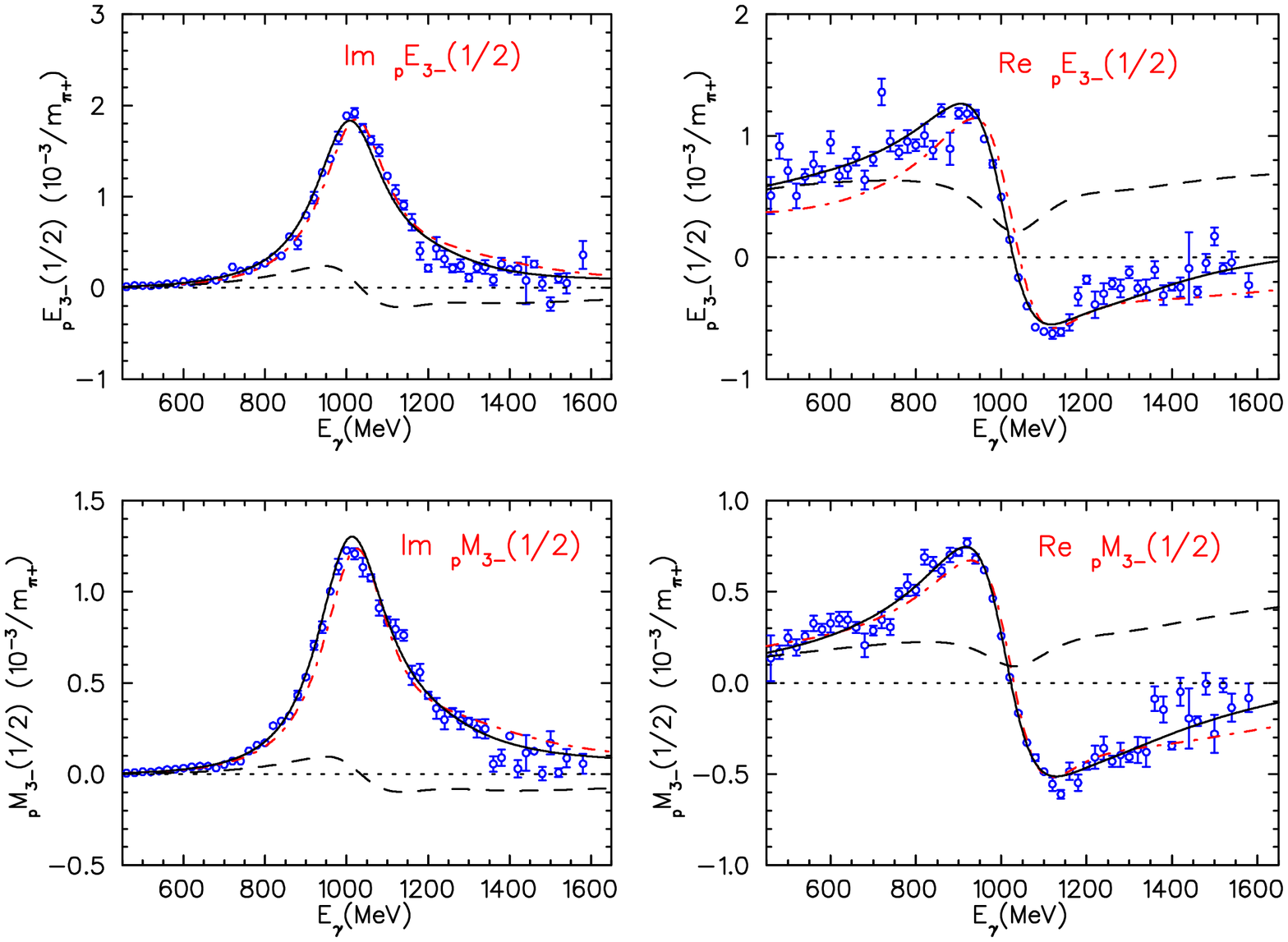, width=10 cm, angle=90}}
\caption{The global solutions of MAID2007 (solid lines) and GWU/SAID~\cite{SAID06} (red
dashed-dotted lines) for the multipoles $_pE_{2+}^{1/2}$, $_pM_{2+}^{1/2}$,
$_pE_{3-}^{1/2}$, and $_pM_{3-}^{1/2}$ as function of the photon lab energy
$E_{\gamma}$ in the second and third resonance regions. Further notation as in
Fig.~\ref{fig:6_photo}. The resonance contribution to the $_pE_{2+}^{1/2}$
multipole is very small, and therefore the solid and dashed lines coincide.}
\label{fig:9_photo}
\end{figure*}

\begin{figure*}[ht]
\centerline{\epsfig{file=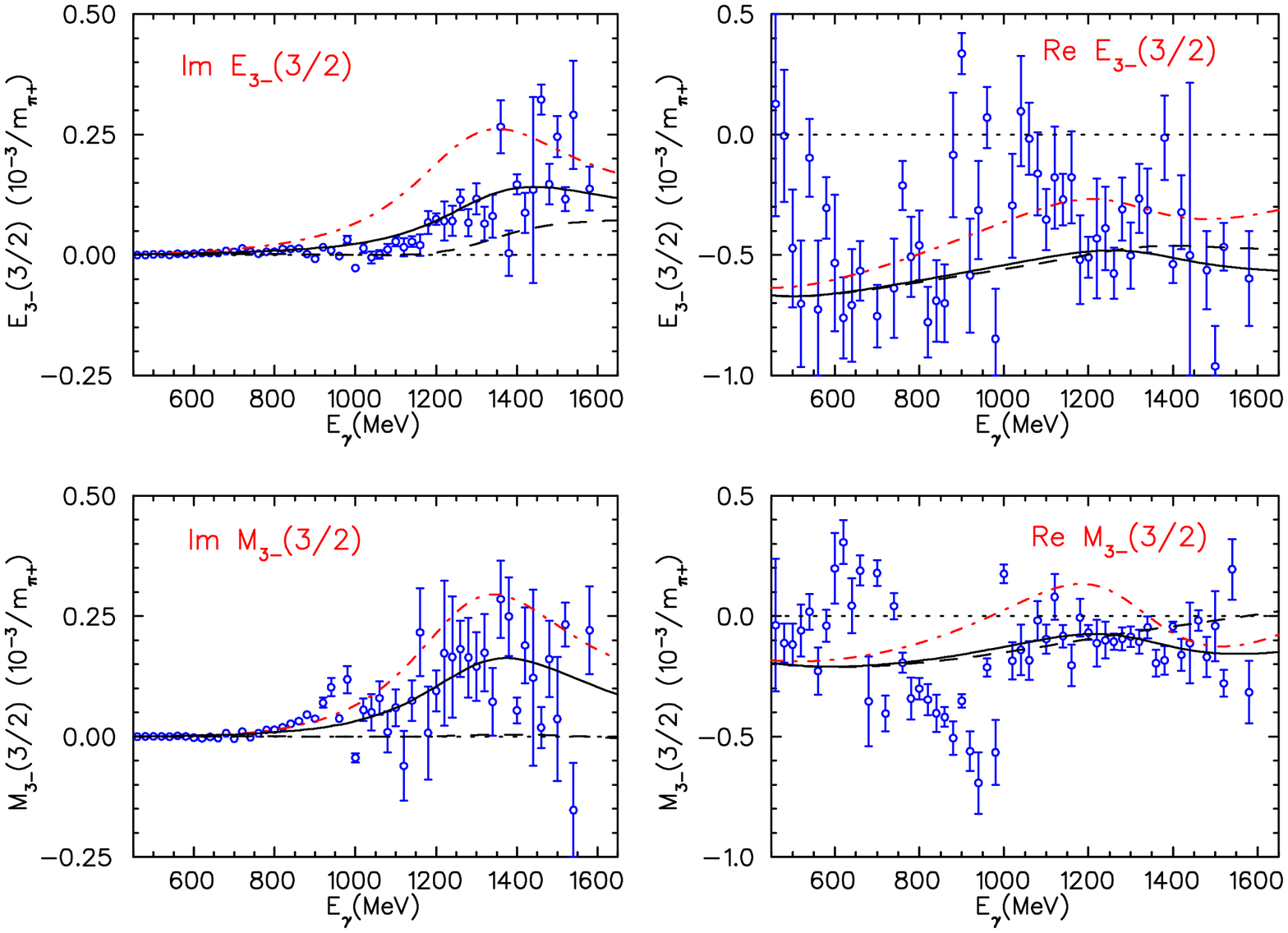, width=10 cm, angle=90}}
\vspace{0.5cm}
\centerline{\epsfig{file=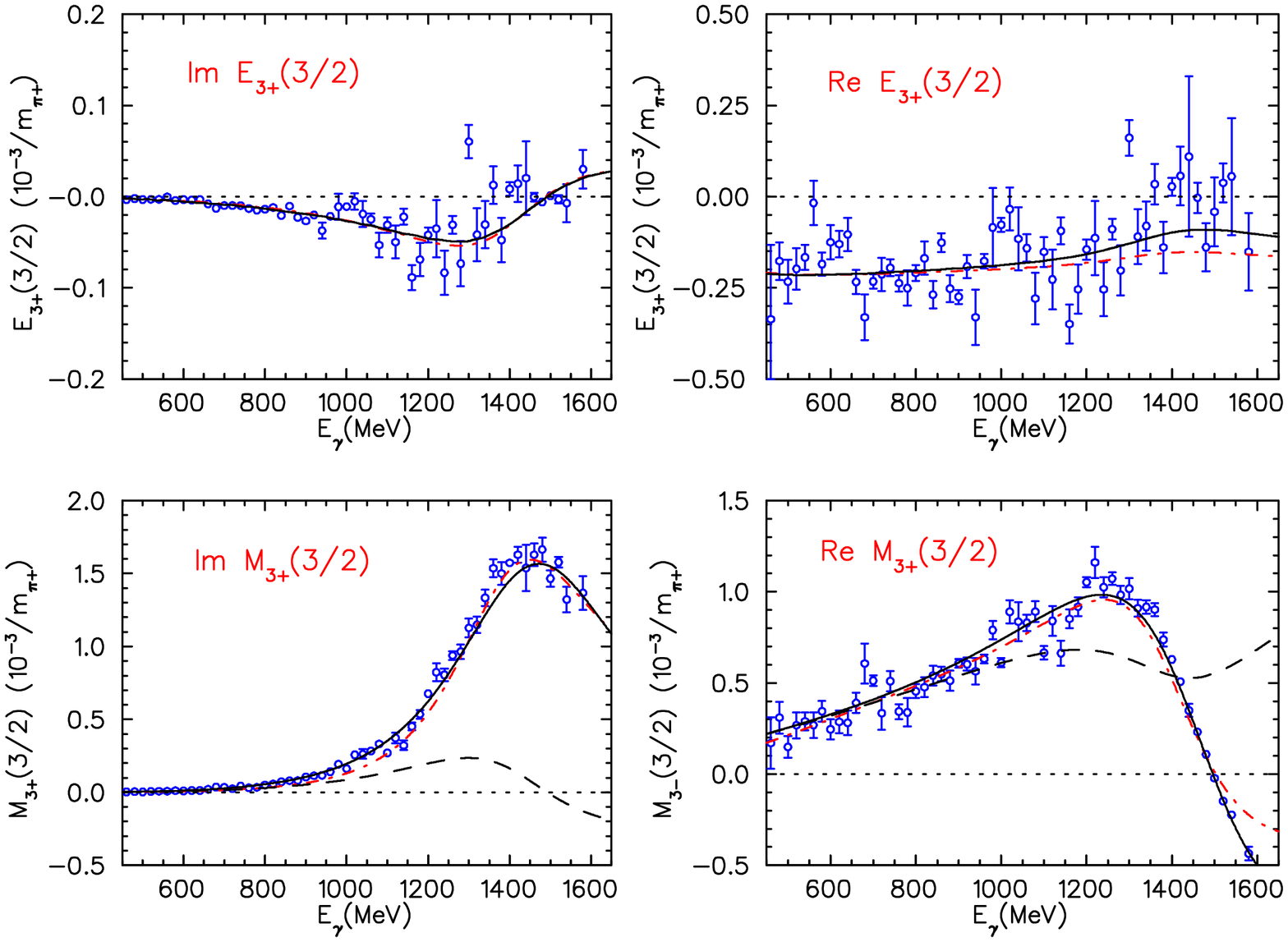, width=10 cm, angle=90}}
\caption{The global solutions of MAID2007 (solid lines) and GWU/SAID~\cite{SAID06} (red
dashed-dotted lines) for the multipoles $E_{3-}^{3/2}$, $M_{3-}^{3/2}$,
$E_{3+}^{3/2}$, and $M_{3+}^{3/2}$ as function of the photon lab energy
$E_{\gamma}$ in the second and third resonance regions. Further notation as in
Fig.~\ref{fig:6_photo}. The resonance contribution to the $E_{3+}^{3/2}$
multipole is very small, and therefore the solid and dashed lines coincide.}
\label{fig:10_photo}
\end{figure*}

\begin{figure*}[ht]
\centerline{\epsfig{file=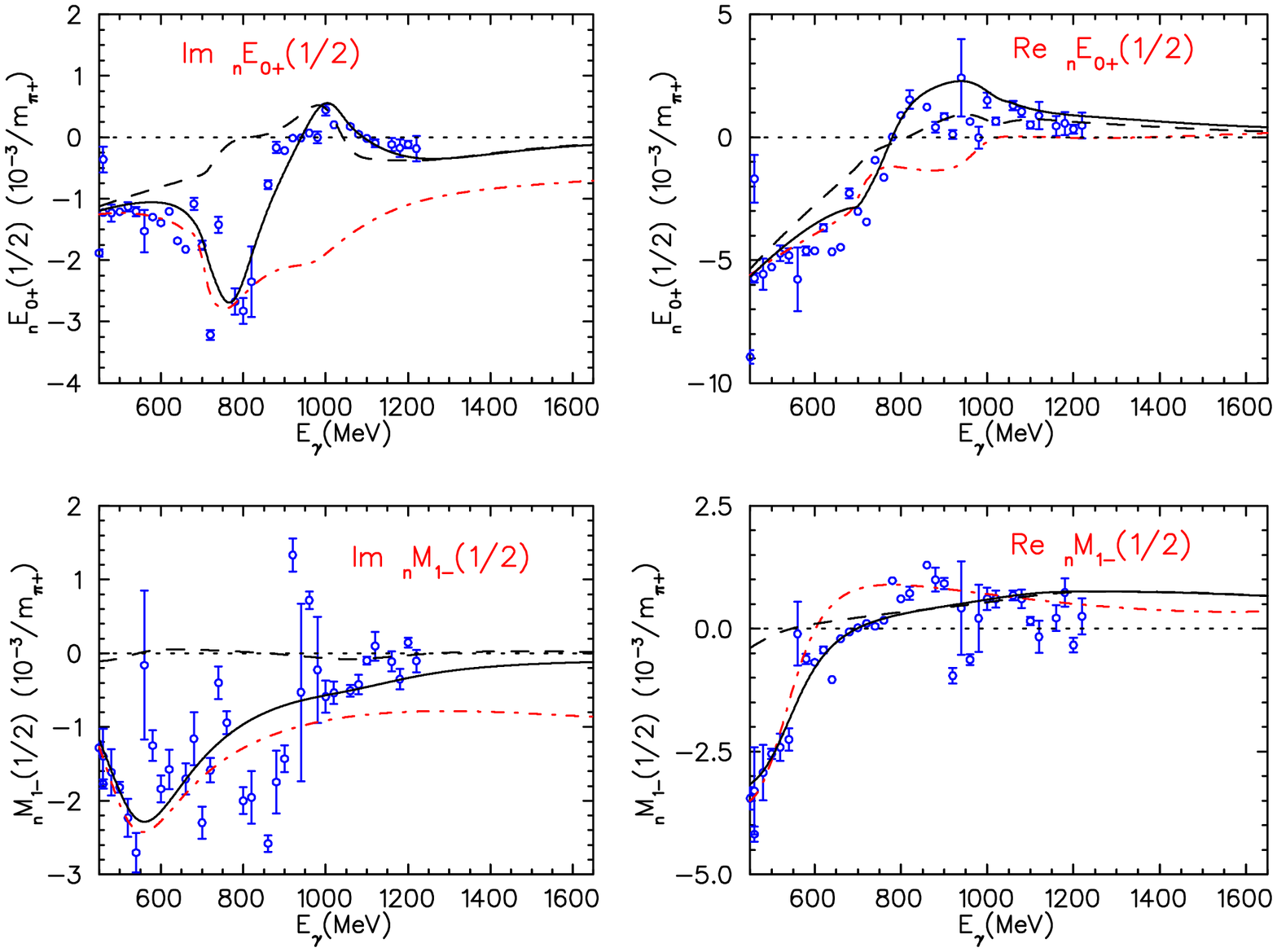, width=10 cm, angle=90}}
\vspace{0.5cm}
\centerline{\epsfig{file=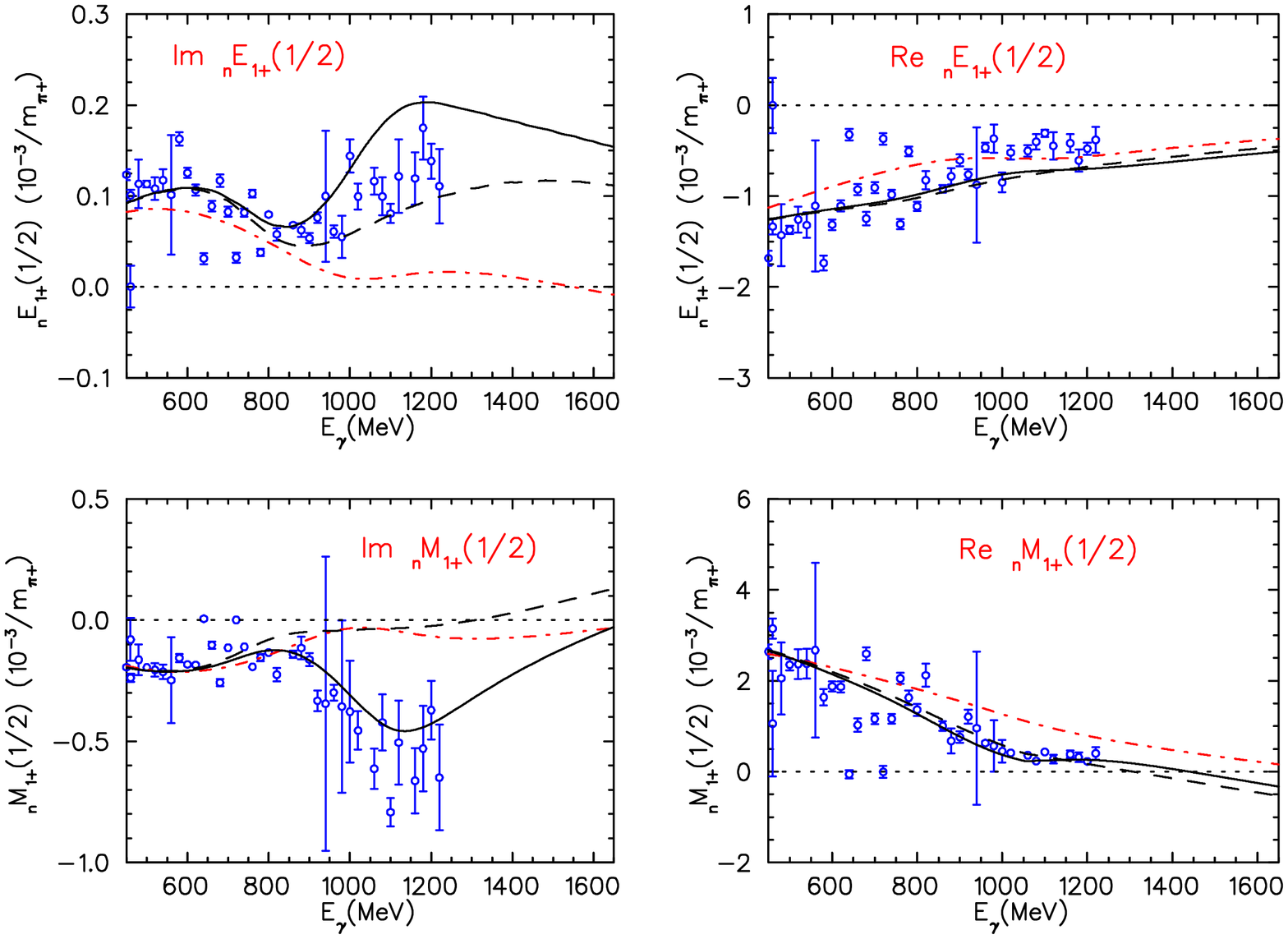, width=10 cm, angle=90}}
\caption{The global solutions of MAID2007 (solid lines) and
GWU/SAID~\cite{SAID06} (red dashed-dotted lines) for the multipoles
$_nE_{0+}^{1/2}$, $_nM_{1-}^{1/2}$, $_nE_{1+}^{1/2}$, and $_nM_{1+}^{1/2}$ as
function of the photon lab energy $E_{\gamma}$ in the second and third
resonance regions. Further notation as in Fig.~\ref{fig:6_photo}.}
\label{fig:11_photo}
\end{figure*}

\begin{figure*}[ht]
\centerline{\epsfig{file=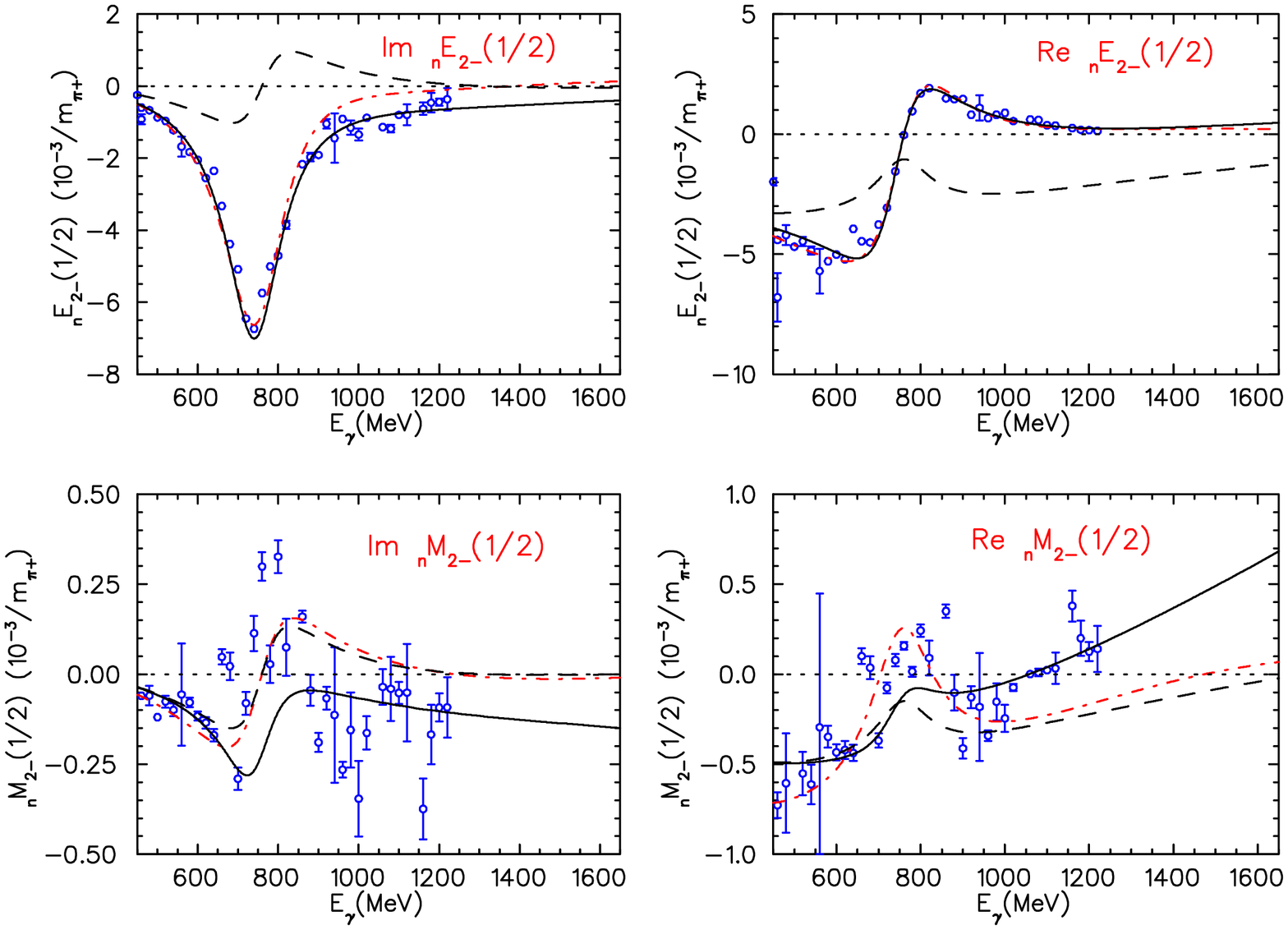, width=10 cm, angle=90}}
\vspace{0.5cm}
\centerline{\epsfig{file=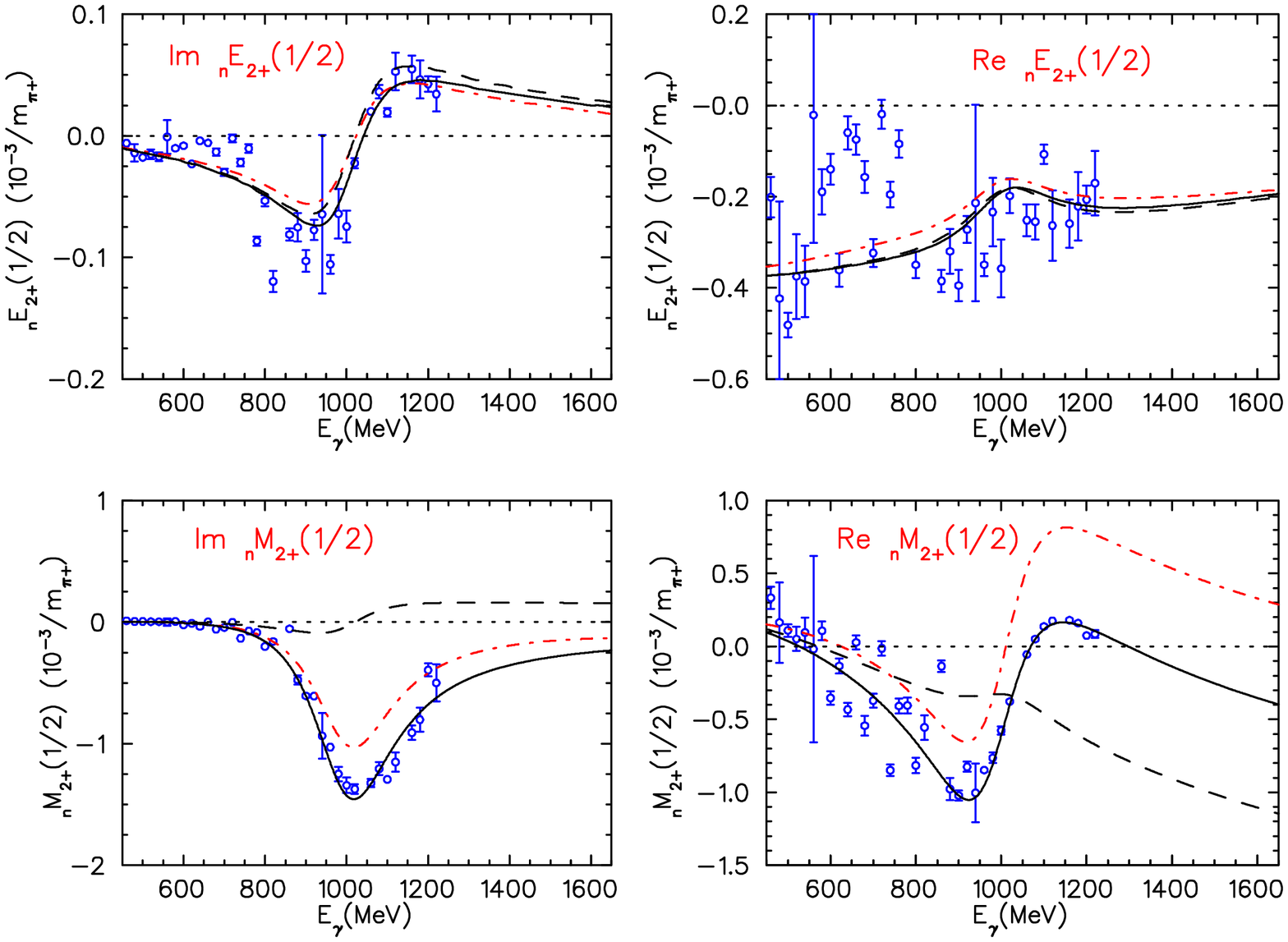, width=10 cm, angle=90}}
\caption{The global solutions of MAID2007 (solid lines) and GWU/SAID~\cite{SAID06} (red
dashed-dotted lines) for the multipoles $_nE_{2-}^{1/2}$, $_nM_{2-}^{1/2}$,
$_nE_{2+}^{1/2}$, and $_nM_{2+}^{1/2}$ as function of the photon lab energy
$E_{\gamma}$ in the second and third resonance regions. Further notation as in
Fig.~\ref{fig:6_photo}.} \label{fig:12_photo}
\end{figure*}

\begin{figure*}[ht]
\centerline{\epsfig{file=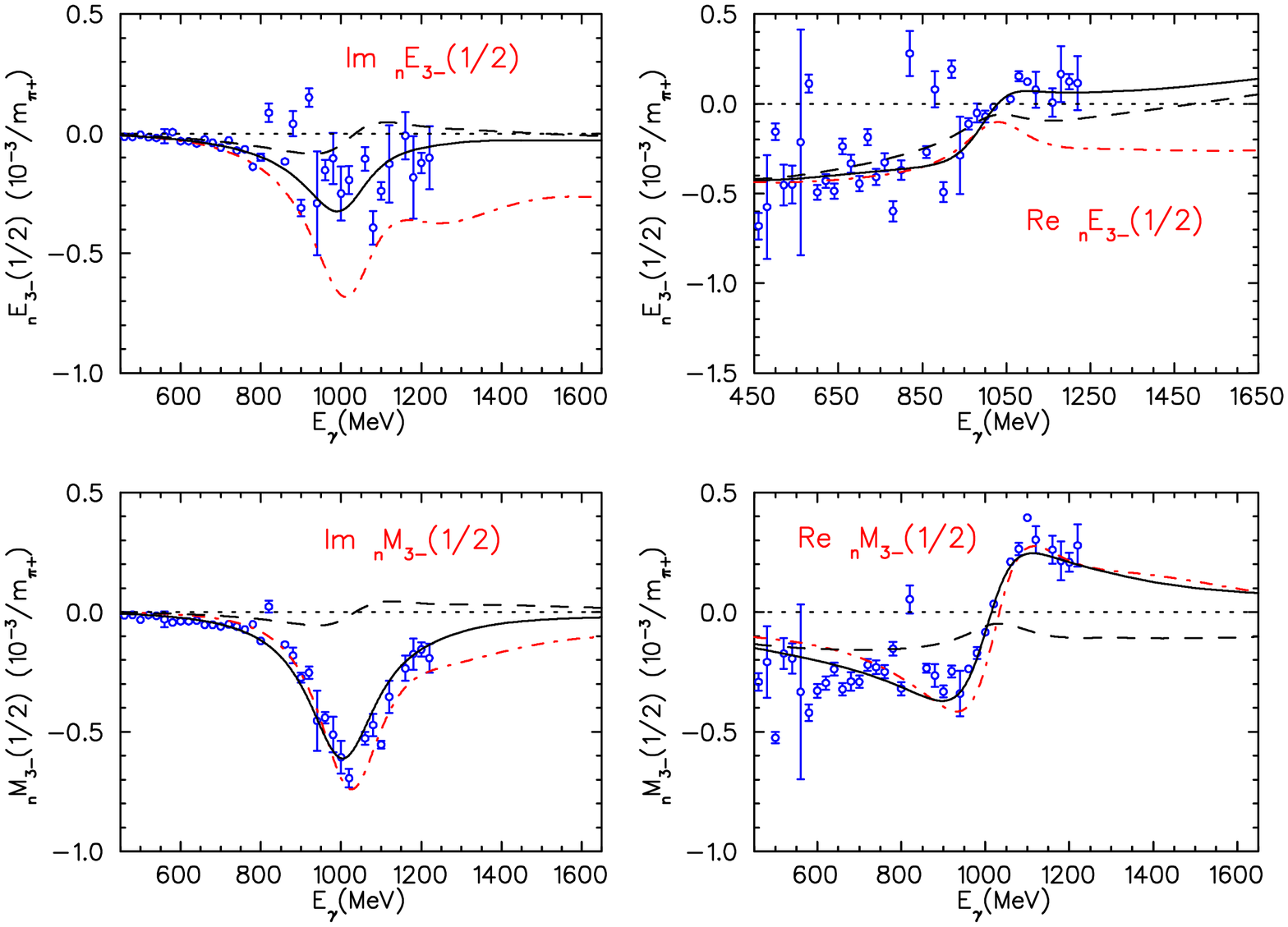, width=10 cm, angle=90}}
\caption{The global solutions of MAID2007 (solid lines) and GWU/SAID~\cite{SAID06} (red
dashed-dotted lines) for the multipoles $_nE_{3-}^{1/2}$ and $_nM_{3-}^{1/2}$
as function of the photon lab energy $E_{\gamma}$ in the second and third
resonance regions. Further notation as in Fig.~\ref{fig:6_photo}.}
\label{fig:13_photo}
\end{figure*}

More substantial discrepancies between the MAID and SAID analyses are found in
the second and third resonance regions. A detailed comparison of the two models
is shown in Figs.~\ref{fig:6_photo}-\ref{fig:13_photo}. As pointed out in Sect.
3.1, it is prerequisite to know the phases of the multipoles in order to get
correct single-energy solutions above the two-pion threshold. In MAID2007 these
phases are determined by Eqs.~(\ref{eq:3.9}), (\ref{eq:3.15}), and
(\ref{eq:3.18}). The SAID analysis is based on the following parametrization of
the partial wave amplitudes:
\begin{eqnarray}
t_{\gamma\pi}\,&=&\,({\rm {Born}}+A)\,(1+i t_{\pi N}) + B\,t_{\pi N}
\nonumber\\
& & + (C+iD)\,({\rm {Im}}\,t_{\pi N}\,-\mid t_{\pi N} \mid^2)\,, \label{eq:4.2}
\end{eqnarray}
where $A$, $B$ $C$ and $D$ are polynomials in the energy with real
coefficients, and $t_{\pi N}$ is the pion-nucleon elastic scattering amplitude
of Eq.~(\ref{eq:3.15}). As seen in the first resonance region, the most serious
differences between MAID and SAID are again found for the real parts of the
multipoles $_pE_{0+}^{1/2}$ and $_pE_{2-}^{1/2}$. We have checked the phases of
these multipoles by independent calculations on the basis of dispersion
relations~\cite{DR02,aznauryan}. The result confirmed our phase relations.
Concerning the small amplitudes, the most sizable differences between SAID and
MAID are in the $M_{1-}^{3/2}$ $_pM_{1+}^{1/2}$ and $_pE_{1+}^{1/2}$ multipoles.
In the neutron channel, the largest differences are in the multipoles $_nE_{0+}^{1/2}$,
$_nE_{3-}^{1/2}$, $_nE_{1+}^{1/2}$, and $_nM_{1+}^{1/2}$. In the last two cases
this is due to the large contribution from the $P_{13}(1720)$ resonance which
is not found in the SAID analysis (see Table~\ref{tab:helicity_p}).\\

\begin{figure*}[ht]
\centerline{\epsfig{file=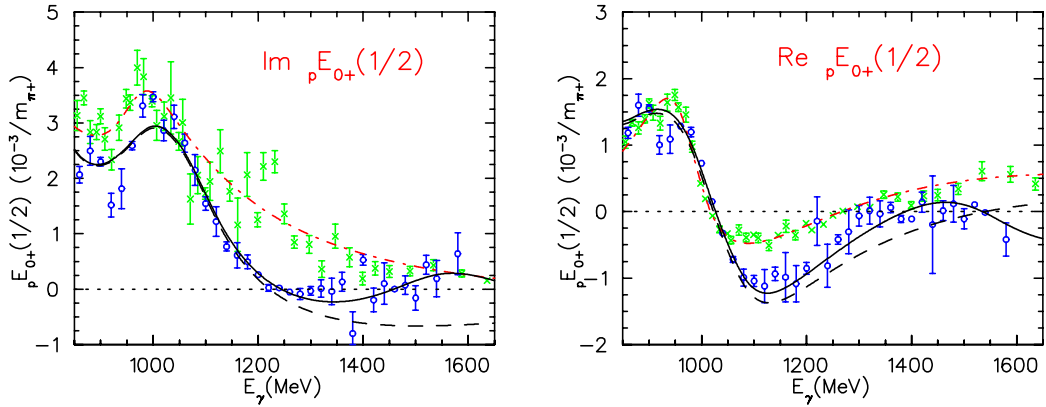, width=5.3 cm, angle=90} }
\caption{The contribution of a third $S_{11}$ resonance in the $_pE_{0+}^{1/2}$ multipole
with $M_R$=1950 MeV, $\Gamma_R$=200 MeV, single-pion branching ratio
$\beta_\pi$=0.4, and helicity amplitude $A_{1/2}$=0.028~GeV$^{-1/2}$. The solid
and dashed lines are our global solutions with and without this resonance,
respectively. The red dashed-dotted lines represent the global SAID solution.
The blue open circles and green crosses are the single-energy solutions of
MAID2007 and the SAID, respectively. This resonance is included in MAID2007.}
\label{fig:14_photo}
\end{figure*}

\begin{figure*}[ht]
\centerline{\epsfig{file=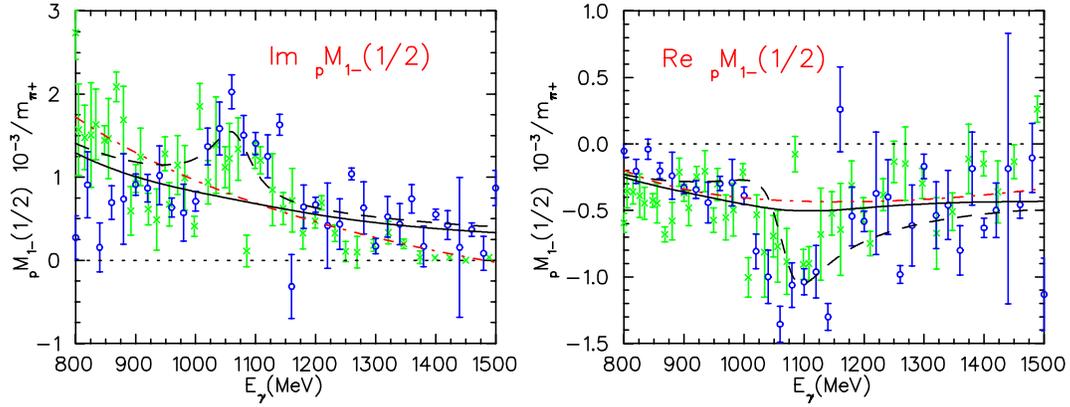, width=5.3 cm, angle=90}}
\caption{The contribution of a second $P_{11}$ resonance in the $_pM_{1-}^{1/2}$ multipole
with $M_R$=1700 MeV, $\Gamma_{tot}$=47 MeV, single-pion branching ratio
$\beta_\pi$=0.1, and helicity amplitude $A_{1/2}=-0.024$~GeV$^{-1/2}$. The
solid and dashed lines are our global solutions without and with this
resonance, respectively. Further notation as in Fig.~\ref{fig:14_photo}. The
$P_{11}(1700)$ is {\emph {not}} included in MAID2007.}
\label{fig:15_photo}
\end{figure*}

Let us finally discuss the possible contributions of the weaker resonances. As
discussed in Ref.~\cite{S11}, the two additional $S_{11}$ resonances found with
masses of about 1800 and 2000 MeV might also show up in pion photoproduction.
This conclusion was mainly based on the single-energy solution of the SAID
group. As illustrated by Fig.~\ref{fig:14_photo}, our present analysis requires
only one additional $S_{11}$ resonance with mass $M_R\approx$~1950 MeV, and our
single-energy solution shows no resonance at $M_R\approx$~1800 MeV. Of course,
the solution of the problem is certainly correlated with the way how the
resonance and background contributions are separated. As demonstrated both in
Ref.~\cite{S11} and by Fig.~\ref{fig:6_photo}, the background is very important
for this particular channel. In this context we recall that we use the same
form of the unitarized background contribution (Born, $\omega$ and $\rho$
exchange) for all the partial waves. Another interesting topic deserving
further experimental and theoretical studies, concerns the Roper or $P_{11}$
channel. As clearly seen in Fig.~\ref{fig:15_photo}, both our and the SAID
analysis yield a second resonance structure of the $_pM_{1-}^{1/2}$ multipole
at $E_{\gamma}\approx$~1070~MeV or $W\approx$~1700~MeV. However, our analysis
yields a very small width of $\Gamma_{\rm {tot}}\approx$~30-60~MeV, whereas the
PDG lists $\Gamma_{\rm {tot}}\approx$~50-250~MeV for this resonance. Moreover,
also the helicity amplitude differs, whereas our result is $A_{1/2}\approx
-0.024$~GeV$^{-1/2}$, the PDG lists $0.009\pm 0.022$~GeV$^{-1/2}$. Of course,
these numbers do strongly depend on the values for the single-pion branching
ratio. On the other hand, we do not anticipate large effects from different
definitions of the background in this channel, because the background
contribution is very small in the resonance region (see
Fig.~\ref{fig:7_photo}).
\section{Partial-wave analysis of pion electroproduction}
In most of the pion electroproduction experiments the five-fold differential
cross section was measured. However, different conventions exist for the
partial cross sections, and therefore we recall the definitions used in MAID.
For an unpolarized target the cross sections written as the product of the
virtual-photon flux factor $\Gamma_v$ and the virtual photon cross section
$d\sigma_v/d\Omega_{\pi}$~\cite{DT92},
\begin{equation}
\frac{d\sigma}{d\Omega_f^L\,dE_f^L\,d\Omega_{\pi}} = \Gamma_v\,
\frac{d\sigma_v}{d\Omega_{\pi}}\,, \label{eq:5.1}
\end{equation}
\begin{eqnarray} \frac{d\sigma_v}{d\Omega_{\pi}} & = &
\frac{d\sigma_T}{d\Omega_{\pi}}  + \epsilon \,\frac{d\sigma_L}{d\Omega_{\pi}} +
\sqrt{2\epsilon (1+\epsilon)}\,\,\frac{d\sigma_{LT}}{d\Omega_{\pi}}\,
\cos{\Phi_{\pi}}\label{eq:5.2} \\
&+& \epsilon\,\frac{d\sigma_{TT}}{d\Omega_{\pi}}\, \cos{2\Phi_{\pi}} +
h\sqrt{2\epsilon (1-\epsilon)}\,\,\frac{d\sigma_{LT'}}
{d\Omega_{\pi}}\,\sin{\Phi_{\pi}}\,,\nonumber
\end{eqnarray}
where $\epsilon$ and $h$ describe the polarizations of the virtual photon and
the electron, respectively. We further note that the hadronic kinematics is
expressed in the c.m. system, whereas the electron and virtual photon
kinematics is written in the lab frame, as indicated by $L$ in the following
variables: the initial and final electron energies $E_i^L$ and $E_f^L$,
respectively, the electron scattering angle $\theta_L$, the photon energy
$\omega_L=E_i^L-E_f^L$, and the photon three-momentum ${\bf k}_L$. With these
definitions the virtual photon flux and the transverse photon polarization take
the form
\begin{equation}
\epsilon=\frac {1} {1 + 2\frac{{\bf k}^2_L}{Q^2}\tan^2 \frac{\theta_L}{2}},\;
\Gamma_v = \frac{\alpha_{\rm {em}}}{2\pi^2}\
\frac{E_f^L}{E_i^L}\,\frac{K}{Q^2}\ \frac{1}{1-\epsilon}\,. \label{eq:5.3}
\end{equation}
As in our previous notation~\cite{DT92}, the flux is denoted by the photon
``equivalent energy'' in the lab frame, $K=K_H=(W^2-m^2)/2m$ as originally
introduced by Hand~\cite{Hand}. Another definition was given by
Gilman~\cite{Gilman} who used $K=K_G=\mid{{\bf k}_L}\mid$.\\

The first two terms on the r.h.s. of Eq.~(\ref{eq:5.2}) are the transverse
($T$) and longitudinal ($L$) cross sections. They do not depend on the pion
azimuthal angle $\Phi_{\pi}$. The third and fifth terms describe
longitudinal-transverse interferences ($LT$, $LT'$). They contain an explicit
factor $\sin\theta_{\pi}$ and therefore are vanishing along the axis of
momentum transfer. The same is true for the fourth term, a
transverse-transverse interference ($TT$) proportional to $\sin^2\theta_{\pi}$.
It is useful to express these 5 cross sections in terms of hadronic response
functions depending only on 3 independent variables, i.e.,
$R_i=R_i(Q^2,W,\theta_{\pi})$. The corresponding relations take the form
\begin{eqnarray}\nonumber
\frac{d\sigma_T}{d\Omega_{\pi}}=\frac{q}{k_W}\,R_T\,,\,
\frac{d\sigma_{TT}}{d\Omega_{\pi}}=\frac{q}{k_W}\,R_{TT}\,,\,
\frac{d\sigma_L}{d\Omega_{\pi}}=\frac{q}{k_W}
\frac{Q^2}{\omega_{\gamma}^2}\,R_L\,,
\end{eqnarray}
\begin{equation}
\frac{d\sigma_{LT}}{d\Omega_{\pi}}=\frac{q}{k_W}\,
\frac{Q}{\omega_{\gamma}}\,R_{LT}\,,\,
\frac{d\sigma_{LT'}}{d\Omega_{\pi}}=\frac{q}{k_W}\,
\frac{Q}{\omega_{\gamma}}\,R_{LT'}\,. \label{eq:5.4}
\end{equation}
As a result of this equation, the longitudinal ($L$) and
longitudinal-transverse ($LT$ and $LT'$) response functions must be
proportional to $\omega^2_{\gamma}$ and $\omega_{\gamma}$,
respectively, in order to avoid non-physical singularities at the
energy for which the c.m. virtual photon energy passes through zero.
The 5 response functions may be expressed in terms of 6 independent
CGLN amplitudes $F_1,...,F_6$~\cite{CGLN}, or in terms of the
helicity amplitudes $H_1,...,H_6$, which are linear combinations of
the CGLN amplitudes. The relevant expressions can be found in
Refs.~\cite{DT92,KDT}.\\
\subsection{Data base for pion electroproduction and fit procedure}
The main part of our data base for pion electroproduction includes the
compilation of the GWU/SAID group~\cite{SAID} in 2000 and recent data from Bonn
and JLab (see Table~\ref{database}). Altogether this base contains about 70000
data points within the energy range 1.074~GeV$< W < 2~$GeV and photon
virtuality range 0.1~GeV$^2 \leq Q^2 \leq $6~GeV$^2$. In addition we have analyzed
high precision data from Bates~\cite{Mer01,Stave06},
Mainz~\cite{Pos01,Elsner06}, and JLab~\cite{Kelly,Lav04}. Our fitting procedure
was as follows. In a first step we fitted the data sets at constant values of
$Q^2$ (single-$Q^2$ fit). This procedure is similar to the partial-wave
analysis for pion photoproduction except for the additional longitudinal
couplings of the resonances. Second, we introduced a smooth $Q^2$ evolution of
the e.m. transition form factors and parameterized the 3 helicity amplitudes
accordingly. In a combined fit with the complete electroproduction data base
and information from the single-$Q^2$ fits we finally constructed the
$Q^2$-dependent solution (super-global fit). This new solution (MAID2007) was
then compared with the previous solution (MAID2003) in terms of $\chi^2$ as
presented in Table~\ref{database}. In most cases the new fit improves the
description of the data, in particular for the $n\pi^+$ channel.
\begin{table}[htbp]
\caption{\label{database} The number of data points, $N_{\rm {data}}$, and the
$\chi^2$ value per data point obtained with MAID2003 and MAID2007.}
\begin{center}
\begin{tabular}{|c|c|c|c|}
\hline
  Ref. & $W$ (MeV)       &$N_{\rm {data}}$    &  $\chi^2/N_{\rm {data}}$ (2003) \\
channel   & $Q^2$ (GeV$^2$) & observables  &  $\chi^2/N_{\rm {data}}$ (2007) \\
\hline
SAID00 & 1074-1895  &   13152     &  3.238 \\
 $p\pi^0$   &  0.1-4.3   &  d$\sigma$, ...  &  3.172 \\
\hline
SAID00 & 1125-1975  &   5464     &  3.297 \\
 $n\pi^+$   &  0.117-4.4   &  d$\sigma$, ...  &  4.188 \\
\hline
Bonn02~\cite{Ban02} & 1153-1312  &   4914     &  1.378 \\
 $p\pi^0$   &  0.63   &  d$\sigma$  &  1.400 \\
\hline
CLAS02~\cite{Joo02} & 1110-1680  &  31810     &  1.907 \\
 $p\pi^0$   &  0.4-1.80   &  d$\sigma$  &  1.952 \\
\hline
CLAS03~\cite{Joo03} & 1100-1660  &  223     &  4.881 \\
 $p\pi^0$   &  0.4-0.65   &  d$\sigma_{LT'}$ & 3.490 \\
\hline
CLAS04~\cite{Joo04} & 1100-1660  &  224     &  4.879 \\
 $n\pi^+$   &  0.4-0.65   &  d$\sigma_{LT'}$ & 2.196 \\
\hline
CLAS06~\cite{Egi06} & 1110-1570  &  4179     &  10.04 \\
 $n\pi^+$   &  0.3-0.60   &  d$\sigma$ & 4.954 \\
\hline
CLAS06~\cite{Ungaro06} & 1110-1390  &  8491     & 1.691 \\
 $p\pi^0$   &  3.0-6.0   &  d$\sigma$ & 1.335 \\
\hline
 total & 1074-1975  &  68457     & 2.724 \\
 $p\pi^0$, $n\pi^+$ & 0.1-6.0   &  d$\sigma$, ... & 2.437 \\
\hline \hline
 SAID00 & 1253-1976  &   799     & 2.100 \\
 $p\pi^-$ & 0.54-1.36 &  d$\sigma$ & 2.264 \\
\hline
\end{tabular}
\end{center}
\end{table}
\subsection{ Results for the $\Delta(1232)$ form factors}
\begin{figure}[ht]
\centerline{\epsfig{file=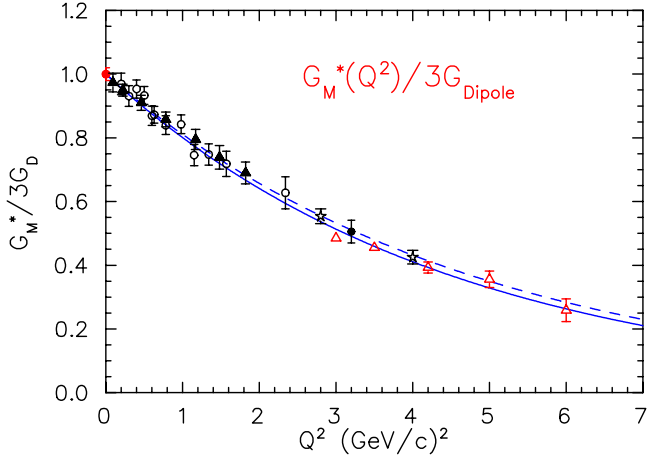, width=6 cm, angle=90}}
\caption{The $Q^2$ dependence of the magnetic form factor $G_M^*$ for the $N\Delta (1232)$
transition divided by $3\,G_D(Q^2)$. The solid and dashed blue lines are the
results of MAID2007 and MAID2003, respectively. The red open triangles
represent the new JLab data of Ungaro {\emph {et al.}}~\cite{Ungaro06}. See
Ref.~\cite{DMT} for the notation of the other data points.}
\label{fig:gmstar}
\end{figure}

In the literature the e.m. properties of the $N \Delta(1232)$ transition are
described by either the magnetic ($G_M^*$), electric ($G_E^*$), and Coulomb
($G_C^*$) form factors or the helicity amplitudes $A_{1/2}$, $A_{3/2}$, and
$S_{1/2}$, which can be derived from the reduced e.m. amplitudes $\bar{\mathcal
A}_\alpha$ as defined by Eq.~(\ref{eq:3.18}). It is worthwhile pointing out
that these amplitudes are related to the multipoles over the full energy
region, that is, they are the primary target of the fitting procedure. The form
factors and helicity amplitudes are then obtained by evaluating the reduced
e.m. amplitudes at the resonance position $W=M_\Delta$=1232~MeV. The respective
relations take the following form:
\begin{eqnarray}
G_M^*(Q^2) &=&  -c_\Delta (A_{1/2}+\sqrt{3} A_{3/2})=2 c_\Delta \,
\bar{\mathcal A}_M^\Delta(M_\Delta,Q^2)\,,\nonumber\\
G_E^*(Q^2) &=&  \,\,\, c_\Delta (A_{1/2}-\frac{1}{\sqrt{3}}
A_{3/2})=-2
c_\Delta \,\bar{\mathcal A}_E^\Delta(M_\Delta,Q^2)\,,\nonumber\\
G_C^*(Q^2) &=& \sqrt{2} c_\Delta \frac{2M_{\Delta}}{k_\Delta}
S_{1/2}=-2 c_\Delta \, \frac{2M_{\Delta}}{k_\Delta}\bar{\mathcal
A}_S^\Delta(M_\Delta,Q^2)
\,,\nonumber\\
\mbox{with} \;\; c_\Delta &=& \left( \frac{m^3 k_W^\Delta}{4\pi\alpha_{\rm
{em}} M_\Delta k_\Delta^2} \right)^{1/2}\,, \label{eq:5.5}
\end{eqnarray}
and where  $k_\Delta=k_\Delta (Q^2)=k(M_{\Delta},Q^2)$ and
$k_W^\Delta=k(M_\Delta,0)$ are the virtual photon momentum and the photon
equivalent energy at resonance. Because the $\Delta(1232)$ is very close to an
ideal resonance, the real part of the amplitudes vanishes for
$W=M_{\Delta}$ and the form factors can be directly expressed by the imaginary
parts of the corresponding multipoles at the resonance position,
\begin{eqnarray}
G^{\ast}_M(Q^2) &=& b_\Delta\, {\rm {Im}}\{M_{1+}^{(3/2)}(M_\Delta,Q^2)\}
\,,\nonumber\\
G^{\ast}_E(Q^2) &=& -b_\Delta\, {\rm {Im}}\{E_{1+}^{(3/2)}(M_\Delta,Q^2)\}
\,,\label{eq:5.6}\\
G^{\ast}_C(Q^2) &=& -b_\Delta \frac{2M_{\Delta}}{k_{\Delta}}\,
{\rm {Im}}\{S_{1+}^{(3/2)}(M_\Delta, Q^2)\}\,,\nonumber\\
\mbox{where}\;\; b_\Delta &=& \left( \frac{8\, m^2\, q_\Delta\,
\Gamma_\Delta}{3\,\alpha_{\rm {em}}\, k_\Delta^2} \right)^{1/2} \,, \nonumber
\end{eqnarray}
and with $\Gamma_{\Delta}=115$ MeV and $q_{\Delta}=q(M_{\Delta})$
the pion momentum at resonance. The above definition of the form
factors is due to Ash~\cite{Ash67}. The form factors of Jones and
Scadron~ \cite{Jon73} are obtained by multiplying our form factors
with $\sqrt{1+Q^2/(M_N +M_\Delta)^2}$. We note that the form factor
$G_C^{\ast}$ differs from our previous work~\cite{Tiat03} by the
factor $2M_\Delta/k_{\Delta}$ in Eq.~(\ref{eq:5.6}). With these
definitions all 3 transition form factors remain finite at
pseudo-threshold (Siegert limit). In the literature, the following
ratios of multipoles have been defined:
\begin{eqnarray}
R_{EM} &=& -\frac{G_E^{\ast}}{G_M^{\ast}}=\frac{A_{1/2}-\frac{1}{\sqrt{3}}
A_{3/2}}{A_{1/2}+\sqrt{3} A_{3/2}}\,,\label{eq:5.7}\\
R_{SM} &=&
-\frac{k_\Delta}{2M_{\Delta}}\frac{G_C^{\ast}}{G_M^{\ast}}=\frac{\sqrt{2}
S_{1/2}}{A_{1/2}+\sqrt{3} A_{3/2}}\,. \label{eq:5.8}
\end{eqnarray}
\indent In MAID2003 the $Q^2$ dependence of the e.m. amplitudes $\bar{\mathcal
A}_\alpha^\Delta$ was parameterized as follows:
\begin{equation}
\bar{\mathcal A}_\alpha^\Delta(W,Q^2)=A_\alpha^0 (1 + \beta_\alpha
Q^{2n_\alpha}) \frac{ k}{k_W} e^{-\gamma_\alpha Q^2}G_D(Q^2), \label{eq:5.9}
\end{equation}
where $G_D(Q^2)=1/(1+Q^2/0.71\,{\rm {GeV}}^2)^2$ is the dipole form factor.
MAID2007 follows this parametrization for the magnetic and electric amplitudes,
although with somewhat different values of the parameters (see
Table~\ref{tab:2}). In order to fulfill the Siegert theorem, we have however
changed the description of the Coulomb amplitude as specified below.
\begin{table}[htbp]
\caption{Parameters for the $N \Delta$ amplitudes given by Eqs.~(\ref{eq:5.9})
and~(\ref{eq:5.16}). The amplitudes $A_\alpha^0$ are in units 10$^{-3}$
GeV$^{-1/2}$, the parameters $\beta$ and $\gamma$ in GeV$^{-2}$. For the
Coulomb amplitude in MAID2007 we use Eq.~(\ref{eq:5.16}) with $d$=4.9.}
\label{tab:2}
\begin{center}
\begin{tabular}{|c|ccc|c|}
\hline
  & M & E & S & model\\
\hline
  $A_\alpha^0$ & 300 & -6.50 & -19.50    &   2003\\
                & 300  & -6.37 & -12.40 & 2007\\
\hline
 $\beta_\alpha$        & 0     & -0.306& 0.017 & 2003\\
                &  0.01 & -0.021&  0.12 & 2007\\
\hline
$\gamma_\alpha$        &  0.21  & 0.21 &  0.21  & 2003\\
                &  0.23  & 0.16 &  0.23  & 2007\\
\hline
$n_\alpha$        &  1  &  1 &  3  & 2003\\
                &   1  & 1 &  ---  & 2007\\
\hline
\end{tabular}
\end{center}
\end{table}
The results of MAID2003 and MAID2007 for $G_M^{\ast}(Q^2)$ are compared in
Fig.~\ref{fig:gmstar}. Because our single-$Q^2$ analysis follows the global fit
closely, it is not shown in the figure.  We find an excellent agreement with
the data, which also include the new high-$Q^2$ data of the JLab/CLAS
Collaboration~\cite{Ungaro06}. At this point a word of caution is in order.
Because the form factors are extracted from the multipoles by
Eq.~(\ref{eq:5.6}), they are proportional to $\sqrt {\Gamma_{\Delta}}$.  The
MAID fit to the experimental data yields $\Gamma_{\Delta}$=130 MeV, which is
different from the usually assumed value of about 115 MeV. Therefore, in order
to compare with form factor values of other analyses, we scale our predicted
form factor with $\sqrt{115/130}$. As shown in Fig.~\ref{fig:gmstar},
$G_M^{\ast}(0)/3$ will then take the usual value of 1 to an accuracy of 1\%.
From this number we can determine the $N\rightarrow\Delta$ magnetic transition
moment, $\mu_{N\Delta}=3.46\pm 0.03$, in units of the nuclear magneton.
\subsection{Siegert theorem and ratios $R_{EM}$ and $R_{SM}$}
Let us next discuss our results for the $R_{EM}$ and $R_{SM}$ ratios. In all
previous solutions these ratios were nearly constant for $Q^2<1$~GeV$^2$.
However, calculations in effective field theories~\cite{Gel99,Pas05} and
dynamical models~\cite{DMT,KY99,SL01} indicated a rapid rise of $R_{SM}$ for
$Q^2 \rightarrow +0$.  This dependence is rather
model-independent, because it reflects the behavior of the multipoles at
physical threshold (pion momentum ${\bf {q}} \rightarrow 0)$ and
pseudothreshold (Siegert limit, photon momentum ${\bf {k}}\rightarrow
0)$~\cite{Dre06}. The longitudinal and Coulomb multipoles are related by gauge
invariance, ${\bf {k}} \cdot \vec J = \omega_\gamma \rho$, which leads to
\begin{equation}
|{\bf {k}}| \, L_{\ell \pm}^I (W, Q^2)= \omega_\gamma \, S_{\ell \pm}^I (W,
Q^2)\,. \label{eq:5.10}
\end{equation}
Since the photon c.m. energy  $\omega_\gamma$ vanishes for $Q^2=Q^2_0=W^2-m^2$,
the longitudinal multipole must have a zero at that momentum transfer, $L_{\ell
\pm}^I(W,Q^2_0)=0$. Furthermore, gauge invariance implies that the longitudinal
and Coulomb multipoles take the same value in the real photon limit, $L_{\ell
\pm}^I(W,Q^2=0)=S_{\ell \pm}^I(W,Q^2=0)$. Finally, the multipoles obey the
following model-independent relations at physical threshold (${\bf
{q}}\rightarrow 0$) and pseudothreshold (${\bf {k}}\rightarrow 0$):
\begin{eqnarray}
(E_{\ell+}^I, L_{\ell+}^I) &\rightarrow& k^\ell q^\ell \;\; (\ell \ge 0)\, \nonumber\\
(M_{\ell+}^I, M_{\ell-}^I) &\rightarrow& k^\ell q^\ell\;\; (\ell \ge 1)\, \label{eq:5.11}\\
(L_{\ell-}^I) &\rightarrow& kq \;\;\;\;\; (\ell = 1)  \nonumber\\
(E_{\ell-}^I, L_{\ell-}^I) &\rightarrow& k^{\ell-2} q^\ell\;\; (\ell \ge 2)\
\,. \nonumber
\end{eqnarray}
According to Eq.~(\ref{eq:5.10}) the Coulomb amplitudes acquire an additional
factor $k$ at pseudothreshold, i.e., $S_{\ell \pm}^I \sim k L_{\ell \pm}^I$.
This limit is reached at $Q^2=Q_{\rm{pt}}^2=-(W-m)^2$ (pseudo-threshold), and
because no direction is defined for ${\bf {k}} = 0$, the electric and
longitudinal multipoles are no longer independent at this point,
\begin{equation}
E_{\ell+}^I/L_{\ell+}^I \rightarrow 1 \;\;  \mbox{and} \;\; E_{\ell-}^I/
L_{\ell-}^I \rightarrow -\ell/(\ell-1)\;\; \mbox{if} \;\;  k \rightarrow 0.
\label{eq:5.12}
\end{equation}
In the case of the $N \Delta$ multipoles, Eq.~(\ref{eq:5.12}) yields the
following relation in the limit ${\bf {k}}\rightarrow 0$: $L_{1+}^{3/2}
\rightarrow E_{1+}^{3/2} \rightarrow {\mathcal O}(k)$ and consequently
$S_{1+}^{3/2}= kE_{1+}^{3/2}/\omega_\gamma\rightarrow {\mathcal O}(k^2)$. Although
the pseudo-threshold is reached at the unphysical point
$Q_{\rm{pt}}^2=-(M_{\Delta}-m)^2 \approx -0.084$~GeV$^2$, it still influences
the multipoles near $Q^2=0$ because of the relatively small excitation energy
of the $\Delta(1232)$. In particular we get the following relation for
$Q^2\rightarrow Q_{\rm{pt}}^2$:
\begin{equation}
R_{SM}=\frac{S_{1+}^{(3/2)}}{M_{1+}^{(3/2)}}
=\frac{k}{\omega_\gamma}\frac{E_{1+}^{(3/2)}}{M_{1+}^{(3/2)}} \rightarrow
\frac{k}{M_\Delta-m}R_{EM}\,. \label{eq:5.13}
\end{equation}
With increasing value of $Q^2$, the Siegert relation fails to describe the
experimental data. Moreover, it contains a singularity at $\omega_\gamma$=0,
which occurs in $\Delta(1232)$ electroproduction already at $Q^2=0.64$~GeV$^2$.
However, we obtain a good overall description by using the idea of
Ref.~\cite{Buchmann} that the ratio $R_{SM}$ is related to the (elastic) form
factors of the neutron,
\begin{equation}
R_{SM}(Q^2)=\frac{m \, k_\Delta(Q^2)\,G_E^n(Q^2)} {2\, Q^2 \,
G_M^n(Q^2)}\,. \label{eq:5.14}
\end{equation}
This relation gives the necessary proportionality to the photon momentum at
small $Q^2$, describes the experimental value of the ratio over a wide range of
$Q^2$, and yields an asymptotic behavior consistent with the prediction of
perturbative QCD that $R_{SM}$ should approach a constant for
$Q^2\rightarrow\infty$. This leads to the following simple parametrization:
\begin{equation}
R_{SM}(Q^2)=-\frac{k_\Delta(Q^2)}{8m}\,\frac{a}{1+d\tau}\,, \label{eq:5.15}
\end{equation}
with $\tau=Q^2/(4m^2)$, and the parameters $a$ and $d$ to be determined by a
fit to the data. On the basis of this ansatz, the Coulomb coupling has been
modified as follows:
\begin{eqnarray}
\bar{\mathcal A}_S^\Delta(W,Q^2) =A_S^0 \frac{1 + \beta_S Q^2}{1+d\tau}
\frac{k^2}{k_W\,k_W^\Delta} e^{-\gamma_S Q^2}G_D(Q^2), \label{eq:5.16}
\end{eqnarray}
with parameters given in Table~\ref{tab:2}. This leads to the multipole ratio
\begin{equation}
R_{SM}(Q^2)=\frac{A_S^0}{A_M^0}\,\frac{1}{1+d\tau}\left(\frac{1+\beta_S Q^2}
{1+\beta_M Q^2}\right)\,\frac{k_\Delta}{k_W^\Delta}\,. \label{eq:5.17}
\end{equation}
By construction this ratio vanishes in the Siegert limit, $Q^2\rightarrow
Q_{\rm{pt}}^2$, and approaches a (negative) constant for $Q^2\rightarrow
\infty$ in agreement with perturbative QCD. However, a word of caution has to
be added at this point. The polynomials and gaussians used to fit the data in
the range of low and intermediate virtualities, $Q^2<10$~GeV$^2$, should not be
expected to
yield realistic extrapolations to the higher values of $Q^2$.\\

The correct Siegert limit is even more important for pion $S$-wave production
in the threshold region, in which case the pseudo-threshold comes as close as
$Q^2_{\rm{pt}}=-m_{\pi}^2\approx -0.02$~GeV$^2$. The term describing the pion
cloud contribution has therefore been parameterized as follows:
\begin{equation}
L_{0+}^{\rm {corr}}(W,Q^2)=\frac{\omega_\gamma}{\omega_{\rm{pt}}}\,
e^{-\beta(Q^2-Q_{pt}^2)} \,E_{0+}^{\rm {corr}}(W,Q^2)\,, \label{eq:5.18}
\end{equation}
where $\omega_{\rm{pt}}^2=-Q_{\rm{pt}}^2=(W-m)^2$. From a fit to $\pi^0$
electroproduction data near threshold~\cite{Wei07}, we obtain $\beta=10$
GeV$^{-2}$. In the future we intend to study pion electroproduction near
threshold in more detail.\\

\begin{figure*}[ht]
\centerline{\epsfig{file=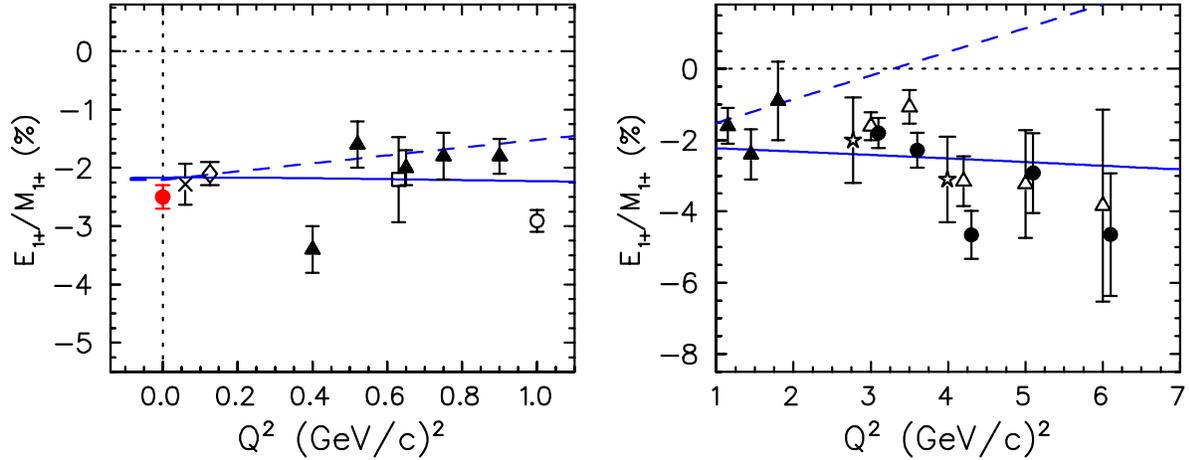, width=6 cm, angle=90}}
\caption{The $Q^2$ dependence of the ratio $R_{EM}$ at the $\Delta(1232)$ resonance. The
blue solid and dashed lines are the super-global solutions from MAID2007 and
MAID2003, respectively. The data points are from Refs.~\cite{Ban02} (open
square), \cite{Joo02} (solid triangles), \cite{Mer01} (open diamond),
\cite{Stave06} (cross), \cite{Kelly} (open circle), \cite{Fro99} (asterisks),
and \cite{Ungaro06} (open triangles). The green solid circle at $Q^2=0$ in the
left panel is from Ref.~\cite{Beck97}, and the black solid circles in the right
panel are obtained by our single-$Q^2$ analysis of the data from
Ref.~\cite{Ungaro06}.}
\label{fig:emratios}
\end{figure*}

\begin{figure*}[ht]
\centerline{\epsfig{file=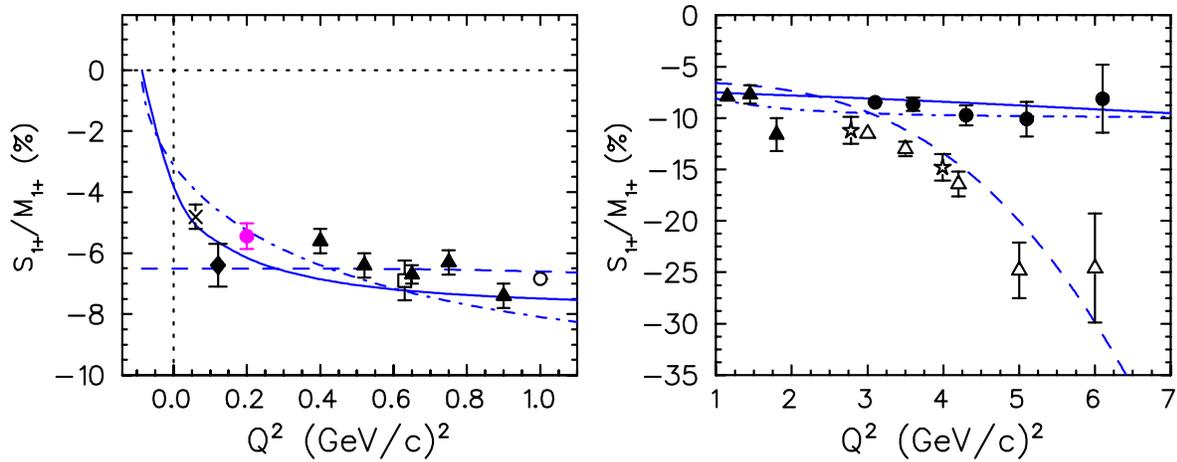,width=6.0 cm,angle=90}}
\caption{The $Q^2$ dependence of the ratio $R_{SM}$ at the $\Delta(1232)$ resonance
position. The blue solid and dashed lines are the MAID2007 and MAID2003
super-global solutions, respectively, the dashed-dotted line is obtained using
Eq.~(\ref{eq:5.15}) with $a=0.9$ and $d=1.75$. The data point of
Ref.~\cite{Pos01} (diamond) at $Q^2=0.1$~GeV$^2$ is practically identical to
the Bates result~\cite{Mer01}, the full circle at $Q^2=0.2$~GeV$^2$ is from
Ref.~\cite{Elsner06}. See Fig.~\ref{fig:emratios} for the notation of the
further data points.}
\label{fig:smratios}
\end{figure*}

Figures~\ref{fig:emratios} and \ref{fig:smratios} display the super-global
solutions of MAID2003 (dashed lines) and MAID2007 (solid lines) for the ratios
$R_{EM}$ and $R_{SM}$ in comparison with other analyses. Different from our
previous solution, the ratio $R_{EM}$ of MAID2007 stays always below the zero
line, in agreement with the original analysis of the data~\cite{Ungaro06,Fro99}
and also with the dynamical model of Sato and Lee~\cite{SL01} who concluded
that $R_{EM}$ remains negative and tends towards more negative values with
increasing $Q^2$. This indicates that the predicted helicity conservation at
the quark level is irrelevant for the present experiments. We also analyzed the
new data of Ref.~\cite{Ungaro06} in the range of 3~GeV$^2\leq Q^2 \leq
6$~GeV$^2$ and found slightly decreasing values of $R_{EM}$ from our
single-$Q^2$ analysis. In this analysis we varied both the $\Delta$ and the
Roper multipoles. For the ratio $R_{SM}$ both the super-global and the
single-$Q^2$ solutions yield ratios that asymptotically tend to a negative
constant. This result is in good agreement with the prediction of
Ref.~\cite{Buchmann} (dash-dotted curve in Fig.~\ref{fig:smratios}) but
disagrees with our previous solution and with the analysis of
Ref.~\cite{Ungaro06}. As discussed before, the new solution has a large slope
at small $Q^2$  as a consequence of the Siegert theorem. The following
Fig.~\ref{helicity_P33} displays the $Q^2$ dependence of the helicity
amplitudes for the $N \Delta(1232)$ transition. Our single-$Q^2$ fit is in
excellent agreement with the super-global solution, except for the values of
$S_{1/2}$ at $Q^2=0.4$ and 0.525 GeV$^2$.
\subsection{Results for the higher resonances}
Above the two-pion threshold we can no longer apply the two-channel unitarity
and consequently the Watson theorem does not hold. Therefore, the background
amplitude of the partial waves does not vanish at resonance as was the case for
the $\Delta(1232)$ resonance. As an immediate consequence the resonance-background
separation becomes more model-dependent. In MAID2007 we choose to separate the
background and resonance contributions according to the K-matrix approximation.
Furthermore, we recall that the absolute values of the helicity amplitudes are
correlated with the values used for the total resonance width $\Gamma_R$ and
the single-pion branching ratio $\beta_\pi$. On the experimental side, the data
at the higher energies are no longer as abundant as in the $\Delta$ region.
However, the large data set recently obtained by the CLAS collaboration (see
Table~\ref{database}) enabled us to determine the transverse and longitudinal
helicity couplings as functions of $Q^2$ for all the 4-star resonances below
1700~MeV. These data are available in the kinematical region of
$1100~\mbox{MeV}<W<1680~\mbox{MeV}$ and $0.4~\mbox{GeV}^2<Q^2<1.8~\mbox{GeV}^2$.\\

The helicity amplitudes for the Roper resonance $P_{11}(1440)$ are shown in
Fig.~\ref{helicity_P11}. Our latest super-global solution (solid lines) is in
reasonable agreement with the single-$Q^2$ analysis. The figure shows a zero
crossing of the transverse helicity amplitude at $Q^2\approx 0.7$~GeV$^2$ and a
maximum at the relatively large momentum transfer $Q^2\approx 2.5$~GeV$^2$. The
longitudinal Roper excitation rises to large values around $Q^2\approx
0.5$~GeV$^2$ and in fact produces the strongest longitudinal amplitude we can
find in our analysis. This answers the question raised by Li and
Burkert~\cite{Burk92} whether the Roper resonance is a radially excited 3-quark
state or a quark-gluon hybrid, because in the latter case the longitudinal
coupling should vanish completely. From the global fit we find the following
parametrization for the $Q^2$ dependence of the Roper amplitudes for the proton
and neutron channels:
\begin{eqnarray}
A_{1/2}^p(Q^2) &=& A_{1/2}^{0,p}(1 -1.22\, Q^2-0.55 \,Q^8)\,
e^{-1.51 Q^2}\,,\nonumber\\
S_{1/2}^p(Q^2) &=& S_{1/2}^{0,p}\,(1 + 40\, Q^2 +1.5\,Q^8)\,e^{-1.75Q^2}\,,
\label{eq:5.19}\\
A_{1/2}^n(Q^2) &=& A_{1/2}^{0,n}(1 +0.95\, Q^2)\,e^{-1.77 Q^2}\,,\nonumber\\
S_{1/2}^n(Q^2) &=& S_{1/2}^{0,n} \,(1 + 2.98\, Q^2)\,e^{-1.55 Q^2}\,,
\label{eq:5.20}
\end{eqnarray}
where $Q^2$ should be inserted in units of GeV$^2$. The numerical values of the
helicity amplitudes for real photons are given in Table~\ref{tab:3}. At $Q^2$=0
the fit yields a large neutron value for the Coulomb amplitude $S_{1/2}^0$, but
with increasing $Q^2$ the proton and neutron amplitudes become comparable.\\

\begin{table}[htbp]
\caption{Helicity amplitudes for the $P_{11}(1440)$ resonance at
$Q^2$=0 in units 10$^{-3}$~GeV$^{-1/2}$.} \label{tab:3}
\begin{center}
\begin{tabular}{|c|ccc|ccc|}
\hline
&   & $A_{1/2}^0$ &  &  & $S_{1/2}^0$ & \\
 &  proton  & neutron &  & proton & neutron & \\
\hline
$P_{11}(1440)$ & -61.4 & 54.1 &  & 4.2 &  -41.5 & \\
\hline
\end{tabular}
\end{center}
\end{table}
For all the higher resonances the transverse and longitudinal helicity
amplitudes are simply parameterized by the form
\begin{equation}
A_{\lambda}(Q^2) = A_\lambda^0\,(1 +\alpha\, Q^2) e^{- \beta Q^2}\,.
\label{eq:5.21}
\end{equation}
The values of  the fit parameters $A_\lambda^0$, $\alpha$ and $\beta$ are
listed in Tables~\ref{tab:4} and \ref{tab:5}. In the following
Fig.~\ref{helicity_S11} we present the results for the $S_{11}(1535)$. As is
also known from $\eta$ electroproduction, the transverse form factor falls off
very slowly. At a virtuality of $Q^2\approx 3$~GeV$^2$ this resonance is much
stronger than the $\Delta (1232)$ or the $D_{13} (1520)$ and only comparable to
the Roper. However, due to its much smaller width as compared to the Roper, the
$S_{11}$ dominates over the Roper at large $Q^2$. This result is in agreement
with the inclusive electroproduction cross section on the proton, which clearly
shows the dominance of the $\Delta(1232)$ at small momentum transfer whereas at
the larger momentum transfers the second resonance region takes over.\\

\begin{figure*}[ht]
\centerline{\epsfig{file=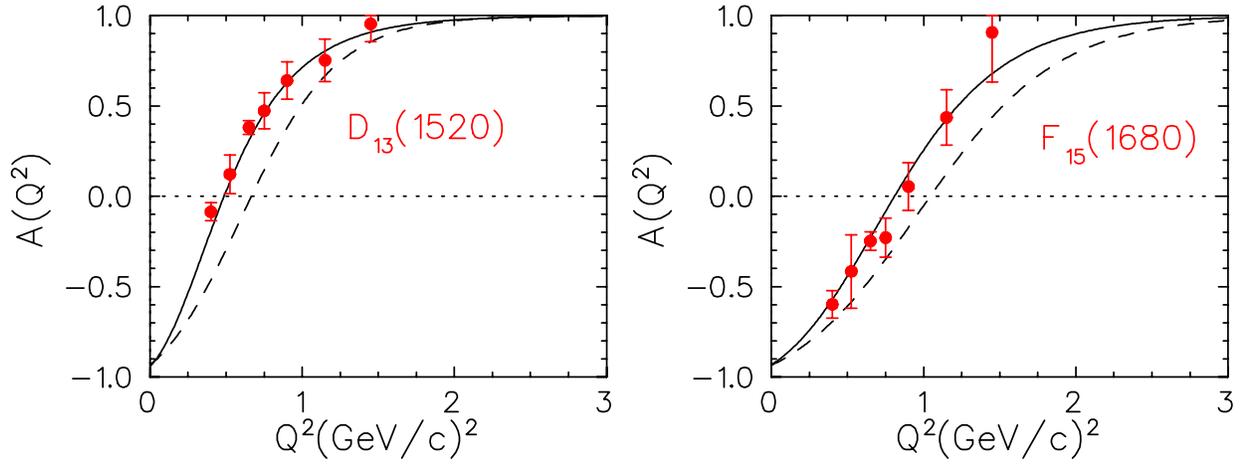, width=6.0 cm, angle=90}}
\caption{The helicity asymmetry ${\mathcal {A}}(Q^2)$ of
Eq.~(\ref{eq:5.22}) for the $D_{13}(1520)$ and $F_{15}(1680)$
resonances of the proton. The solid and dashed curves are the
super-global MAID2007 and MAID2003 solutions, respectively. The data
are the results of our single-$Q^2$ fits to the CLAS data (red full
circles, see Table~\ref{database} for references).} \label{fig:asym}
\end{figure*}

In Fig.~\ref{aznar} we compare our results to those of Aznauryan
\emph{et al.}~\cite{Azna05}
who used a similar set of the CLAS data in the second resonance region. Our
super-global solutions (solid lines) agree generally quite well with the
JLab-Yerevan analysis, which was performed with both an isobar model and
dispersion analysis.
\begin{table}[htbp]
\caption{The proton parameters for the higher resonances: $\alpha$ and $\beta$
as defined by Eq.~(\ref{eq:5.21}), in units GeV$^{-2}$,  and $S_{1/2}^0$, the
longitudinal amplitude at $Q^2=0$, in units 10$^{-3}$ GeV$^{-1/2}$. The values
for the transverse amplitudes $A_{1/2,3/2}^0$ are determined by the real photon
physics and listed in Table~\ref{tab:helicity_p}.} \label{tab:4}
\begin{center}
\begin{tabular}{|c|c|c|c|c|}
\hline
\multicolumn{1}{|c|}{} & $A_{1/2}$ &  $A_{3/2}$ &   $S_{1/2}$  & $S_{1/2}^0$   \\
proton& $\alpha$\hspace{5mm}$\beta$ & $\alpha$\hspace{5mm}$\beta$ & $\alpha$\hspace{5mm}$\beta$ &  \\
\hline
$D_{13}(1520)$  & 7.77\hspace{2mm}1.09 & 0.69\hspace{2mm}2.10 & 4.19\hspace{2mm}3.40 & -63.6 \\
$S_{11}(1535)$  & 1.61\hspace{2mm}0.70 & ---\hspace{2mm}--- & 23.9\hspace{2mm}0.81 & -2.0 \\
$S_{31}(1620)$  & 1.86\hspace{2mm}2.50 & ---\hspace{2mm}--- & 2.83\hspace{2mm}2.00 & 16.2 \\
$S_{11}(1650)$  & 1.45\hspace{2mm}0.62 & ---\hspace{2mm}--- & 2.88\hspace{2mm}0.76 & -3.5 \\
$D_{15}(1675)$  & 0.10\hspace{2mm}2.00 & 0.10\hspace{2mm}2.00 & 0.00\hspace{2mm}0.00 & 0.00 \\
$F_{15}(1680)$  & 3.98\hspace{2mm}1.20 & 1.00\hspace{2mm}2.22 & 3.14\hspace{2mm}1.68 & -44.0 \\
$D_{33}(1700)$  & 1.91\hspace{2mm}1.77 & 1.97\hspace{2mm}2.20 & 0.00\hspace{2mm}0.00 & 0.00 \\
$P_{13}(1720)$  & 1.89\hspace{2mm}1.55 & 16.0\hspace{2mm}1.55 & 2.46\hspace{2mm}1.55 & -53.0 \\
\hline
\end{tabular}
\end{center}
\end{table}
\begin{table}[htbp]
\caption{The neutron parameters for the higher resonances. The values for the
transverse amplitudes $A_{1/2,3/2}^0$ are given in Table~\ref{tab:helicity_n}.
Further notation as in Table~\ref{tab:4}.} \label{tab:5}
\begin{center}
\begin{tabular}{|c|c|c|c|c|}
\hline
\multicolumn{1}{|c|}{} & $A_{1/2}$ &  $A_{3/2}$ &   $S_{1/2}$  & $S_{1/2}^0$   \\
neutron& $\alpha$\hspace{5mm}$\beta$ & $\alpha$\hspace{5mm}$\beta$ & $\alpha$\hspace{5mm}$\beta$ &  \\
\hline
$D_{13}(1520)$  & -0.53\hspace{2mm}1.55 & 0.58\hspace{2mm}1.75 & 15.7\hspace{2mm}1.57 & 13.6 \\
$S_{11}(1535)$  & 4.75\hspace{2mm}1.69 & ---\hspace{2mm}--- & 0.36\hspace{2mm}1.55 & 28.5 \\
$S_{11}(1650)$  & 0.13\hspace{2mm}1.55 & ---\hspace{2mm}--- & -0.50\hspace{2mm}1.55 & 10.1 \\
$D_{15}(1675)$  & 0.01\hspace{2mm}2.00 & 0.01\hspace{2mm}2.00 & 0.00\hspace{2mm}0.00 & 0.00 \\
$F_{15}(1680)$  & 0.00\hspace{2mm}1.20 & 4.09\hspace{2mm}1.75 & 0.00\hspace{2mm}0.00 & 0.00 \\
$P_{13}(1720)$  & 12.7\hspace{2mm}1.55 & 4.99\hspace{2mm}1.55 & 0.00\hspace{2mm}0.00 & 0.00 \\
\hline
\end{tabular}
\end{center}
\end{table}
The following Fig.~\ref{helicity_D13_F15} displays our super-global and
single-$Q^2$ fits for the $D_{13}(1520)$ and $F_{15}(1680)$ resonances. The
figure demonstrates that (I) the helicity non-conserving amplitude $A_{3/2}$
dominates for real photons and (II) with increasing values of $Q^2$, $A_{3/2}$
drops faster than the helicity conserving amplitude $A_{1/2}$. As a consequence
the asymmetry
\begin{equation}
{\mathcal {A}}(Q^2)=\frac{\mid A_{1/2} \mid^2 - \mid A_{3/2} \mid^2} {\mid
A_{1/2} \mid^2 + \mid A_{3/2} \mid^2} \label{eq:5.22}
\end{equation}
changes rapidly from values close to $-1$ to values near $+1$ over a small
$Q^2$ range. As is seen in Fig.~\ref{fig:asym}, the asymmetry crosses the zero
line at $Q^2\approx 0.5$~GeV$^2$ for the $D_{13}(1520)$ resonance and at
$Q^2\approx 0.8$~GeV$^2$ for the $F_{15}(1680)$. As a comparison,
the asymmetry $A$ for the $\Delta(1232)$ resonance is practically
constant over this $Q^2$ range with a value $\approx -0.5$.
This again shows the special role of the $\Delta$ resonance,
where the helicity conservation is not observed.

\section{Conclusion}
Using the world data base of pion photo- and electroproduction and recent data
from Bates/MIT, ELSA/Bonn, MAMI/Mainz, and Jefferson Lab, we have extracted the
longitudinal and transverse helicity amplitudes of nucleon resonance excitation
for all the 4-star resonances below $W=2$~GeV. For this purpose we have
extended our unitary isobar model MAID and parameterized the $Q^2$ dependence
of the transition amplitudes. The comparison between such super-global solutions
with the corresponding single-$Q^2$ fits gives us confidence in the obtained
helicity couplings for the $P_{33}(1232)$, $P_{11}(1440)$, $S_{11}(1535)$,
$D_{13}(1520)$, and $F_{15}(1680)$ resonances, even though the model
uncertainty is still quite large, particularly for the longitudinal amplitudes.\\

For the higher 4-star and all 3-star resonances the situation is less clear.
This deplorable situation reflects the fact that a model-independent analysis
requires precision data over a large kinematical range. In some cases
double-polarization experiments will be helpful, as has already been shown for
pion photoproduction. Furthermore, charged pion electroproduction data are
needed with the same quantity and quality as for neutral pions, in order to
resolve the ambiguities in the isospin structure, in particular for the
$S_{11}$ and $S_{31}$ resonances. While we have mostly discussed the
electroproduction from proton targets, also the existing neutron data have been
analyzed. The latter are of course less abundant, and moreover no new neutron
data have been reported over the recent years. Because the isospin symmetry is
most likely on safe grounds in the resonance region, only the electromagnetic
neutron couplings with isospin $1/2$ are still lacking. In spite of the
discussed problems, we have implemented a super-global solution also for the
neutron amplitudes in MAID07.\\

Pion photo- and electroproduction are invaluable tools to study the resonance
structure of the nucleon. With the advent of the new c.w. electron
accelerators, new precision experiments have afforded a host of new data to
unravel this structure in the first and second resonance regions. In
particular, electroproduction has provided new insights in the spatial
distribution of the nucleon-resonance transition densities. However, in order
to get the whole picture, several challenges remain:
\begin{itemize}
\item {Dedicated experiments to investigate the higher energy region,
which have to include an intense study of the polarization degrees of freedom.
Experience has shown that even the physics of the $\Delta~(1232)$ requires a
full-fledged program to measure the spin observables in order to understand the
background of the non-resonating multipoles.}
\item {A fresh approach to also determine the excitation spectrum of the neutron.
As an example, the comparison of the Roper or $P_{11}~(1440)$ helicity
amplitudes for proton and neutron will shed light on the structure of this
enigmatic resonance.}
\item {The open question of the excitation spectrum in the third resonance region
and above deserves further studies in both theory and experiment. This includes
``missing'' and ``exotic'', e.g., 5-quark resonances as well as more mundane
second and third resonances in a multipole, which show up in a particular
analysis and not in another one.}
\end{itemize}
In conclusion we hope that MAID2007, just as other approaches based on
partial-wave analysis, dynamic models, coupled-channels calculations, and
dispersion theory, will contribute to settle the mentioned issues and thus to
improve our still somewhat vestigial knowledge of the nucleon's resonance
structure.\\

\centerline{\bf Acknowledgment} This work was supported by the Deutsche
Forschungsgemeinschaft through the SFB~443, by the joint project
NSC/DFG 446 TAI113/10/0-3 and  by the joint Russian-German Heisenberg-Landau program.\\
\newpage
\clearpage
%
\begin{figure*}[ht]
\centerline{\epsfig{file=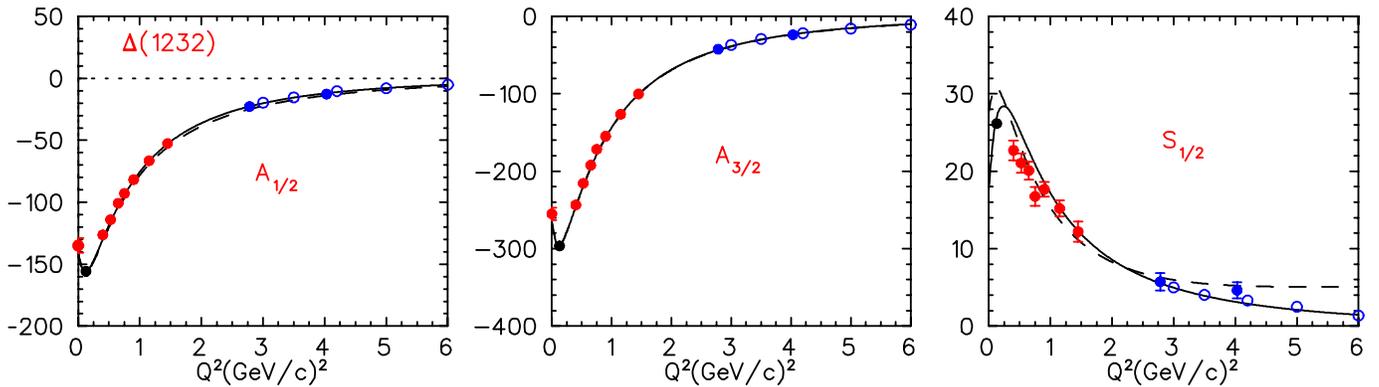, width=5.0 cm, angle=90}}
\caption{The $Q^2$ dependence of the 3 helicity amplitudes for the $\Delta(1232)$ resonance,
in units 10$^{-3}$~GeV$^{-1/2}$. The solid and dashed lines are the MAID2007
and MAID2003 super-global solutions, respectively. The data points are from our
single-$Q^2$ fits to the $\pi^0$ and $\pi^+$  CLAS data (red full and blue open
circles, see Table~\ref{database} for references), Ref.~\cite{Fro99} (blue full
circles), Ref.~\cite{Mer01} (black full circles at $Q^2=0.127$~GeV$^2$), and
Ref.~\cite{PDG06} (green full circles at $Q^2=0$).}\label{helicity_P33}
\end{figure*}
\begin{figure*}[ht]
\centerline{\epsfig{file=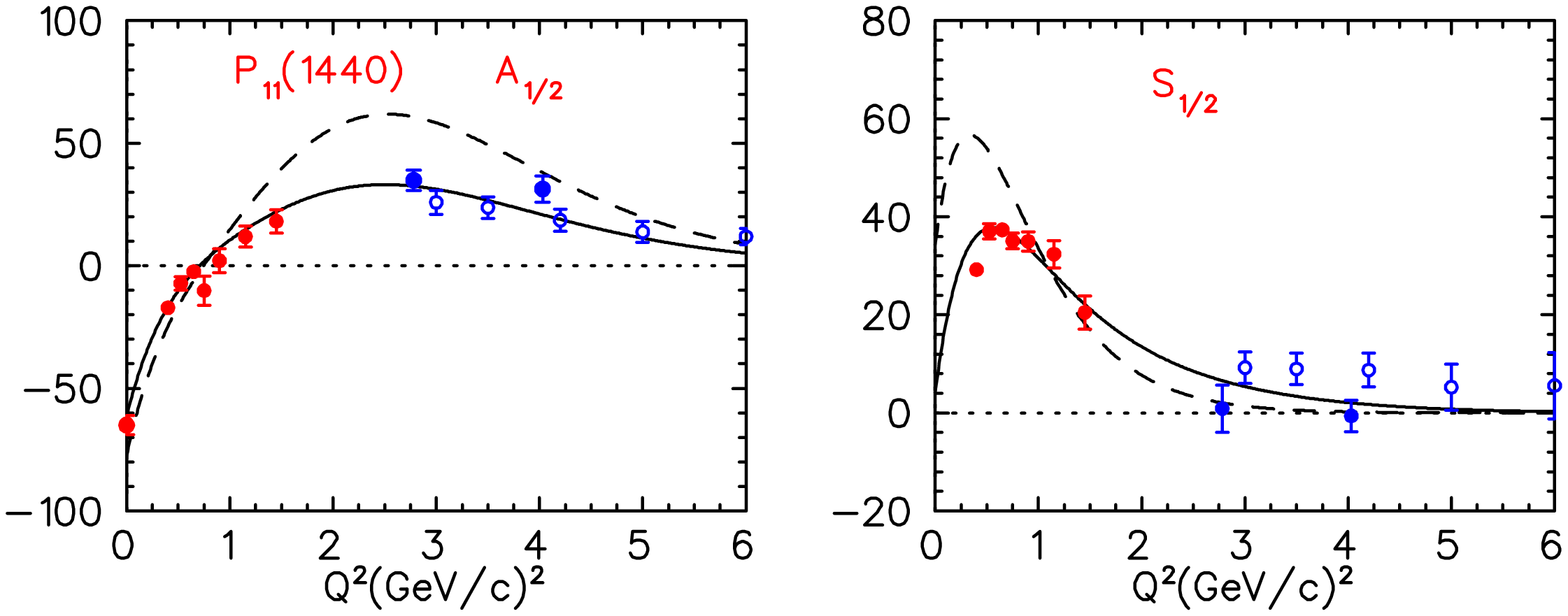,width=6.0 cm,angle=90}}
\caption{The $Q^2$ dependence of the helicity amplitudes for the $P_{11}(1440)$ resonance of
the proton. Further notation as in Fig.~\ref{helicity_P33}.}
\label{helicity_P11}
\end{figure*}
\begin{figure*}[ht]
\centerline{\epsfig{file=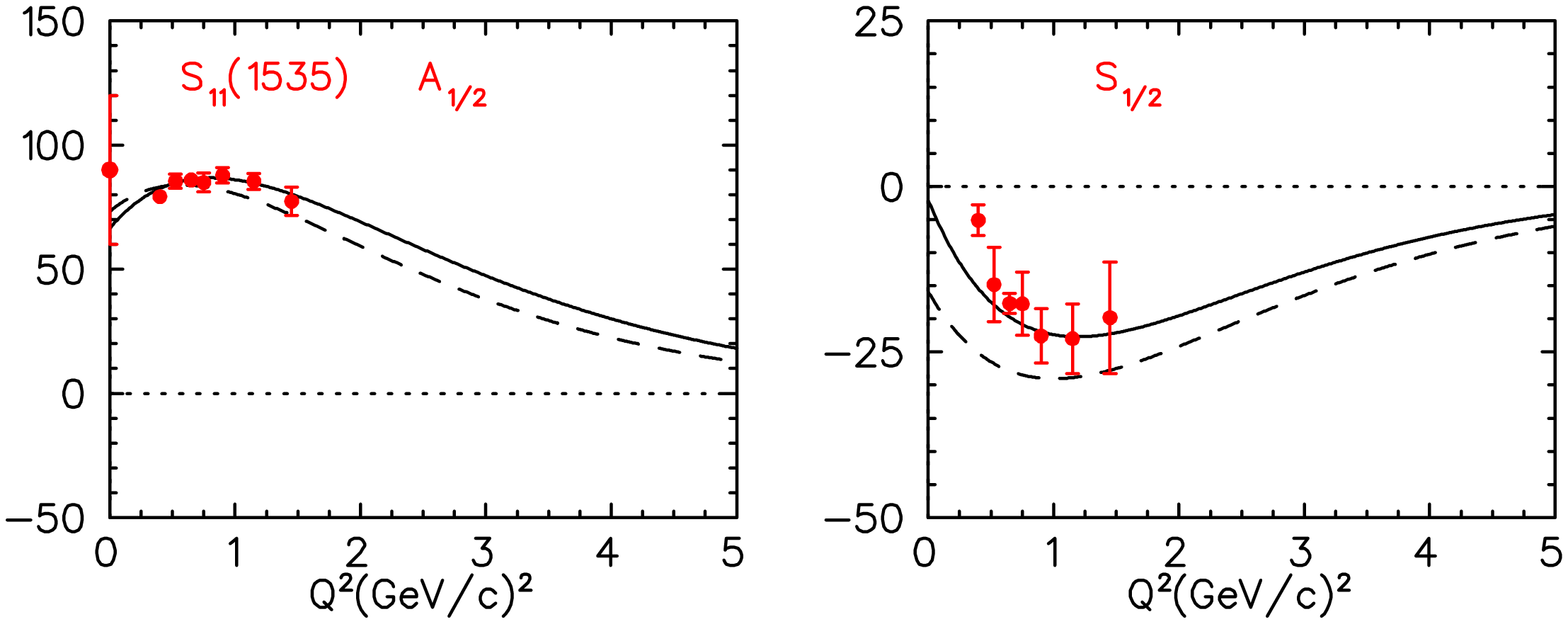,width=6.0 cm,angle=90}}
\caption{The $Q^2$ dependence of the helicity amplitudes for the $S_{11}(1535)$ resonance of
the proton. Further notation as in Fig.~\ref{helicity_P33}.}
\label{helicity_S11}
\end{figure*}
\begin{figure*}[ht]
\centerline{\epsfig{file=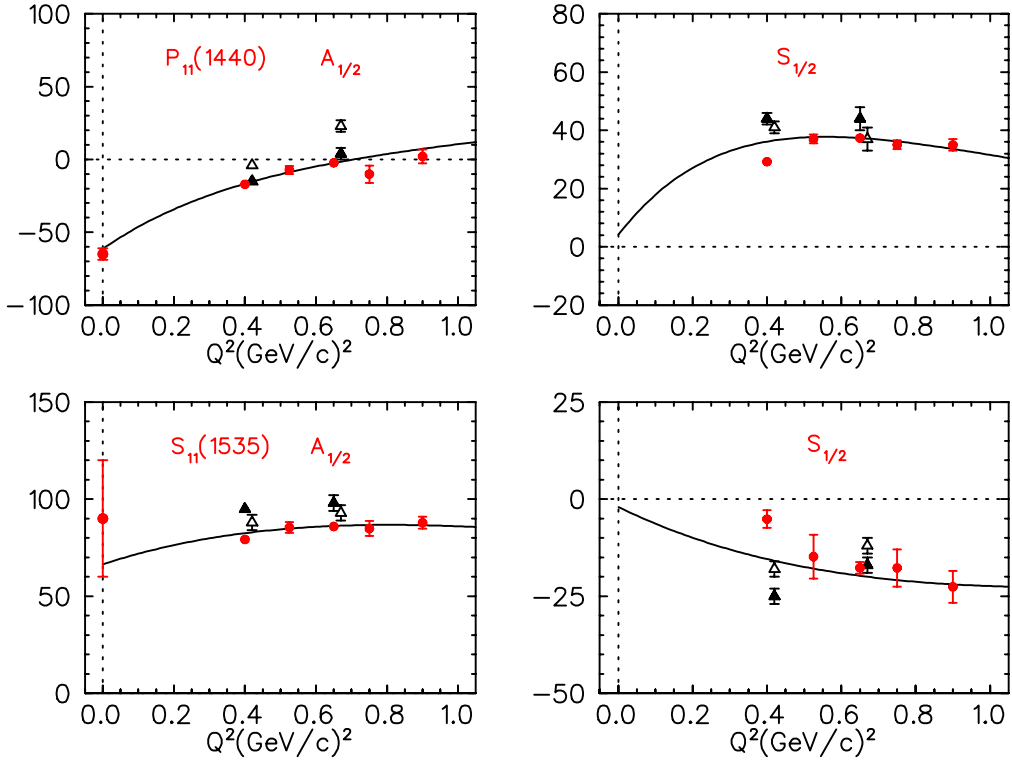, width=10 cm, angle=90} }
\caption{The $Q^2$ dependence of the helicity amplitudes for the $P_{11}(1440)$
and $S_{11}(1535)$ resonances of the proton. The MAID2007 super-global analysis
(solid lines) and the single-$Q^2$ fits (red full circles with error bars) are
compared to the results of Aznauryan~\cite{Azna05} obtained from a similar data
set within an isobar model (full triangles) and dispersion theory (open
triangles). Further notation as in Fig.~\ref{helicity_P33}.}
\label{aznar}
\end{figure*}
\begin{figure*}[ht]
\centerline{\epsfig{file=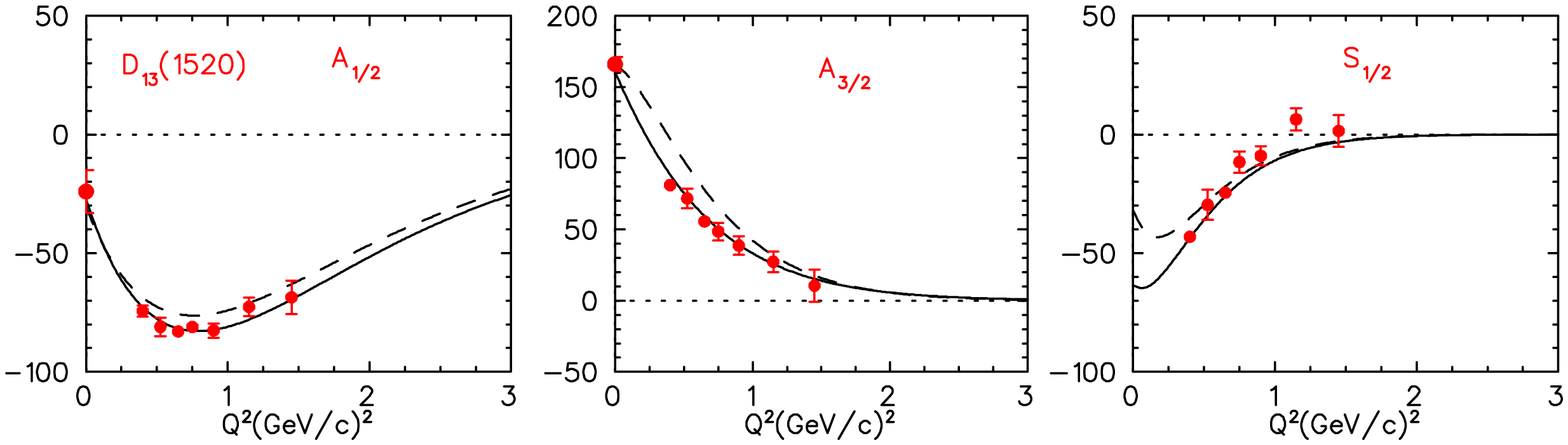, width=5. cm, angle=90}}
\vspace{0.5cm}
\centerline{\epsfig{file=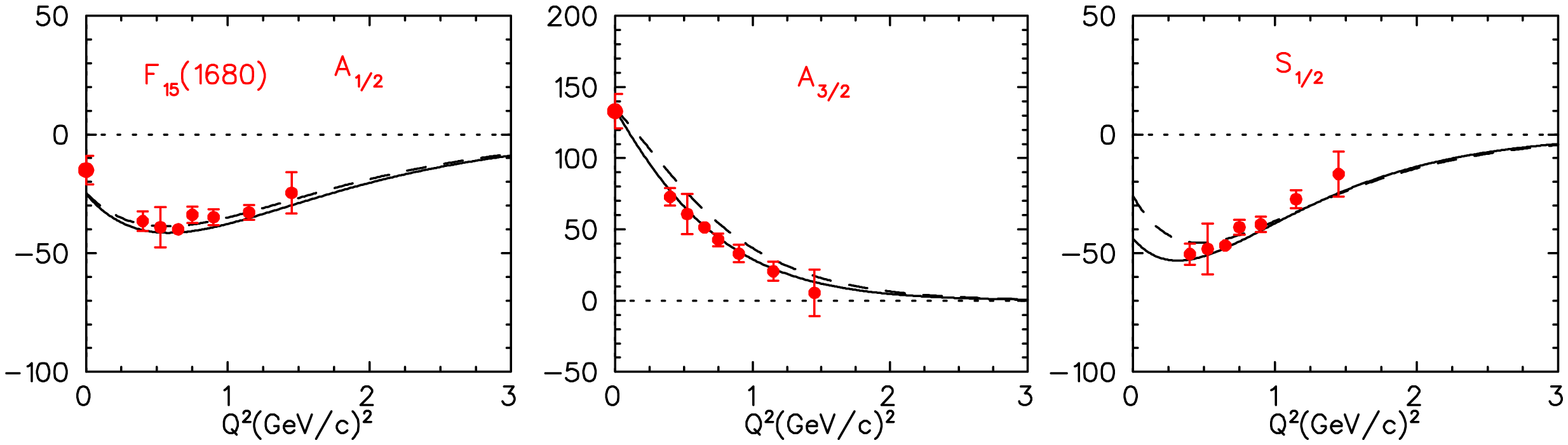, width=5. cm, angle=90} }
\caption{The $Q^2$ dependence of the helicity amplitudes for the $D_{13}(1520)$ and
$F_{15}(1680)$ resonances of the proton. Notation as in
Fig.~\ref{helicity_P33}.}
\label{helicity_D13_F15}
\end{figure*}
\end{document}